\documentclass[aip,amsmath,amssymb,preprint,author-year]{revtex4-1}
\usepackage{graphicx}
\usepackage{dcolumn}
\usepackage{bm}
\usepackage[utf8]{inputenc}
\usepackage[T1]{fontenc}
\usepackage{mathptmx}
\usepackage{etoolbox}
\usepackage{stackengine}

\makeatletter
\def\@email#1#2{%
 \endgroup
 \patchcmd{\titleblock@produce}
  {\frontmatter@RRAPformat}
  {\frontmatter@RRAPformat{\produce@RRAP{*#1\href{mailto:#2}{#2}}}\frontmatter@RRAPformat}
  {}{}
}%
\makeatother

\begin{document}
\renewcommand\stacktype{L}

\preprint{AIP/123-QED}

\title{Diagonalizing the Born-Oppenheimer Hamiltonian via Moyal
  Perturbation Theory, Nonadiabatic Corrections and Translational
  Degrees of Freedom}
\author{Robert Littlejohn}
\email{robert@wigner.berkeley.edu} \affiliation{Department of Physics,
  University of California, Berkeley, California 94720 USA}
\author{Jonathan Rawlinson}
\email{jonathan.rawlinson@manchester.ac.uk} \affiliation{School of
  Mathematics, University of Manchester, Manchester UK} 
\author{Joseph
  Subotnik} \email{subotnik@sas.upenn.edu} \affiliation{Department of
  Chemistry, University of Pennsylvania, Philadelphia, PA, USA}

\date{\today}

\newcommand{\Abar}{{\bar A}}
\newcommand{\Abarhat}{\hat{\mbox{$\Abar$}}}
\newcommand{\Ahat}{{\hat A}}
\newcommand{\Amat}{{\mathsf{A}}}
\newcommand{\angmompaper}{I}
\newcommand{\angmompaperSecV}{V}
\newcommand{\Aspace}{{\mathcal{A}}}
\newcommand{\avec}{{\mathbf{a}}}
\newcommand{\Avec}{{\mathbf{A}}}
\newcommand{\Bhat}{{\hat B}}
\newcommand{\Bspace}{{\mathcal{B}}}
\newcommand{\bra}[1]{\langle#1\vert}
\newcommand{\braket}[2]{\langle#1\vert#2\rangle}
\newcommand{\Bregion}{{\mathcal{B}}}
\newcommand{\Bregionbar}{{\overline{\mathcal{B}}}}
\newcommand{\BS}{{\mathcal{B}}}
\newcommand{\Bvec}{{\mathbf{B}}}
\newcommand{\bvec}{{\mathbf{b}}}
\newcommand{\Chat}{{\hat C}}
\newcommand{\codim}{{\mathop{\textrm{codim}}}}
\newcommand{\Complexes}{\mathbb{C}}
\newcommand{\Cbar}{{\bar C}}
\newcommand{\CS}{{\mathcal{C}}}
\newcommand{\Ctilde}{{\tilde C}}
\newcommand{\etavec}{\bm{\eta}}
\newcommand{\Fvec}{{\mathbf{F}}}
\newcommand{\Ghat}{{\hat G}}
\newcommand{\Hbar}{{\bar H}}
\newcommand{\Hbarhat}{\hat{\mbox{$\Hbar$}}}
\newcommand{\Hhat}{{\hat H}}
\newcommand{\Hspace}{{\mathcal{H}}}
\newcommand{\Ihat}{{\hat I}}
\newcommand{\Integers}{\mathbb{Z}}
\newcommand{\Ivec}{{\mathbf{I}}}
\newcommand{\iquat}{{\mathbf{i}}}
\newcommand{\jquat}{{\mathbf{j}}}
\newcommand{\Jbar}{{\bar J}}
\newcommand{\Jvec}{{\mathbf{J}}}
\newcommand{\ket}[1]{\vert#1\rangle}
\newcommand{\ketbra}[2]{\vert#1\rangle\langle#2\vert}
\newcommand{\kquat}{{\mathbf{k}}}
\newcommand{\Khat}{{\hat K}}
\newcommand{\Ktilde}{{\tilde K}}
\newcommand{\Kvec}{{\mathbf{K}}}
\newcommand{\Levels}{{I}}
\newcommand{\Lvec}{{\mathbf{L}}}
\newcommand{\mathspan}{\mathop{\rm span}}
\newcommand{\matrixelement}[3]{\langle#1\vert#2\vert#3\rangle}
\newcommand{\MOI}{{\mathsf{M}}}
\newcommand{\Ne}{{N_e}}
\newcommand{\Nn}{{N_n}}
\newcommand{\nvechat}{{\hat{\mathbf{n}}}}
\newcommand{\omegavec}{\bm{\omega}}
\newcommand{\Pbar}{{\bar{P}}}
\newcommand{\phat}{{\hat p}}
\newcommand{\Phat}{{\hat P}}
\newcommand{\phivechat}{{\hat{\bm{\phi}}}}
\newcommand{\Proj}{{\mathcal{P}}}
\newcommand{\psibar}{{\bar\psi}}
\newcommand{\Pvec}{{\mathbf{P}}}
\newcommand{\Pvecbar}{\bar{\mathbf{P}}}
\newcommand{\Pvechat}{{\hat{\mathbf{P}}}}
\newcommand{\pvec}{{\mathbf{p}}}
\newcommand{\pvecbar}{{\bar{\mathbf{p}}}}
\newcommand{\Quaternions}{\mathbb{H}}
\newcommand{\Qvec}{{\mathbf{Q}}}
\newcommand{\qvec}{{\mathbf{q}}}
\newcommand{\Reals}{\mathbb{R}}
\newcommand{\Region}{{\mathcal{R}}}
\newcommand{\Regionbar}{{\overline{\mathcal{R}}}}
\newcommand{\rhohat}{{\hat\rho}}
\newcommand{\rvec}{{\mathbf{r}}}
\newcommand{\rvechat}{{\hat{\mathbf{r}}}}
\newcommand{\Rvec}{{\mathbf{R}}}
\newcommand{\Rvecbar}{\bar{\mathbf{R}}}
\newcommand{\scalarprod}[2]{\langle #1,#2\rangle}
\newcommand{\sigmavec}{\bm{\sigma}}
\newcommand{\Sspace}{{\mathcal{S}}}
\newcommand{\Svec}{{\mathbf{S}}}
\newcommand{\svec}{{\mathbf{s}}}
\newcommand{\thetavec}{\bm{\theta}}
\newcommand{\thetavechat}{{\hat{\bm{\theta}}}}
\newcommand{\tr}{{\mathop{\textrm{tr}}}}
\newcommand{\Uhat}{{\hat U}}
\newcommand{\Vvec}{{\mathbf{V}}}
\newcommand{\xbar}{{\bar x}}
\newcommand{\xdot}{{\dot x}}
\newcommand{\xhat}{{\hat x}}
\newcommand{\xivec}{\bm{\xi}}
\newcommand{\xpecval}[1]{\langle #1\rangle}
\newcommand{\xtilde}{{\tilde x}}
\newcommand{\xvec}{{\mathbf{x}}}
\newcommand{\Xvec}{{\mathbf{X}}}
\newcommand{\Xvecbar}{\bar{\mathbf{X}}}
\newcommand{\Xvechat}{{\hat{\mathbf{X}}}}
\newcommand{\Xvectilde}{\tilde{\mathbf{X}}}
\newcommand{\ytilde}{{\tilde y}}
\newcommand{\yvechat}{{\hat{\mathbf{y}}}}
\newcommand{\Yvec}{{\mathbf{Y}}}
\newcommand{\Yvectilde}{\tilde{\mathbf{Y}}}
\newcommand{\ztilde}{{\tilde z}}
\newcommand{\Zvectilde}{\tilde{\mathbf{Z}}}

\begin{abstract}
This article describes a method for calculating higher order or
nonadiabatic corrections in Born-Oppenheimer theory and its
interaction with the translational degrees of freedom.  The method
uses the Wigner-Weyl correspondence to map nuclear operators into
functions on the classical phase space and the Moyal star product to
represent operator multiplication on those functions.  The result is a
power series in $\kappa^2$, where $\kappa =(m/M)^{1/4}$ is the usual
Born-Oppenheimer parameter.  The lowest order term is the usual
Born-Oppenheimer approximation while higher order terms are
nonadiabatic corrections.  These are needed in calculations of
electronic currents, momenta and densities.  The method was applied to
Born-Oppenheimer theory by \cite{LittlejohnWeigert93}, in a treatment
that notably produced the correction $K_{22}$ to the Born-Oppenheimer
Hamiltonian (see {\em infra}).  Recently \cite{MatyusTeufel19} have
applied an improved and more elegant version of the method to
Born-Oppenheimer theory, and have calculated the Born-Oppenheimer
Hamiltonian for multiple potential energy surfaces to order
$\kappa^6$.  One of the shortcomings of earlier methods is that the
separation of nuclear and electronic degrees of freedom takes place in
the context of the exact symmetries (for an isolated molecule) of
translations and rotations, and these need to be a part of the
discussion.  This article presents an independent derivation of the
Moyal expansion in molecular Born-Oppenheimer theory, with special
attention to the translational degrees of freedom.  We show how
electronic currents and momenta can be calculated within the framework
of Moyal perturbation theory; we derive the transformation laws of the
electronic Hamiltonian, the electronic eigenstates, and the derivative
couplings under translations; we discuss in detail the rectilinear
motion of the molecular center of mass in the Born-Oppenheimer
representation; and we show how the elimination of the translational
components of the derivative couplings leads to a unitary
transformation that has the effect of exactly separating the
translational degrees of freedom.  Our discussion is framed in terms
of dressing transformations, that is, unitary transformations that
change the physical meanings of the operators of the theory.  We show,
for example, that in the Born-Oppenheimer representation, the operator
that looks like the momentum of the nuclei actually includes, from a
physical standpoint, the momentum of the electrons.  Moreover, this
result is exact.  In regard to hybrid classical-quantum models such as
used in surface hopping calculations, we note that the theory
presented here will be useful in developing new semiclassical
approximations going beyond the Born-Oppenheimer approximation, as
well as taking into account a glaring error of existing schemes,
namely, the failure to conserve linear and angular momentum.

\end{abstract}

\maketitle

\section{Introduction}
\label{intro}

In the Born-Oppenheimer theory of molecules the Hamiltonian becomes an
infinite-dimensional matrix of nuclear operators once an electronic
basis is chosen.  Moyal perturbation theory provides a method of
block-diagonalizing this matrix to obtain effective dynamics on one or
a small number of potential energy surfaces.  This is an elegant way
of deriving Born-Oppenheimer dynamics on a small number of surfaces in
a manner that is more rigorous than the usual approaches, with other
advantages enumerated below.

By ``Moyal perturbation theory'' we refer to the use of the
Wigner-Weyl correspondence and the Moyal star product to
block-diagonalize matrices of operators.  It would be more descriptive
to call it ``Wigner-Weyl-Moyal-van~Vleck'' perturbation theory, but we
will use the shorter designation.  The Wigner-Weyl
correspondence maps quantum operators into functions on a classical
phase space, and the Moyal star product is used to represent operator
multiplication on those functions.  The Wigner-Weyl-Moyal formalism is
fairly well known, and for convenience is summarized in
Appendix~\ref{Moyal}, along with literature citations.  The
application in this article of Moyal perturbation theory to
Born-Oppenheimer problems is based on the treatment of
\cite{WeigertLittlejohn93}, with many improvements.

Moyal perturbation theory is an efficient way of obtaining higher
order or nonadiabatic corrections in Born-Oppenheimer theory.  Such
corrections are needed in cases where the lowest order contribution
vanishes, notably in the calculation of electronic currents,
densities, linear and angular momenta, and matrix elements needed for
radiative transitions.  There is an extensive literature on such
problems, including \cite{MeadMoscowitz67, Nafie83, Stephens85,
StephensLowe85, FreedmanNafie86, BuckinghamFowlerGalwas87, Nafie92,
Nafie97, Barthetal09, OkuyamaTakatsuka09, Patchkovskii12, Diestler12,
Diestler12a, Diestler13, Diestleretal13, Bredtmannetal15,
PohlTremblay16, AlbertHaderEngel17, SchauppAlbertEngel18,
SchauppEngel19, Nafie20, SchauppEngel20,
SchauppRenziehausenBarthEngel21, SchauppEngel22a, SchauppEngel22}.  In
recent years the method of exact factorization has been developed and
promoted (\cite{AbediMaitraGross10, AbediMaitraGross12, Cederbaum13,
Cederbaum14, Cederbaum15, ParasharSajeevGhosh15,
JeckoSutcliffeWoolley15, RequistGross16, Requistetal16,
CurchodAgostini17, RequistProettoGross17, AgostiniCurchod18, Lorin21,
MartinazzoBurghardt23}).  A notable application of this method is the
calculation of higher order or nonadiabatic corrections
(\cite{ScherrerVuilleumierSebastiani13, Scherreretal15,
SchildAgostiniGross16}).  Other approaches to higher order corrections
include \cite{PachukiKomasa08, Matyus18, MatyusTeufel19,Requist23}.
Of these, \cite{MatyusTeufel19, Requist23} have used a version of what
we are calling Moyal perturbation theory.

The advantages of Moyal perturbation theory, as we see them,  for
accessing nonadiabatic corrections and for related topics are the
following.  The method is clean and elegant, allowing one to visualize
the structure of the theory as a systematic procedure that can be
carried to any order in the Born-Oppenheimer parameter
$\kappa=(m/M)^{1/4}$.  The method automatically incorporates terms in
the effective Hamiltonian, such as $K_{22}$ that is discussed below,
that should be there but are usually neglected.  The method
automatically accommodates large-amplitude motions, such as occur in
scattering, isomerization and photoexcitation problems, and is not
restricted to small-amplitude expansions about an equilibrium.  It
works in simple cases such as motion on a single surface in the
electrostatic model, as well as in generalizations including multiple
surfaces, fine-structure and spin.  Many of the first order results
presented in this article, for example, the first order corrections to
the wave function and formulas for electronic momenta and currents,
are known in the case of the electrostatic model on a single surface,
but generalizations to multiple surfaces and spin are incomplete.
These are of current interest, for example, in hybrid
classical-quantum models used in surface hopping
(\cite{MannouchRichardson23, RunesonManolopoulos23, Bianetal21,
Bianetal23}).  The method interacts well with the exact symmetries of
the system, including overall translations and rotations.  The case of
rotational invariance was treated by
\cite{LittlejohnRawlinsonSubotnik23}, and translational invariance is
covered in this article.  The method allows one to make simple
statements about the conservation of linear and angular momentum that
are valid to all orders in $\kappa$.  The method leads to a
perspective in which Born-Oppenheimer theory consists of a series of
dressing transformations, that is, unitary transformations that change
the physical meanings of the variables employed, for example,
``dressing'' nuclear dynamical variables to include the effects of the
electrons.  The notion of the Born-Oppenheimer transformation as a
dressing transformation was introduced by \cite{Cederbaum04}, and we
have developed the idea considerably.  We feel that it makes an
important contribution to perspectives on Born-Oppenheimer theory.

For example, in a recent study (\cite{LittlejohnRawlinsonSubotnik23})
we showed that in the usual Born-Oppenheimer representation the
operator that looks like the orbital angular momentum of the nuclei
actually includes the orbital angular momentum of the electrons.  This
is in the electrostatic model; in fine-structure models the electron
spin is included as well.  These statements are exact in the sense of
being valid to all orders of the Born-Oppenheimer expansion parameter
$\kappa$.  For another example, in this article we display (in
subsection~\ref{eliminatingFCM}) a unitary transformation that has the
effect of separating the translational degrees of freedom, after which
the the operator that looks like the nuclear center of mass is
actually, from a physical standpoint, the molecular center of mass,
including the electrons.

Finally, there are questions regarding three steps that may be
distinguished in the Born-Oppenheimer theory of isolated molecules,
namely, the separation of the translational degrees of freedom, the
separation of the rotational degrees of freedom, and the expansion in
the Born-Oppenheimer parameter $\kappa$.  Note that the third step
amounts to the separation of the nuclear and electronic degrees of
freedom.  For example, there is the question of the advantages and
disadvantages of carrying out these steps in various orders.  We feel
that treating the third step by Moyal perturbation theory clarifies
the analysis of this question.  

The commutativity of the first and third steps (the separation of
translations and the $\kappa$-expansion) is explicitly addressed in
this article.  It is sometimes stated that the Born-Oppenheimer
approximation violates the separation of translational degrees of
freedom (\cite{Cederbaum13}), but this does not mean that
translational invariance or the associated conservation laws are lost,
simply that they take on different forms in the different
representations.  Some of these are very interesting.

Moyal perturbation theory was first developed by \cite{Blount62}, who
applied it to various problems in condensed matter physics as well as
to the Foldy-Wouthuysen transformation
(\cite{Blount62a,FoldyWouthuysen50,BjorkenDrell64}), the latter of
which is the block-diagonalization of the Dirac Hamiltonian.
The method was rediscovered by
\cite{LittlejohnFlynn91,LittlejohnFlynn91a}, who applied it to the WKB
theory of multi-component wave equations and to several other problems
in adiabatic or semiclassical theory, including spin-orbit coupling
and multi-dimensional Landau-Zener transitions
(\cite{LittlejohnFlynn92,LittlejohnFlynn93}).  Later
\cite{LittlejohnWeigert93} applied the method to the adiabatic motion
of a neutral spinning particle in an inhomogeneous magnetic field,
otherwise a Stern-Gerlach apparatus.  That problem is formally almost
identical to the Born-Oppenheimer approximation in molecules, in which
the spin of the neutral particle corresponds to the electronic degrees
of freedom, while the spatial variables correspond to the nuclear
degrees of freedom.  A notable outcome of that work was the production
of a second-order contribution to the Hamiltonian, what we call
$K_{22}$ in this article.  This term had been missing in the analysis
by other authors (\cite{AharonovStern92}) of the Stern-Gerlach problem.  

Later \cite{WeigertLittlejohn93} applied Moyal theory to the
Born-Oppenheimer approximation in molecules, producing among other
things the contribution $K_{22}$ to the Hamiltonian.  This term is
notable for the fact that for large-amplitude motions it is of the
same order of magnitude as the well known diagonal Born-Oppenheimer
correction, but it has received much less attention than the latter.

Later \cite{Panatietal02} and \cite{Teufel03} developed the functional
analysis of Moyal perturbation theory, making several applications and
innovations.  One of the latter is the breaking of the problem into
two parts, the computation of the formally invariant subspace and the
representation of the dynamics on that subspace.  Another is the use
of operator-valued Weyl symbols, which leads to a compact and largely
coordinate- and basis-independent formalism.  This formalism was
applied by \cite{MatyusTeufel19} to Born-Oppenheimer theory, producing
an expansion of the Born-Oppenheimer Hamiltonian on one or a few
potential energy surfaces out to order $\kappa^6$ (described as
``third order'').

The term $K_{22}$ is discussed more fully in Sec.~\ref{K22discussion}
below.  Its history and its role in Born-Oppenheimer theory have been
discussed by \cite{Matyus18}, whose citations place its discovery in
the 1960's, although several of the earlier references are specific to
small vibrations about an equilibrium in diatomic or triatomic
molecules.  To put this term into general language, applying to a
molecule with any number of atoms, it is necessary to work in lab or
Cartesian coordinates.  In addition, the importance of this term grows
with energy, and it is more important for large-amplitude motions than
for small vibrations.  Therefore to treat this term in generality it
is necessary not to expand about an equilibrium position.  As far as
we can see the first authors to derive this term in such generality
were \cite{Moodyetal89}, who used a path-integral formalism for
treating the Born-Oppenheimer approximation and its higher order
corrections and who noted that this term modifies the kinetic-energy
metric, making it non-Euclidean at second order.  In addition, there
is the independent discovery by \cite{Goldhaber05}, who derived
$K_{22}$ in a discussion that links Berry's phase, geometry, and
averaging techniques in quantum field theory.  In this article we will
show that the contribution $K_{22}$ to the Hamiltonian is necessary in
the Born-Oppenheimer representation to account for the proper
(rectilinear) evolution of the center of mass of the molecule.

The original treatments of Born-Oppenheimer theory
(\cite{BornOppenheimer27,BornHuang54}) and large parts of the vast
literature on this subject envision an expansion in powers of
$\kappa=(m/M)^{1/4}$, in which the nuclear displacements from an
equilibrium position are small, of order $\kappa$, while nuclear
momenta are of order $\kappa^{-1}$.  We call this the
``small-amplitude'' ordering.  It was pointed out by
\cite{LittlejohnFlynn92}, however, that large-amplitude motions such
as occur in scattering theory required a different ordering, in which
nuclear momenta are of order $\kappa^{-2}$.  Later \cite{Mead06}
discussed the large-amplitude ordering in detail and its contrast to
the small-amplitude ordering.  Another analysis has been given
recently by \cite{MatyusTeufel19}, and we have returned to the subject
in Sec.~\ref{BOordering} below.  To summarize the situation, Moyal
perturbation theory is an expansion in powers of $\kappa^2$, it
accommodates large-amplitude motions, and it is equivalent to a
semiclassical theory in the nuclear degrees of freedom, that is, it is
an expansion in the ``nuclear $\hbar$'' (but not the electronic).

The outline of this paper is as follows.  Section~\ref{setup} contains
the setup of the problem, including notation, the relevant
Hamiltonians, the electronic bases that will be used (adiabatic,
diabatic, etc.), the coordinate transformations that bring about the
exact elimination of the translational degrees of freedom, and the
several representations we will use.  In this article the word
``representation'' is used in a particular sense, implying a space of
wave functions, a mapping between physical states (for example, the
ground state of the molecule or an eigenstate of energy and angular
momentum) and wave functions in that space, and a mapping between
physical observables (energy, momentum, etc.) and specific linear
operators that act on those wave functions.  Different representations
are connected by unitary operators.

In Sec.~\ref{singlesurface} we describe Moyal perturbation theory in a
manner that is hopefully as painless as possible, based on the
approach of \cite{LittlejohnFlynn91, LittlejohnWeigert93,
WeigertLittlejohn93}.  Much the same territory has been covered by
\cite{MatyusTeufel19}, based on the formalism of
\cite{Panatietal02,Teufel03}.  In comparison, our formalism is more
coordinate-based and less abstract.  Our discussion of the relation
between the generator of the dressing transformation and the usual
Landau-Zener transmission probability seems to be new, as is our
presentation in subsection~\ref{transfwavefun} of the second order,
nonadiabatic corrections to the wave function.  For simplicity in this
section we treat only single-surface problems.

In Sec.~\ref{generalizationapplication} we generalize parts of the
single-surface calculation in Sec.~\ref{singlesurface} to multiple
surfaces, and we review how the Moyal theory is applied to the
calculation of electronic momenta and currents.   We offer an approach
in which the higher order corrections are obtained, not by finding
corrections to the wave function, but by transforming (``dressing'')
the operators. 

In Sec.~\ref{translationaldofsec} we begin by treating a number of
issues regarding the translational degrees of freedom, in a manner
similar to our treatment of rotations in
\cite{LittlejohnRawlinsonSubotnik23}.  In particular, we establish
phase conventions for the electronic basis states, both adiabatic and
diabatic, in effect showing how they transform under translations.
We also give the transformation laws for the electronic Hamiltonian
and the derivative couplings under translations.  These results are
placed in the geometrical context of the translational fiber bundle,
which we explain as a certain decomposition of the nuclear
configuration space.   A notable consequence of this work is that in
the Born-Oppenheimer representations, both the original and all the
dressed versions of it, the operator that looks like the total
momentum of the nuclei actually includes, from a physical standpoint,
the momentum of the electrons.   A similar statement regarding angular
momentum was a principal point of
\cite{LittlejohnRawlinsonSubotnik23}.

We then address several topics concerning the translational degrees of
freedom and Moyal perturbation theory.  First we study the molecular
center of mass (including the electrons) and show how its uniform,
rectilinear motion is expressed in the dressed Born-Oppenheimer
representation.  It turns out that the term $K_{22}$ in the
Hamiltonian is required for this to work out right.  Next we study the
translational components of the electronic current and the derivative
couplings, in which the resolvent operator can be eliminated and
results obtained in closed form.  The existence of the translational
components of the derivative couplings leads to some apparent
paradoxes, which we resolve by engineering a unitary transformation
that removes them.  This transformation is revealed by the use of
Moyal perturbation theory, but ultimately is expressed in closed form
(not as a power series).  This produces an operator that has the
effect of exactly separating the translational degrees of freedom.  As
far as we can tell this is a new perspective on the separation of the
translational degrees of freedom, in that the procedure can be seen as
a dressing transformation. 

Finally, in Sec.~\ref{conclusions} we present some conclusions and
plans for future work.

A note on notation and terminology.  Starting in
subsection~\ref{generatorG1} we use hats on nuclear operators, for
example $\Pvechat_\alpha$, to distinguish them from classical
quantities or $c$-numbers.  See the introduction to
Sec.~\ref{singlesurface}.  The Hamiltonian $H_{\rm BO}$ is the usual
(scalar) Born-Oppenheimer Hamiltonian on a single surface; see
(\ref{HBOdef}).  An overbar is used to indicate fully dressed
operators, for example the Hamiltonian $\Hbar$ (see (\ref{Hbardef}))
or $\pvecbar_i$.  The scalar Hamiltonian $K$ is the diagonal element
of $\Hbar$, the effective Hamiltonian that replaces $H_{\rm BO}$ (see
above (\ref{K22def})).  In this article, ``diabatic'' means what is
sometimes called ``quasi-diabatic,'' that is, there is no implication
that the diabatic basis eliminates the derivative couplings (which is
usually impossible anyway).

\section{Preliminaries}
\label{setup}

In this section we describe the setup of the problem and establish
notation.  We adopt the electrostatic model for the electronic
Hamiltonian (\ref{Hedef}).  For simplicity, we treat the nuclei as
distinguishable, spinless particles.

\subsection{Hamiltonians and Notation}
\label{Hamiltonians}

We consider a molecule with $N$ nuclei.  The molecular Hamiltonian in
the lab frame is
\begin{equation}
	H = \sum_{\alpha=1}^N
	\frac{\Pvec_\alpha^2}{2M_\alpha} 
	+H_e(x;\rvec,\pvec),
	\label{Hmoldef}
\end{equation}
where the electronic Hamiltonian is
\begin{equation}
	H_e(x;\rvec,\pvec) = \sum_{i=1}^{N_e} 
	\frac{\pvec_i^2}{2m_e} 
	+ V_{\rm Coul}(\Xvec,\rvec).
	\label{Hedef}
\end{equation}
Here $N_e$ is the number of electrons; $\Xvec_\alpha$,
$\Pvec_\alpha=-i\hbar\nabla_\alpha$ and $M_\alpha$ are respectively
the nuclear positions, momenta and masses; and $\rvec_i$ and
$\pvec_i=-i\hbar\nabla_i$ are respectively the electron positions and
momenta.  Gradient operators are $\nabla_\alpha=\partial
/\partial\Xvec_\alpha$ and $\nabla_i=\partial /\partial\rvec_i$.  All
positions and momenta are taken with respect to a lab frame.  Indices
$\alpha$, $\beta$ label nuclei and range over $1,\ldots,N$, while
indices $i$, $j$ label electrons and range over $1,\ldots,N_e$.  The
electron mass is $m_e$.  The bare symbols $\Xvec$ and $\rvec$ stand
for the collections $(\Xvec_1,\ldots,\Xvec_N)$ and
$(\rvec_1,\ldots,\rvec_{N_e})$, respectively, and similarly for the
momenta $\Pvec$ and $\pvec$. The potential $V_{\rm Coul}$ contains all
the Coulomb interactions among all the particles, nuclei and
electrons.  The nuclear configuration space is the space upon which
the $\Xvec_\alpha$ are coordinates; it is $\Reals^{3N}$.  We denote a
point of nuclear configuration space by the symbol $x$, whose
coordinates are $\Xvec=(\Xvec_1,\ldots,\Xvec_N)$.  In our usage, $x$
and $\Xvec$ mean almost the same thing ($x$ is the point, $\Xvec$, the
coordinates of that point).

\subsection{The Adiabatic  Basis and the Working Basis}
\label{elecbases}

We define the ``adiabatic basis'' as the energy eigenbasis of the
electronic Hamiltonian,
\begin{equation}
	H_e(x)\,\ket{ax;k} = \epsilon_k(x)\,\ket{ax;k}.
	\label{axkdef}
\end{equation}
The adiabatic basis vectors in ket language are $\ket{ax;k}$, where
$a$ stands for ``adiabatic,'' $x$ indicates the parametric dependence
on the nuclear configuration, and $k$ is the quantum number (a
sequencing number for energy eigenvalues).  In (\ref{axkdef}) we write
simply $H_e(x)$ for the electronic Hamiltonian, indicating the
$x$-dependence of the operator, which elsewhere is written
$H_e(x;\rvec,\pvec)$.  The energy eigenvalues themselves are
$\epsilon_k(x)$.  The energy eigenfunctions in wave function language
will be denoted by
\begin{equation}
	\phi_{ak}(x;\rvec)=\braket{\rvec}{ax;k}.
	\label{phiakdef}
\end{equation}

Although we shall treat the index $k$ notationally as if it were
discrete, in fact above a threshold the spectrum of $H_e(x)$ becomes
continuous and formal sums over $k$ that employ the adiabatic basis
must be interpreted as discrete sums over the discrete spectrum plus
integrals over the continuous spectrum.  In addition, one should note
that the continuum threshold is a function of $x$.

In the following we shall make use of a basis of electronic states
that we call the ``working basis,'' which we denote simply by
$\ket{x;k}$ (without the $a$).   It is defined to be either the
adiabatic basis or else a substitute, which will be defined as the
need arises.  We define the working basis wave functions by
\begin{equation}
	\phi_k(x;\rvec)=\braket{\rvec}{x;k}
	\label{phikdef}
\end{equation}
(again, without the $a$).

We choose the phase conventions of the adiabatic basis states so that
these states are invariant under time reversal. We also assume that the
coefficients of the linear transformation taking us from the adiabatic
basis to the working basis are real, so that the working basis states
are also invariant under time reversal.  In the electrostatic model,
this just means that the basis wave functions $\phi_{ak}(x;\rvec)$ or
$\phi_k(x;\rvec)$ are real.  

Here and below when we use bra-ket notation for scalar products and
matrix elements, it is understood that we are integrating only over
the electronic coordinates $\rvec$, not the nuclear coordinates,
except as noted.  For example, we have the orthonormality relations,
\begin{equation}
	\braket{x;k}{x;l}=\delta_{kl}.
	\label{orthonorm}
\end{equation}

We introduce the derivative couplings,
\begin{equation}
	\Fvec_{\alpha;kl}(x) = \matrixelement{x;k}{\nabla_\alpha}
	{x;l}=-(\nabla_\alpha\bra{x;k})\ket{x;l},
	\label{Fkldef}
\end{equation}
where the alternative form follows from the orthogonality relations
(\ref{orthonorm}).  The derivative couplings are defined to the extent
that the basis states $\ket{x;k}$ are differentiable with respect to
$x$.  The orthonormality relations (\ref{orthonorm}) imply that
$\Fvec_{\alpha;kl}(x)$, regarded as a matrix in $kl$, is
anti-Hermitian,
\begin{equation}
  \Fvec_{\alpha;kl}(x) = -\Fvec^*_{\alpha;lk}(x).
  \label{FisantiHerm}
\end{equation}
In addition, the invariance of the basis vectors under time reversal
implies that $\Fvec_{\alpha;kl}(x)$ is real, so that as a matrix in
$(kl)$, it is real and antisymmetric.  Thus, the derivative couplings
vanish on the diagonal,
\begin{equation}
	\Fvec_{\alpha;kk}(x)=0.
	\label{Fdiagzero}
\end{equation}

\subsection{Two Representations}
\label{tworepresentations}

We introduce two representations for the quantum state of the
molecule, the ``molecular'' and the ``Born-Oppenheimer.''  These are
described in more detail in Sec.~\angmompaperSecV{} of
\cite{LittlejohnRawlinsonSubotnik23}.  In the molecular 
representation the state of the molecule is represented by a wave
function $\Psi(\Xvec,\rvec)$ that depends on the coordinates of both
nuclei and electrons. In the Born-Oppenheimer representation the state
of the molecule is represented by an infinite-dimensional vector of
purely nuclear wave functions, $\psi_k(\Xvec)$, indexed by $k$ and
defined by
\begin{equation}
	\Psi(\Xvec,\rvec)=\sum_k \psi_k(\Xvec)\,\phi_k(\Xvec;\rvec).
	\label{psikdef}
\end{equation}
The Born-Oppenheimer representation depends on the choice made for the
working basis, and, in particular, on the phase conventions of that
basis.  We denote the association between the two representations of
the state of the molecule by
\begin{equation}
	\Psi(\Xvec,\rvec) \longleftrightarrow \psi_k(\Xvec),
	\label{tworepcorresp}
\end{equation}
where the notation $\longleftrightarrow$ is a reminder that the
relationship is one-to-one and that no information is lost by
transforming to the Born-Oppenheimer representation.  For
time-dependent problems both $\Psi$ and $\psi_k$ depend on $t$, but
the basis kets $\ket{x;k}$ or wave functions $\phi_k(\Xvec;\rvec)$ do
not.

Similarly, let $A$ be an operator acting in the molecular representation,
say, $\Psi^{\,\prime}(\Xvec,\rvec) = (A\Psi)(\Xvec,\rvec)$, and let
$\Psi\longleftrightarrow\psi_k$ and
$\Psi^{\,\prime}\longleftrightarrow\psi^{\,\prime}_k$.   Then in the Born-Oppenheimer
representation $A$ is represented by an infinite-dimensional matrix of
operators $A_{kl}$ that act on nuclear wave functions, such that
\begin{equation}
	\psi^{\,\prime}_k(\Xvec) = \sum_l (A_{kl}\,\psi_l)(\Xvec).
	\label{Akldef}
\end{equation}
We denote the association between these two representations of the
operator by $A\longleftrightarrow A_{kl}$.  The components $A_{kl}$ of
the operator in the Born-Oppenheimer representation are purely nuclear
operators, that can be thought of as functions of $\Xvec$ and
$\Pvec$.

Several examples of this association are discussed in
\cite{LittlejohnRawlinsonSubotnik23}, including the following.  A 
purely multiplicative operator in the molecular representation
depending only on the nuclear coordinates $\Xvec_\alpha$, call it
$f(x)$, corresponds, in the Born-Oppenheimer representation, to a
multiple of the identity matrix, that is, $f(x)\longleftrightarrow
f(x)\,\delta_{kl}$.  In particular, this applies to the nuclear
coordinates, the components of $\Xvec_\alpha$.  A purely electronic
operator in the molecular representation, that is, a function of
$\rvec$ and $\pvec$, corresponds to a purely multiplicative operator
in the Born-Oppenheimer representation, that is, a matrix whose matrix
elements are functions of $\Xvec$ only.  These are nothing but the
matrix elements of the original operator in the working basis.  For
example, in the case of the momentum $\pvec_i$ of one of the
electrons, we have
$\pvec_i\longleftrightarrow\matrixelement{x;k}{\pvec_i}{x;l}$.  This
also applies to electronic operators that are parameterized by $x$,
such as the electronic Hamiltonian.  We define
\begin{equation}
	\matrixelement{x;k}{H_e(x)}{x;l} = W_{kl}(x),
	\label{Wkldef}
\end{equation}
so that $H_e(x) \longleftrightarrow W_{kl}(x)$.  Note that if the
working basis is the adiabatic basis (at least for some range of
indices $k$, $l$), then
\begin{equation}
	W_{kl}(x)=\epsilon_k(x)\,\delta_{kl}.
	\label{Wkladiab}
\end{equation}
The momentum $\Pvec_\alpha=-i\hbar\,\nabla_\alpha$, which in the
molecular representation represents the kinetic momentum of nucleus
$\alpha$, is transformed to the Born-Oppenheimer representation
according to
\begin{equation}
	\Pvec_\alpha \longleftrightarrow \Pvec_\alpha\,\delta_{kl}
	-i\hbar\,\Fvec_{\alpha;kl}(x).
	\label{Pvecalphareps}
\end{equation}

Equation~(\ref{Pvecalphareps}) illustrates an important theme of this
article.   The operator $\Pvec_\alpha$ is the differential operator
$-i\hbar\nabla_\alpha$ in all representations, but it only represents
physically  the kinetic momentum of nucleus $\alpha$ in the molecular
representation.   In the Born-Oppenheimer representation, there is a
correction term, as shown (which, it turns out, is related  to the
electronic momentum).  

Now suppose that we have operators $A$, $B$ and $C$, and that
$A\longleftrightarrow A_{kl}$, $B\longleftrightarrow B_{kl}$ and
$C\longleftrightarrow C_{kl}$.  Then we have some theorems.  First, if
$B=A^\dagger$, then $B_{kl} = (A_{lk})^\dagger$.  Next, if $C=AB$,
then
\begin{equation}
	C_{kl} = \sum_p A_{kp}\,B_{pl}.
	\label{ABmult}
\end{equation}
Next, if $\Psi_1 \longleftrightarrow \psi_{1k}$ and $\Psi_2
\longleftrightarrow \psi_{2k}$, then
\begin{equation}
	\braket{\Psi_1}{\Psi_2}_{Xr} =
	\sum_k \braket{\psi_{1k}}{\psi_{2k}}_X
	\label{Psibraketrule}
\end{equation}
where the subscripts $X$, $Xr$ indicate the variables that are
integrated over in taking the scalar product.  (If no subscript is
given, integration over $\rvec$ alone is implied, as in
(\ref{orthonorm}).)  Finally, matrix elements of operators transform
between the two representations according to
\begin{equation}
	\matrixelement{\Psi_1}{A}{\Psi_2}_{Xr} =
	\sum_{kl} \matrixelement{\psi_{1k}}{A_{kl}}{\psi_{2l}}_X.
	\label{Psimerule}
\end{equation}

As for the molecular Hamiltonian (\ref{Hmoldef}), when transformed to
the Born-Oppenheimer representation, $H \longleftrightarrow
H_{kl}$, it becomes
\begin{equation}
	H_{kl} =
	\sum_{\alpha=1}^N
	\frac{1}{2M_\alpha}\sum_p
	\bigl[\Pvec_\alpha\,\delta_{kp}-i\hbar\,\Fvec_{\alpha;kp}
	(x)\bigr]\cdot
	\bigl[\Pvec_\alpha\,\delta_{pl}-i\hbar\,\Fvec_{\alpha;pl}
	(x)\bigr]
	+W_{kl}(x),
	\label{Hmolkldef}
\end{equation}
where the sum on $p$ indicates a multiplication of matrices of
operators, as in (\ref{ABmult}).  We expand this according to
\begin{equation}
	H_{kl}=H_{0,kl} + \hbar\,
	H_{1,kl} + \hbar^2\, H_{2,kl},
	\label{Hmolexpand}
\end{equation}
where
\begin{subequations}
\label{Hmolndefs}
\begin{eqnarray}
	H_{0,kl} &=& \left(
	\sum_{\alpha=1}^N 
	\frac{\Pvec_\alpha^2}{2M_\alpha}
	\right)\delta_{kl} + W_{kl}(\Xvec),
	\label{Hmol0def}\\
	H_{1,kl} &=&
	-i\sum_{\alpha=1}^N
	\frac{1}{2M_\alpha}
	[\Pvec_\alpha\cdot\Fvec_{\alpha;kl}(\Xvec)
	+\Fvec_{\alpha;kl}(\Xvec)\cdot\Pvec_\alpha], 
	\label{Hmol1def}\\
	H_{2,kl} &=&
	-\sum_{\alpha=1}^N\sum_p
	\frac{1}{2M_\alpha} \Fvec_{\alpha;kp}(\Xvec)\cdot
	\Fvec_{\alpha;pl}(\Xvec).
	\label{Hmol2def}
\end{eqnarray}
\end{subequations}
Note that, since $\Fvec_{\alpha;kl}(x)$ is real, the two terms in the
square brackets in (\ref{Hmol1def}) are Hermitian conjugates of each
other, so the sum is Hermitian.  Overall, we have $(H_{1,kl})^\dagger
= H_{1,lk}$, so that $H_1$, as a matrix of nuclear operators, is
Hermitian.  

The molecular Schr\"odinger equation, which in the molecular
representation is  
\begin{equation}
	H\,\Psi(\Xvec,\rvec)=E\,\Psi(\Xvec,\rvec),
	\label{molScheqnmolrep}
\end{equation}
or, in its time-dependent version,
\begin{equation}
	H\,\Psi(\Xvec,\rvec,t) = i\hbar
	\frac{\partial\Psi(\Xvec,\rvec,t)}{\partial t},
	\label{molScheqnmolreptdep}
\end{equation}
becomes, in the Born-Oppenheimer representation,
\begin{equation}
	\sum_l H_{kl} \, \psi_l(\Xvec) = E\, \psi_k(\Xvec),
	\label{molScheqnBOrep}
\end{equation}
or, in its time-dependent version,
\begin{equation}
	\sum_l H_{kl}\,\psi_l(\Xvec,t) = i\hbar
	\frac{\partial \psi_k(\Xvec,t)}{\partial t}.
	\label{molScheqnBOreptdep}
\end{equation}.

\subsection{Born-Oppenheimer Ordering}
\label{BOordering}

The standard Born-Oppenheimer ordering parameter is
$\kappa=(m_e/M)^{1/4}$, where $M$ is an average or typical nuclear
mass (\cite{BornOppenheimer27,BornHuang54}).  By this definition,
$\kappa$ has the numerical value of roughly $10^{-1}$.  A different
approach is to scale the nuclear masses by $\kappa^{-4}$, that is, to
replace the masses $M_\alpha$ in the Hamiltonian by
$M_\alpha/\kappa^4$, where $\kappa$ is a scaling parameter that ranges
from $\kappa=0$ (for complete adiabatic separation of the length and
time scales) to $\kappa=1$ (which gives the physical values).
Actually, the limit $\kappa\to0$ is singular since the nuclear
de~Broglie wavelength goes to zero and the nuclear wave function is no
longer defined.  But many physical quantities are well defined in the
limit $\kappa\to0$.  By making $\kappa$ a variable the Hamiltonian
becomes parameterized by $\kappa$, and thereby represents a family of
dynamical systems.  We can think of a curve through the space of
Hamiltonians, parameterized by $\kappa$.  Then we have a solvable or
at least simplified system as $\kappa\to0$.

In the following we shall adopt the second approach, in which $\kappa$
is a variable.  Then we can speak of the order of magnitude of various
quantities in terms of the power of $\kappa$ by which they scale as
$\kappa\to0$.  For example, since the electronic Hamiltonian does not
depend on the nuclear masses, it is independent of $\kappa$ and
everything associated with it (the energy eigenvalues $\epsilon_k(x)$,
the derivative couplings $\Fvec_{\alpha;kl}(x)$, etc.), are
independent of $\kappa$ and therefore of order $\kappa^0=1$.

To say a quantity is of order $\kappa^0=1$ in this sense does not
necessarily imply that its numerical value is close to unity, since
there may be other reasons (besides the values of the nuclear masses
or momenta) why it is large or small.  For example, in a scattering
problem $\Xvec_\alpha$ may take on large values, independent of
$\kappa$, and the derivative couplings in the adiabatic basis diverge
as a degeneracy or conical intersection is approached.  In terms of
the $\kappa$ scaling, however, these quantities are of order unity.
On the other hand, if there is no other reason why a given quantity
should be large or small, and if it is of order $\kappa^n$ in the
scaling sense, then it should have a numerical value in atomic units
of roughly $10^{-n}$.

In addition to the scaling of the nuclear masses, we must also
indicate how the dynamical variables scale with $\kappa$ as we take
$\kappa\to0$.  For small vibrations about an equilibrium, the
displacement from equilibrium $\Xvec_\alpha - \Xvec_{0,\alpha}$ is of
order $\kappa^1=\kappa$, as discussed by \cite{BornHuang54}, while the
nuclear momentum $\Pvec_\alpha$ is order $\kappa^{-1}$ and the nuclear
velocity $\Vvec_\alpha=\Pvec_\alpha/M_\alpha$ is of order $\kappa^3$.
Thus, the time require to execute an oscillation of amplitude of order
$\kappa$ is $\kappa^{-2}$, and the vibrational frequency is of order
$\kappa^2$.  In this analysis we are imagining that as we move through
the space of dynamical systems by taking $\kappa\to0$, we are also
changing the space of wave functions or quantum states that we are
considering.  For the problem of small vibrations, we take this limit
in such a way that the vibrational quantum number is constant, which
causes the spatial extent of the wave function to shrink in proportion
to $\kappa$.  A nice discussion of this (small-amplitude) ordering is
given by \cite{MeadMoscowitz67}.

Modern applications such as isomerization, scattering and
photoexcitation problems involve motions in which the nuclei are not
restricted to a small range around an equilibrium, but rather move
over distances of the order of an atomic unit or larger.  For these
applications a different assumption about the scaling of the dynamical
variables is required.  The modified scaling assumptions required for
such applications have been discussed in detail by \cite{Mead06}.  See
also \cite{MatyusTeufel19}.  In the following we add to this
discussion and show how it applies to our approach to Moyal
perturbation theory.

We shall call the modified scaling law  the ``large-amplitude''
scaling.  In this scaling we imagine that the kinetic or potential
energy of the nuclei remains constant as $\kappa\to0$.  For example,
if a nucleus climbs half way up a potential well when $\kappa=1$, we
require that it continue to do so as we take $\kappa\to0$.  We do this
because we want the qualitative features of the simplified problem as
$\kappa\to0$ to be the same as those of the physical problem at
$\kappa=1$.  This means, for example, that bound state quantum
numbers, to the extent those are meaningful for large-amplitude
motions, go to infinity as $\kappa\to0$.

If the kinetic energy $\Pvec_\alpha^2/2M_\alpha$ is independent of
$\kappa$ as $\kappa\to0$, as required by the large-amplitude ordering,
then, given the scaling of $M_\alpha$, we conclude that $\Pvec_\alpha$
is order $\kappa^{-2}$ (therefore larger by a factor of $1/\kappa$
than in the small-amplitude scaling).  Nuclear velocities
$\Vvec_\alpha=\Pvec_\alpha/M_\alpha$ are therefore of order $\kappa^2$
and the time required for a nucleus to travel a distance of the order
of one atomic unit is of order $\kappa^{-2}$.  Thus nuclear motions
are executed ever more slowly as $\kappa\to0$, as expected in a system
that can be considered ``adiabatic.''

The spatial scale of an electronic energy eigenfunction
$\phi_{ak}(\Xvec;\rvec)$ for fixed $k$, that is, the electronic
de~Broglie wave length, is of order $\kappa^0=1$, since the electronic
Hamiltonian is independent of the nuclear masses, as explained.  The
same applies to the operator $\nabla_i$.  The same also applies to the
$\Xvec$-dependence of the electronic energy eigenfunctions and the
operator $\nabla_\alpha$ when acting on them.  We shall assume that
the transformation from the adiabatic basis to the working basis is
independent of $\kappa$, so the same scale lengths (in both $\rvec$
and $\Xvec$) apply to the working basis functions
$\phi_k(\Xvec;\rvec)$.  The nuclear wave functions $\psi_k(\Xvec)$ in
the Born-Oppenheimer representation, however, have a de~Broglie wave
length of order $\kappa^2$, since the operator $\Pvec_\alpha=-i\hbar\,
\nabla_\alpha$ is of order $\kappa^{-2}$.  Thus, in the expansion
(\ref{psikdef}) the scale lengths of the two factors on the right
($\psi_k$ and $\phi_k$) with respect to $\Xvec$ are quite different.

In the large-amplitude scaling, the nuclear de~Broglie wave length is
of order $\kappa^2$ times smaller than the scale length of the
environment (the potential, derivative couplings, etc.) in which the
nuclei move.  These are otherwise the conditions for WKB theory, which
would be more useful for molecular dynamics were it not for the usual
difficulties of WKB theory in multidimensional problems (chaotic
orbits and the lack of global, invariant Lagrangian manifolds in phase
space).  Nevertheless, the basic method of this paper, the use of
Moyal perturbation theory, is a part of semiclassical theory.

In this paper we use the large-amplitude scaling exclusively.
For simplicity we will not insert explicit factors of $\kappa^n$ to
indicate the order of various terms.  Instead, the ordering of any
term can be obtained by applying the two rules,
$M_\alpha=O(\kappa^{-4})$ and $\Pvec_\alpha=O(\kappa^{-2})$.

In the large-amplitude ordering, the three terms in (\ref{Hmolndefs}),
$H_{0,kl}$, $H_{1,kl}$ and $H_{2,kl}$, are of order $\kappa^0=1$,
$\kappa^2$ and $\kappa^4$, respectively.  This ordering will be the
basis of the perturbation theory we will carry out in this paper.  It
is not, however, ordinary perturbation theory, which is based solely
on the magnitudes of the unperturbed Hamiltonian and the perturbation,
because the nuclear wave functions also have a dependence on $\kappa$,
that is, the nuclear de~Broglie wave length is of order $\kappa^2$.

\subsection{Translational Degrees of Freedom}
\label{translationaldof}

The Hamiltonian (\ref{Hmoldef}) commutes with translations and thus
conserves total linear momentum.  The manner in which this invariance
and conservation law are manifested is straightforward in the
molecular representation but more obscure in the Born-Oppenheimer
representation.  Therefore one is motivated to carry out the exact
elimination of the translational degrees of freedom before switching
to the Born-Oppenheimer representation.  This is a standard procedure
(\cite{Sutcliffe00}) but it requires some algebra and attention to
notation.

We begin with the nuclear configuration space, the $3N$-dimensional
space upon which the coordinates are $\Xvec_\alpha$,
$\alpha=1,\ldots,N$.  This is the space upon which purely nuclear wave
functions $\psi(\Xvec)$ are defined.  A coordinate transformation on
this space that is useful for separating the translational degrees of
freedom is $\Xvec_\alpha \to (\Xvec_{\rm CM},\Yvec_\kappa)$, where
$\Xvec_{\rm CM}$ is the center of mass of the nuclei,
\begin{equation}
   \Xvec_{\rm CM} = \frac{1}{M_t}\sum_{\alpha=1}^N M_\alpha
	\Xvec_\alpha,
   \label{XCMdef}
\end{equation}
and where $\Yvec_\kappa$, $\kappa=1,\ldots,N-1$, are a set of Jacobi
vectors (see Appendix~\ref{coordinatesnuclearCS}).  Here $M_t$ is the
total nuclear mass,
\begin{equation}
   M_t = \sum_{\alpha=1}^N M_\alpha.
   \label{Mtdef}
\end{equation}
We let the momenta conjugate to $(\Xvec_{\rm CM},\Yvec_\kappa)$ be
$(\Pvec_t,\Qvec_\kappa)$.  Then it turns out that $\Pvec_t$, the
momentum conjugate to $\Xvec_{\rm CM}$, is the total momentum of the
nuclei,
\begin{equation}
  \Pvec_t = \sum_{\alpha=1}^N \Pvec_\alpha.
  \label{Ptdef}
\end{equation}
The other details of this coordinate transformation are summarized in
Appendix~\ref{coordinatesnuclearCS}.

When this transformation is applied to the nuclear kinetic energy, we
find 
\begin{equation}
   \sum_{\alpha=1}^N \frac{\Pvec_\alpha^2}{2M_\alpha} =
   \frac{\Pvec_t^2}{2M_t} + \sum_{\kappa=1}^{N-1}
   \frac{\Qvec_\kappa^2}{2\mu_\kappa},
   \label{nuclearKExfm}
\end{equation}
where $\mu_\kappa$ are reduced masses associated with Jacobi
coordinates $\Yvec_\kappa$.

We require a different coordinate transformation on the molecular
configuration space, the $3(N+N_e)$-dimensional space with coordinates
$(\Xvec_\alpha,\rvec_i)$ upon which the molecular wave function
$\Psi(\Xvec,\rvec)$ is defined.  This can be seen as the composition
of two transformations, $(\Xvec_\alpha,\rvec_i) \to (\Xvec_{\rm
CM},\Yvec_\kappa,\rvec_i) \to (\Rvec_{\rm CM},\Yvec_\kappa,\svec_i)$,
where in the first step we transform the nuclear coordinates as
described above, leaving the electronic coordinates $\rvec_i$ alone,
while in the second step we leave the nuclear Jacobi vectors
$\Yvec_\kappa$ alone while introducing two new sets of coordinates
defined by
\begin{equation}
   \Rvec_{\rm CM}=\frac{M_t\Xvec_{\rm CM} +N_e m_e \rvec_{\rm CM}}
   {M_{\rm mol}}
   \label{RCMdef}
\end{equation}
and
\begin{equation}
    \svec_i=\rvec_i-\Xvec_{\rm CM}.
    \label{sidef}
\end{equation}
Here $M_{\rm mol}$ is the total molecular mass,
\begin{equation}
    M_{\rm mol} = M_t + N_e m_e,
    \label{Mmoldef}
\end{equation}
and $\rvec_{\rm CM}$ is the center of mass of the electrons,
\begin{equation}
    \rvec_{\rm CM} = \frac{1}{N_e}\sum_{i=1}^{N_e} \rvec_i.
    \label{rCMdef}
\end{equation}
Evidently $\Rvec_{\rm CM}$ is the molecular center of mass (not to be
confused with $\Xvec_{\rm CM}$, the nuclear center of mass) and
$\svec_i$ is the position of electron $i$ relative to the nuclear
center of mass.

We denote the momenta conjugate to coordinates $(\Rvec_{\rm
CM},\Yvec_\kappa,\svec_i)$ by $(\Pvec_{\rm CM},\Qvec_\kappa,\qvec_i)$,
where $\Qvec_\kappa$, the momentum conjugate to Jacobi vector
$\Yvec_\kappa$, is the same as produced by the first (purely nuclear)
transformation (see (\ref{Qkappadef})).   As for $\Pvec_{\rm CM}$, the
momentum conjugate to $\Rvec_{\rm CM}$, it turns out to be the total
linear momentum of the molecule,
\begin{equation}
   \Pvec_{\rm CM} = \Pvec_t+\pvec_t,
   \label{PCMdef}
\end{equation}
where $\Pvec_t$, the total nuclear momentum, is given by
(\ref{Ptdef}), and $\pvec_t$, the total electronic momentum, is given
by 
\begin{equation}
   \pvec_t=\sum_{i=1}^{N_e} \pvec_i.
   \label{ptdef}
\end{equation}
The other details of this coordinate transformation are given in
Appendix~\ref{coordinatesmolecularcs}. 

When we transform the molecular Hamiltonian (\ref{Hmoldef}) to the
new coordinates, we obtain
\begin{equation}
   H = \frac{\Pvec_{\rm CM}^2}{2M_{\rm mol}} + 
        H_{\rm TR}(\Yvec,\svec,\Qvec,\qvec),
       \label{HmolTRdef}
\end{equation}
where the ``translation-reduced'' molecular Hamiltonian is
\begin{equation}
  H_{\rm TR}(\Yvec,\svec,\Qvec,\qvec) = \sum_{\kappa=1}^{N-1}
       \frac{\Qvec_\kappa^2}{2\mu_\kappa} + H_{{\rm TR}e}
         (y;\svec,\qvec),
  \label{HTRdef}
\end{equation}
and the translation-reduced electronic Hamiltonian is
\begin{equation}
   H_{{\rm TR}e}(y;\svec,\qvec)=
   \sum_{i=1}^{N_e}\frac{\qvec_i^2}{2m_e}+
   \sum_{i,j=1}^{N_e}\frac{\qvec_i\cdot\qvec_j}{2M_t}+
   V_{\rm Coul}(\Yvec,\svec).
   \label{HTRedef}
\end{equation}
Here $y$ means a point of the translation-reduced configuration space
whose coordinates are $\Yvec_\kappa$, and $V_{\rm Coul}$ is
independent of $\Rvec_{\rm CM}$ because it is translationally
invariant.  The translation-reduced configuration space will be
discussed more fully in subsection~\ref{translationalfiberbundle} (see
also Fig.~\ref{bundle}).  The middle term on the right in
(\ref{HTRedef}) is the mass-polarization term.

\subsection{The Translation-Reduced Molecular and Born-Oppenheimer
  Representations}
\label{TRrepns}

The Hamiltonian $H_{\rm TR}$ in (\ref{HTRdef}) gives us a
``translation-reduced'' molecular representation, in which the wave
function $\Psi_{\rm TR}(\Yvec,\svec)$ is defined by 
\begin{equation}
  \Psi(\Xvec,\rvec) = \Psi(\Rvec_{\rm CM},\Yvec,\svec) =
  \exp(i\Kvec\cdot\Rvec_{\rm CM})\,\Psi_{\rm TR}(\Yvec,\svec),
  \label{PsiTRdef}
\end{equation}
which is an eigenfunction of $\Pvec_{\rm CM}$ with eigenvalue
$\hbar\Kvec$.   This is in the time-independent version; in the
time-dependent version we define $\Psi_{\rm TR}$ by
\begin{equation}
  \Psi(\Xvec,\rvec,t) = \Psi(\Rvec_{\rm CM},\Yvec,\svec,t) =
  \exp[i(\Kvec\cdot\Rvec_{\rm CM}-\hbar \Kvec^2t/2M_{\rm mol})]\,
  \Psi_{\rm TR}(\Yvec,\svec,t).
  \label{PsiTRdeftdep}
\end{equation}
The wave function $\Psi_{\rm TR}$ satisfies a translation-reduced
version of the Schr\"odinger equation,
\begin{equation}
   H_{\rm TR}\,\Psi_{\rm TR}(\Yvec,\svec) = E_{\rm TR} \, 
   \Psi_{\rm TR}(\Yvec,\svec),
   \label{TRScheqn}
\end{equation}
or its time-dependent version,
\begin{equation}
  H_{\rm TR}\, \Psi_{\rm TR}(\Yvec,\svec,t) =
  i\hbar\,\frac{\partial\Psi_{\rm TR}(\Yvec,\svec,t)}{\partial t}.
  \label{TRScheqntdep}
\end{equation}

Once we have achieved the translation-reduced molecular representation
of the system we can expand in an electronic basis and create a
translation-reduced Born-Oppenheimer representation, in which the
state of the system is represented by an infinite-dimensional vector
of nuclear wave functions $\psi_{{\rm TR},k}(\Yvec)$, defined on the
translation-reduced configuration space.  The potential energy
surfaces defined by the eigenvalues $\epsilon_{{\rm TR},k}(y)$ are
not the same as before, because the translation-reduced electronic
Hamiltonian (\ref{HTRdef}) contains mass-polarization terms, in
contrast to the original electronic Hamiltonian (\ref{Hedef}), whose
eigenvalues are $\epsilon_k(x)$. 

\begin{table}
\begin{center}
\begin{tabular}{|l|c@{\extracolsep{10pt}}cc|}\hline
&&&Dressed\\[-2.0ex]
&\raisebox{1.5ex}[0pt]{Molecular}&%
\raisebox{1.5ex}[0pt]{Born-Opp.}&Born-Opp.\\
\hline
& $\Psi(\Xvec,\rvec)$&$\psi_k(\Xvec)$&$\psibar_k(\Xvec)$\\
Lab &$H$&$H_{kl}$&$\Hbar_{kl}$\\
&(\ref{Hmoldef})&(\ref{psikdef}), (\ref{tworepcorresp})&%
(\ref{Hbardef})\\
\hline
&$\Psi_{\rm TR}(\Yvec,\svec)$&$\psi_{{\rm TR},k}(\Yvec)$&%
$\psibar_{{\rm TR},k}(\Yvec)$\\
\raisebox{1.6ex}[0pt]{Translation-}&$H_{\rm TR}$&%
$H_{{\rm TR},kl}$&$\Hbar_{{\rm TR},kl}$\\
\raisebox{3.4ex}[0pt]{Reduced}&(\ref{HTRdef}), (\ref{PsiTRdef})%
, (\ref{PsiTRdeftdep})&---&---\\
\hline
\end{tabular}
\end{center}
\caption{\label{reptable}Six representations of the quantum mechanics 
  of a molecule.  In each case, the relevant wave function and 
  Hamiltonian are shown, as well as references to relevant equations.
  In the molecular representation (first column) the wave function
  $\Psi$ or $\Psi_{\rm TR}$ is a function defined on the combined
  nuclear-electronic configuration space.  In the Born-Oppenheimer
  representation (second column), the wave function $\psi_k$ or 
  $\psi_{{\rm TR},k}$ is an infinite-dimensional vector of wave 
  functions defined on the
  nuclear configuration space.  The same is true of the dressed
  representations (third column); in  this representation, the 
  Hamiltonian is block-diagonal.  Sometimes we will call the
  Born-Oppenheimer representation (second column) the ``original''
  Born-Oppenheimer representation, to distinguish it from the dressed
  representations in the third column.}
\end{table}

Thus we now have several representations of the quantum mechanics of
the molecule, including the original or ``lab'' molecular and
Born-Oppenheimer representations, and now ``translation-reduced''
versions of these.   The situation is summarized in Table~\ref{reptable}.
Once we have achieved the Born-Oppenheimer representation, either the
lab or translation-reduced, the Hamiltonian is an infinite-dimensional
matrix of nuclear operators, which we can block-diagonalize by means
of Moyal perturbation theory.  This carries us to the last column of
Table~\ref{reptable}, the ``dressed'' or block-diagonalized
Born-Oppenheimer representation.   We turn now to a detailed
description of this last step.

\section{Moyal Perturbation Theory on a Single Surface}
\label{singlesurface}

The purpose of this section is to describe in detail the Moyal method
for the block diagonalization of the Born-Oppenheimer Hamiltonian.
For a somewhat different formalism, see \cite{Panatietal02,
MatyusTeufel19}.

For this purpose we can work either with the lab representations or
the translation-reduced.  In some ways it is preferable to carry out
the exact separation of the translational degrees of freedom first,
and then to block-diagonalize the Hamiltonian.  But much of the
current literature, for example, that devoted to surface hopping, uses
the lab representations (so also the treatment of
\cite{MatyusTeufel19}).  Therefore in this section we shall illustrate
Moyal diagonalization of the Born-Oppenheimer Hamiltonian in the lab
representations, effectively working across the first row of
Table~\ref{reptable} from left to right.  For simplicity we will do
this for a single surface problem, deferring multiple surfaces to
subsection~\ref{multiplesurfaces}.  We select one surface $k=n$, so
that $n$ is fixed for the duration of this section
(Sec.~\ref{singlesurface}).

In subsections~\ref{basisprivspace}--\ref{dressing} we use unaccented
symbols for operators, for example, $H$ for the Hamiltonian.  Starting
in subsection~\ref{generatorG1}, however, and continuing to the end of
the paper, it is necessary to distinguish between operators, which are
represented with a hat, and their Weyl symbols, which are $c$-number
functions on the classical phase space and which are denoted without
the hat.  For example, starting in subsection~\ref{generatorG1}, the
Hamiltonian operator is $\Hhat$ while its Weyl symbol is $H$.  Weyl
symbols are explained in Appendix~\ref{Moyal}.

\subsection{Basis Vectors and the Privileged Subspace}
\label{basisprivspace}

We let $\ket{x;n}=\ket{ax;n}$, so that in the one case $k=n$ the
working basis vector and the adiabatic basis vector are the same.  We
define the ``privileged subspace'' of the electronic Hilbert space as
\begin{equation}
	\Sspace(x) = \mathspan\{\ket{ax;n}\}
	\label{S(x)def}
\end{equation}
and we define $\Sspace^\perp(x)$ as the complementary, orthogonal
subspace.  The working basis vectors $\ket{x;k}$ for $k\ne n$ are
either the adiabatic basis vectors $\ket{ax;k}$ for $k\ne n$ or
another choice to be specified.  Both the working basis vectors
$\ket{x;k}$ and the adiabatic basis vectors $\ket{ax;k}$ for $k\ne n$
span $\Sspace^\perp(x)$, so they are linear combinations of one another.
The subspace $\Sspace(x)$ is one-dimensional, while $\Sspace^\perp(x)$
is infinite-dimensional.  The projection operator onto the privileged
subspace is
\begin{equation}
	P(x)=\ketbra{ax;n}{ax;n},
	\label{P(x)def}
\end{equation}
while the complementary projector is $Q(x)=1-P(x)$.
Equation~(\ref{P(x)def}) gives $P(x)$ in the molecular representation;
on transforming to the Born-Oppenheimer representation it becomes
\begin{equation}
	P(x)\longleftrightarrow
	P_{kl} = \delta_{nk}\,\delta_{nl}.
	\label{Pkldef}
\end{equation}

\subsection{The Born-Oppenheimer Approximation}
\label{overviewpertexpn}

Since $\ket{x;n}=\ket{ax;n}$, the diagonal matrix element $W_{nn}$ of
the electronic Hamiltonian satisfies
\begin{equation}
	W_{nn}(x) = \epsilon_n(x).
	\label{Wnnvalue}
\end{equation}
In addition, since $\ket{x;n}$ is orthogonal to $\Sspace^\perp(x)$, we
have $W_{nk}(x)=W_{kn}(x)=0$ for $k\ne n$, and the matrix $W_{kl}(x)$
is block-diagonal.  The blocks in question are the $1\times1$
block and the infinite-dimensional block corresponding to $\Sspace(x)$
and $\Sspace^\perp(x)$, respectively.  An operator that is
block-diagonal does not couple $\Sspace(x)$ with $\Sspace^\perp(x)$.
The terms in the Hamiltonian (\ref{Hmolndefs}) that do couple
$\Sspace(x)$ and $\Sspace^\perp(x)$ are the off-block-diagonal
elements of $H_{1}$ and $H_{2}$, that is, elements with indices $(nk)$
or $(kn)$ for $k\ne n$.  These matrix elements involve the derivative
couplings and are small in the $\kappa$ ordering, either of order
$\kappa^2$ or $\kappa^4$.  In particular, they are small in comparison
to the diagonal element at zeroth order,
\begin{equation}
	H_{0,nn} = \sum_{\alpha=1}^N
	\frac{\Pvec_\alpha^2}{2M_\alpha} + \epsilon_n(x),
	\label{H0nndef}
\end{equation}
which is of order $\kappa^0$.

In the literature the ``Born-Oppenheimer approximation'' usually means
approximating the molecular wave function as
\begin{equation}
	\Psi(\Xvec,\rvec)=\psi(\Xvec)\,\phi_{an}(\Xvec;\rvec),
	\label{BOappx}
\end{equation}
that is, a version of (\ref{psikdef}) with a single term $k=n$ and
employing the adiabatic, energy eigenfunction
$\phi_{an}(\Xvec;\rvec)$.  But if we throw away the small,
off-block-diagonal terms of the molecular Hamiltonian
(\ref{Hmolndefs}), then the molecular Schr\"odinger equation
(\ref{molScheqnmolrep}) has solutions of the form (\ref{BOappx}), that
is, (\ref{molScheqnBOrep}) has solutions in which $\psi_k$ is nonzero
only in the one slot $k=n$.  The same is true of the time-dependent
versions of the molecular Schr\"odinger equation,
(\ref{molScheqnmolreptdep}) and (\ref{molScheqnBOreptdep}), which have
solutions of the form (\ref{BOappx}) in which both
$\Psi(\Xvec,\rvec,t)$ and $\psi(\Xvec,t)$ depend on time but the
adiabatic basis state $\phi_{an}(\Xvec;\rvec)$ does not.

Therefore we shall regard the ``Born-Oppenheimer approximation,'' in
the context of a single-surface problem, as one of throwing away the
off-block-diagonal matrix elements of $H_{kl}$.  In this
approximation, the molecular Schr\"odinger equation
(\ref{molScheqnBOrep}) or (\ref{molScheqnBOreptdep}) on the privileged
subspace $\Sspace(x)$ becomes
\begin{equation}
	H_{\rm BO}(\Xvec,\Pvec)\,\psi(\Xvec) = E\,\psi(\Xvec),
	\label{molScheqnBO}
\end{equation}
or its time-dependent version,
\begin{equation}
	H_{\rm BO}(\Xvec,\Pvec)\,\psi(\Xvec,t) =
	i\hbar\frac{\partial\psi(\Xvec,t)}{\partial t},
	\label{tdepmolScheqnBO}
\end{equation}
where $\psi=\psi_n$ is the one, nonvanishing component of $\psi_k$,
and where $H_{\rm BO} =H_{nn} =H_{0,nn} +\hbar\,H_{1,nn}
+\hbar^2\,H_{2,nn}$ is the diagonal element of the Hamiltonian matrix,
the effective Hamiltonian operator on this subspace.

To write out $H_{\rm BO}$ explicitly we note first that $H_{1,nn}=0$
because of (\ref{Fdiagzero}).  As for $H_{2,nn}$, we take the
divergence of $\Fvec_{\alpha;nn}(x) =\matrixelement{x;n}
{\nabla_\alpha} {x;n}=0$ to obtain the identity,
\begin{equation}
  \nabla_\alpha\cdot\Fvec_{\alpha;nn}(x) =
  (\nabla_\alpha\bra{x;n})\cdot(\nabla_\alpha\ket{x;n})
  +\matrixelement{x;n}{\nabla_\alpha^2}{x;n}=0.
  \label{divFvecidentity}
\end{equation}
Then using (\ref{Fkldef}) we have
\begin{eqnarray}
  \sum_p \Fvec_{\alpha;np}(x)\cdot\Fvec_{\alpha;pn}(x) &=&
  -\sum_p (\nabla_\alpha\bra{x;n})\ket{x;p}\cdot
  \bra{x;p}(\nabla_\alpha\ket{x;n})\nonumber\\
  &=&-(\nabla_\alpha\bra{x;n})\cdot(\nabla_\alpha\ket{x;n})=
  \matrixelement{x;n}{\nabla_\alpha^2}{x;n},
  \label{H2identity}
\end{eqnarray}
where in the first step we use (\ref{Fkldef}) and in the last step,
(\ref{divFvecidentity}).  Altogether, we obtain
\begin{equation}
  H_{\rm BO}=H_{nn} = \sum_{\alpha=1}^N
  \frac{\Pvec_\alpha^2}{2M_\alpha} +\epsilon_n(x)
  -\sum_{\alpha=1}^N\frac{1}{2M_\alpha}
  \matrixelement{x;n}{\nabla_\alpha^2}{x;n}.
  \label{HBOdef}
\end{equation}

This is a standard Hamiltonian in Born-Oppenheimer theory.  It
is the same as the projection of the original molecular Hamiltonian
(\ref{Hmoldef}) onto the subspace $\Sspace(x)$, that is,
\begin{equation}
  P(x)H P(x) \longleftrightarrow \delta_{kn}\,\delta_{nl}\, H_{\rm
    BO}.
  \label{projectHmol}
\end{equation}
The final term in (\ref{HBOdef}) is the ``diagonal Born-Oppenheimer
correction'' (\cite{ValeevSherrill03, GaussTajtiKallayStantonSzalay06,
  MeekLevine16, GheribYeRyabinkinIzmaylov16,
  CulpittPetersTellgrenHelgaker22}). It is otherwise $H_{2,nn}$.  It
appears to be a contribution to the potential energy, but physically
it is a part of the kinetic energy.  Away from conical intersections,
it is small, being of order $\kappa^4$, and many authors take
liberties with it (throwing it away, or throwing part of it away,
etc.).  In the adiabatic basis it diverges at conical intersections,
for the same reason as the derivative couplings
(\cite{MatsunagaYarkony98,Yarkony97a}), but in our treatment of
single-surface problems we shall work in regions far enough from
conical intersections that this term is small.  See
Sec.~\ref{energydens} for more details.  Near conical intersections we
must use a multi-surface approach (see
subsection~\ref{multiplesurfaces}). 

The terms in the molecular Hamiltonian that we have thrown away to
obtain the Born-Oppenheimer approximation are of order $\kappa^2$ or
higher and thus are small compared to the diagonal element $H_{0,nn}$
(see (\ref{H0nndef})).  On the other hand, they are large in
comparison to $H_{2,nn}$, the diagonal Born-Oppenheimer correction.
Therefore there is a question of whether $H_{2,nn}$ is even the
correct, second order contribution to the Hamiltonian when the wave
function has a single component $\psi_n$.  In fact, a perturbative
treatment of the off-block-diagonal terms, one that transforms them
away rather than throwing them away, generates a contribution at
second order, what we call $K_{22}$, that is of the same order in
$\kappa$ as $H_{2,nn}$.  See Sec.~\ref{K22discussion}.

\subsection{Dressing Transformations}
\label{dressing}

As an alternative to the Born-Oppenheimer approximation, that is, just
throwing away terms in the molecular Hamiltonian (\ref{Hmolndefs}), we
will transform them away by means of a series of unitary
transformations.  Since the terms to be eliminated are small, each
unitary transformation is close to the identity and can be
conveniently represented as the exponential of another operator, the
generator.

In the first step we transform the Hamiltonian by the unitary operator
$U_1=\exp(\hbar G_1)$, where for convenience we split off a factor of
$\hbar$ from the anti-Hermitian generator $G_1$.  The transformed
Hamiltonian $H'$ can be written as an exponential series of iterated
commutators,
\begin{equation}
	H'=U_1HU_1^\dagger= e^{\hbar G_1} \, H\, e^{-\hbar G_1}
	=H + \hbar [G_1,H] + \frac{\hbar^2}{2!}[G_1,[G_1,H]] 
	+ \ldots
	\label{HtoH'}
\end{equation}
Although we are writing $H$, $G_1$ and $U_1$ without matrix indices,
we will work in the Born-Oppenheimer representation and all such
symbols should be understood to be matrices of nuclear operators, for
example, $H_{kl}$, $G_{1,kl}$, etc.  Likewise, commutators of
operators are interpreted as matrix commutators, for example,
\begin{equation}
	[G_1,H]_{kl} = \sum_p \bigl(G_{1,kp}\,H_{pl} - H_{kp}\,
        G_{1,pl}\bigr).
	\label{commutatormatrix}
\end{equation}
The products that appear in (\ref{commutatormatrix}) are operator
products, as in (\ref{ABmult}), and the ordering must be respected.

The strategy will be to choose $G_1$ so as to eliminate the
off-block-diagonal terms in $H$ to order $\kappa^2$.  As it turns out,
this means that $G_1$ itself is of order $\kappa^2$.  This leaves
behind off-block-diagonal terms in $H'$ that are of order $\kappa^4$.
To eliminate these, we perform a second unitary transformation,
$H''=U_2\,H'\,U_2^\dagger$, where $U_2=\exp(\hbar^2 G_2)$.  Operator
$G_2$ is a second generator that turns out to be of order $\kappa^4$.
Proceeding this way we envision an infinite sequence of unitary
operators,
\begin{equation}
	U=\ldots U_3U_2U_1,
	\label{Udef}
\end{equation}
and a transformed Hamiltonian $\Hbar$ that is formally free of
off-block-diagonal terms to all orders,
\begin{equation}
	\Hbar=UHU^\dagger = \ldots U_3U_2U_1\,H\,U_1^\dagger
	U_2^\dagger U_3^\dagger\ldots
	\label{Hbardef}
\end{equation}

These unitary transformations create a sequence of new representations
for the state of the system, what we will call ``dressed''
representations.  Thus we can distinguish between the ``original''
(that is, undressed) Born-Oppenheimer representation, and successive
dressed versions of it.  See Table~\ref{reptable}. If the wave
function is $\psi$ in the original Born-Oppenheimer representation
(where $\psi$ is short for the infinite-dimensional vector $\psi_k$),
then the wave function in the final, dressed representation is
$\psibar=U\psi$ (that is, it is the vector $\psibar_k=\sum_l
U_{kl}\psi_l$).  Then the Schr\"odinger equation in the final, dressed
representation is $\Hbar\psibar=E\psibar$ or its time-dependent
version, where $\Hbar$ has no off-block-diagonal terms.  Therefore
there are solutions $\psibar$ that are nonzero only in the one slot
$k=n$, and the effective Hamiltonian for this component is the
diagonal element $\Hbar_{nn}$.  This is the (scalar) Hamiltonian
driving this one component, which takes the place of $H_{\rm BO}$
within dressing theory.  We now explain the dressing transformations
in detail.

\subsection{First Order Transformation and Weyl Symbols}
\label{generatorG1}

In this subsection we begin by assuming that the working basis is the
adiabatic basis, that is, $\ket{x;k}=\ket{ax;k}$ for all $k$ (not just
$k=n$).  This assumption causes (\ref{Wkladiab}) to hold for all $k$,
$l$.  It simplifies the derivation somewhat but leads to difficulties
that we will explain when we encounter them.

In this subsection and henceforth we must change notation in order to
distinguish between nuclear operators and their corresponding Weyl
symbols.  See Appendix~\ref{Moyal} for the definition of the Weyl
symbol of an operator and for the invertible mapping between operators
and their corresponding Weyl symbols.  As in Appendix~\ref{Moyal}, we
will place hats on nuclear operators such as $\Xvechat_\alpha$ or
$\Pvechat_\alpha$, while omitting the hats on the corresponding
classical quantities or Weyl symbols $\Xvec_\alpha$ or $\Pvec_\alpha$.
The notation $\xhat$ or $\phat$ of Appendix~\ref{Moyal} is to be
identified with $\Xvechat$ or $\Pvechat$ here.  We use the Weyl symbol
correspondence only for nuclear operators, not electronic ones, so we
omit the hats on purely electronic operators such as the electronic
Hamiltonian $H_e$.

Thus, we should go back and put hats on operators in equations such as
(\ref{Hmolndefs}) to conform with the new notation.  We will not do
this, but we note that when the (hatted, operator) versions of
(\ref{Hmolndefs}) are converted into Weyl symbols, we obtain
\begin{subequations}
\label{HmolndefsWeyl}
\begin{eqnarray}
	H_{0,kl} &=& \left(
	\sum_{\alpha=1}^N 
	\frac{\Pvec_\alpha^2}{2M_\alpha}
	\right)\delta_{kl}+W_{kl}(\Xvec),
	\label{Hmol0defWeyl}\\
	H_{1,kl} &=&
	-i\sum_{\alpha=1}^N
	\frac{1}{M_\alpha}
	\Pvec_\alpha\cdot\Fvec_{\alpha;kl}(\Xvec),
	\label{Hmol1defWeyl}\\
	H_{2,kl} &=&
	-\sum_{\alpha=1}^N\sum_p
	\frac{1}{2M_\alpha} \Fvec_{\alpha;kp}(\Xvec)\cdot
	\Fvec_{\alpha;pl}(\Xvec),
	\label{Hmol2defWeyl}
\end{eqnarray}
\end{subequations}
where now $W_{kl}(x)=\delta_{kl}\,\epsilon_k(x)$ on account of
our assumptions about the working basis.  These look almost the same
as (\ref{Hmolndefs}) but now they are symbol equations and note that
the expression for $H_{1,kl}$ has simplified.  In fact, this is an
example of (\ref{symmf(x)psymbol}), illustrating the convenient manner
in which Weyl symbols handle operator ordering.  Weyl symbols are
$c$-numbers, and can be multiplied in any order.
Equations~(\ref{HmolndefsWeyl}) contain exactly the same information
as (\ref{Hmolndefs}); no information is lost on transforming to Weyl
symbols.

Likewise, in the case of (\ref{HtoH'}), to conform with the new
notation we replace $H$, $H'$ and $G_1$ by $\Hhat$, $\Hhat'$ and
$\Ghat_1$, respectively, indicating operators (actually, matrices of
operators).  Then converting these operator equations into symbols, we
have the symbol equation,
\begin{equation}
	H'=H+\hbar[G_1,H]_* +\frac{\hbar^2}{2}[G_1,[G_1,H]_*]_*
	+\ldots,
	\label{H'symbolseries}
\end{equation}
where the notation $[G_1,H]_*$ is explained in Appendix~\ref{Moyal}.
Note in particular the definition (\ref{stardef}) of the $*$-product
and the discussion in subsection~\ref{matricesWeyl} of commutators
of matrices of operators, especially (\ref{matrixcommutexpand}).

Let us now write out the expansion of the symbol of the molecular
Hamiltonian, $H=H_0+\hbar\, H_1+\hbar^2\, H_2$, where the individual
terms are given by (\ref{HmolndefsWeyl}), and substitute it into
(\ref{H'symbolseries}).  We obtain
\begin{equation}
  H'=H_0 +\hbar\bigl(H_1+[G_1,H_0]_*\bigr) +\hbar^2\left(
    H_2 + [G_1,H_1]_* +\frac{1}{2}[G_1,[G_1,H_0]_*]_*\right)
    + \ldots
    \label{H'symbolseries1}
\end{equation}
The commutators shown contain an implicit $\hbar$-ordering.  To expand
this out it is convenient to define
\begin{equation}
  T_1 = [G_1,H_0]_* = T_{10} + \hbar \,T_{11} + \ldots,
  \label{T1def}
\end{equation}
where the $1$-subscript on $T_1$ is a reminder that this term is
associated with generator $G_1$, and where the expansion follows
(\ref{matrixcommutexpand}).  In particular, we have
\begin{subequations}
\label{T1ndefs}
\begin{eqnarray}
  T_{10,kl} &=& [G_1,H_0]_{kl}=\sum_p(G_{1,kp}\,H_{0,pl} 
- H_{0,kp}\,G_{1,pl}),\\
  \label{T10def}
  T_{11,kl} &=& \frac{i}{2} \sum_p
  \bigl(\{G_{1,kp},H_{0,pl}\} - \{H_{0,kp},G_{1,pl}\}\bigr),
  \label{T11def}
\end{eqnarray}
\end{subequations}
where $T_{10}=[G_1,H_0]$ is the ordinary commutator of matrices of
Weyl symbols (without the $*$) and where the brackets in
(\ref{T11def}) are Poisson brackets.  Then (\ref{H'symbolseries1})
becomes
\begin{equation}
  H'=H_0 +\hbar\left(H_1+T_{10}\right) +\hbar^2\left(
    H_2 + [G_1,H_1]+T_{11}+\frac{1}{2}[G_1,T_{10}]\right)
    + \ldots
    \label{H'symbolseries2}
\end{equation}  
where we have dropped the $*$-subscript in commutators at order
$\hbar^2$ since the ordinary, matrix commutator is the leading term in
$\hbar$ (see subsection~\ref{matricesWeyl}) and the higher order
corrections are beyond the order to which we are working.
  
The Moyal expansion of the $*$-product (\ref{Moyalstarexpansion})
obviously generates a power series in $\hbar$.  As pointed out at the
end of subsection~\ref{BOordering}, however, each power of $\hbar$ in
the $\hbar$-ordering of the molecular Hamiltonian (\ref{Hmolndefs}) is
coupled to one power of $\kappa^2$.  That is, terms $H_0$, $H_1$ and
$H_2$ are of order $\kappa^0$, $\kappa^2$ and $\kappa^4$,
respectively.  Likewise, in the Moyal expansion (\ref{A*Bfewterms}),
which is nominally an expansion in $\hbar$, each successive bracket
involves one higher derivative with respect to momentum (see
(\ref{fewMoyalbrackets})).  In this work the operators of interest are
polynomials in the nuclear momentum $\Pvechat_\alpha$, which is of
order $\kappa^{-2}$, and their Weyl symbols are corresponding
polynomials in $\Pvec_\alpha$.  Therefore each momentum derivative
kills one factor of $\Pvec_\alpha$ and raises the order of $\kappa$ by
2.  In other words, the Moyal expansion couples every power of $\hbar$
with a power of $\kappa^2$.  The result is that every term of order
$\hbar^n$ that we encounter in the Moyal expansion behaves in the
large-amplitude scaling as of order $\kappa^{2n}$.  In the following
when we say, ``first order,'' we shall mean, order $\hbar$ or
$\kappa^2$, while ``second order'' means order $\hbar^2$ or
$\kappa^4$, etc.

The fact that the $\hbar$ ordering is closely related to the $\kappa$
ordering should not be surprising, in view of the close connection
between Born-Oppenheimer theory, especially in the large-amplitude
ordering, with semiclassical theory (for example, the fact that the
nuclear de~Broglie wavelength is of order $\kappa^2$).  But it is a
semiclassical theory only in the nuclear variables, not the
electronic.

\subsection{The Generator $G_1$}
\label{G1resolvent}

Returning now to (\ref{H'symbolseries2}) we wish to choose $G_1$ to
kill the off-block-diagonal elements of $H_1$, so that we will have
$H'_{1,nk}=H'_{1,kn}=0$ for $k\ne n$.  Thus, we require
\begin{equation}
  [G_1,H_0]_{nk} = -H_{1,nk},
  \qquad (k\ne n).
  \label{G1criterion}
\end{equation}
Of the two terms in $H_0$ (see (\ref{Hmol0defWeyl})) the kinetic
energy is a multiple of the identity matrix which commutes with the
matrix $G_1$, while for the potential energy we use (\ref{Wkladiab})
since we are working (for the time being) in the adiabatic basis.
Then the commutator in (\ref{G1criterion}) is easy to evaluate and we
obtain,
\begin{equation}
  G_{1,nk}(\epsilon_k-\epsilon_n) = -H_{1,nk},
  \qquad (k\ne n),
  \label{G1nkcond}
\end{equation}
or,
\begin{equation}
  G_{1,nk}= -i\sum_{\alpha=1}^N \frac{1}{M_\alpha}
  \frac{\Pvec_\alpha\cdot\Fvec_{\alpha;nk}(\Xvec)}
  {\epsilon_n(\Xvec)-\epsilon_k(\Xvec)},
  \qquad (k\ne n).
  \label{G1soln}
\end{equation}

If we carry out the same procedure for the $(kn)$-block, we obtain
\begin{equation}
  G_{1,kn}= -i\sum_{\alpha=1}^N \frac{1}{M_\alpha}
  \frac{\Pvec_\alpha\cdot\Fvec_{\alpha;kn}(\Xvec)}
  {\epsilon_k(\Xvec)-\epsilon_n(\Xvec)},
  \qquad (k\ne n),
  \label{G1knsoln}
\end{equation}
so that $G_{1,kn} = -G_{1,nk}^*$, as required if $\Ghat_1$ is to be
anti-Hermitian.  These are symbol equations; if we want the operator
$\Ghat_1$ then we can convert back to operators.  For example, we find
\begin{equation}
  \Ghat_{1,nk}= -i\sum_{\alpha=1}^N\frac{1}{2M_\alpha}\left[
  \Pvechat_\alpha\cdot\frac{\Fvec_{\alpha;nk}(\Xvechat)}
  {\epsilon_n(\Xvechat)-\epsilon_k(\Xvechat)}+
  \frac{\Fvec_{\alpha;nk}(\Xvechat)}
  {\epsilon_n(\Xvechat)-\epsilon_k(\Xvechat)}
  \cdot\Pvechat_\alpha\right],
  \qquad (k\ne n).
  \label{Ghat1soln}
\end{equation}
in which the ordering of products must be respected.  It will be
appreciated that Weyl symbols are easier to work with than
operators.

Equation~(\ref{G1criterion}) is all we require of $G_1$; since it does
not determine the on-block components of $G_1$, we set these to zero,
that is, $G_{1,nn}=0$ and $G_{1,kl}=0$ for $k\ne n$ and $l\ne n$.  The
result is that $G_1$ is purely off-block-diagonal.

\subsection{Energy Denominators, the Excluded Region, and Landau-Zener}
\label{energydens}

The solution (\ref{G1soln}) diverges where the energy denominator
$\epsilon_n(x)-\epsilon_k(x)$ goes to zero.  Since $k\ne n$ this can
only happen if the energy denominator goes to zero for the nearest
neighbor, $k=n\pm 1$.  We define the ``degeneracy manifold'' as the
subset of nuclear configuration space where
$\epsilon_n(x)=\epsilon_k(x)$ for $k=n\pm1$, that is, where level $n$
is degenerate with one of its immediate neighbors.  (If $n$ is the
ground state, then it has only one neighbor, the first excited state.)
Degeneracy manifolds are usually called ``seams.''  As $x$ approaches
the degeneracy manifold, $G_1$ diverges and the perturbation expansion
(\ref{HtoH'}) breaks down.  Actually, not only does $G_1$ diverge, but
so do the derivative couplings and therefore $H_1$ itself.  This is
shown by the Feynman-Hellman formula,
\begin{equation}
  \Fvec_{\alpha;nk}(x) = -\frac{\matrixelement{ax;n}{\nabla_\alpha
    H_e(x)}{ax;k}}{\epsilon_n(x)-\epsilon_k(x)},
   \qquad (k\ne n),
   \label{FvecFH}
\end{equation}
which applies in the adiabatic basis, as shown.  Substituting this
into (\ref{G1soln}), we see that $G_1$ diverges as
$(\epsilon_n-\epsilon_k)^{-2}$ as we approach the degeneracy manifold.

If the nearest neighbor energy denominator
$\epsilon_n(x)-\epsilon_k(x)$ in (\ref{G1soln}) or in (\ref{FvecFH})
is of order $\kappa^0=1$, then the size of the generators $G_{1,nk}$
is determined by the factor $\Pvec_\alpha/M_\alpha$, the nuclear
velocity, which is of order $\kappa^2$.  In this case the generators
$G_{1,nk}$ (all of them, for all $k$) are of order $\kappa^2$, and
therefore small.  But the nearest neighbor $\epsilon_n(x)
-\epsilon_k(x)$ is of order unity if the distance from the degeneracy
manifold, call it $\Delta x$, is also of order unity.  (This is true
in the sense of the $\kappa$-scaling.)  At such distances, the
generators $G_{1,nk}$ are small and the first few terms of the series
(\ref{HtoH'}) are rapidly decreasing.  This is all we require of the
series for practical purposes; we do not require convergence.  Thus,
Moyal perturbation theory, applied to a single-surface problem for the
adiabatic separation of the privileged surface $n$ from the others, is
valid when $\Delta x$ is of order $\kappa^0=1$.

It is also valid for smaller $\Delta x$, but not too much.  As we
approach the degeneracy manifold, the nearest-neighbor $G_{1,nk}$
grows and when it becomes of order unity the series (\ref{HtoH'})
breaks down, that is, the terms are no longer rapidly decreasing.
Since $G_{1,nk}$ goes as $(\epsilon_n-\epsilon_k)^{-2}$, the breakdown
occurs when the energy difference is of order $\kappa$, which implies
that $\Delta x$ is also of order $\kappa$.  This defines, at least in
terms of its $\kappa$-scaling, the region that must be excluded for
the validity of single-surface methods.  Inside the excluded region it
will be necessary to use multiple-surface methods, which are discussed
in subsection~\ref{multiplesurfaces}.

Since the nuclear de~Broglie wave length is of order $\kappa^2$, the
excluded region, which is within a distance of order $\kappa$ from the
degeneracy manifold, contains on the order of $\kappa^{-1}$ de~Broglie
wave lengths.  The actual number in a practical situation depends on
the magnitudes of various matrix elements, the actual value of the
nuclear momentum and other quantities.  But making a first stab based
on the $\kappa$-scaling alone, we estimate that the excluded region
extends roughly 0.1 atomic unit away from the degeneracy manifold, and
that it contains roughly 10 nuclear de~Broglie wavelengths in this
distance.  This is all in the large-amplitude scaling.

In the case of small molecules, it is possible to visualize the
excluded region geometrically, but it is necessary to distinguish
between the nuclear configuration space and ``shape space.''  Nuclear
configuration space is the space $\Reals^{3N}$ upon which the nuclear
positions in the lab frame, $\Xvec_\alpha$, are coordinates.  Nuclear
configuration space contains information about the center-of-mass
position and the orientation of the molecule.  Shape space, on the
other hand, is the $(3N-6)$-dimensional space in which the
center-of-mass position and orientation of the molecule are ignored.
Coordinates on shape space are translationally and rotationally
invariant quantities such as bond lengths and angles.  In the case of
triatomics ($N=3$), shape space is topologically one half of
$\Reals^3$ (\cite{LittlejohnReinsch95}), while the codimension~2
degeneracy manifold is a curve in this space
(\cite{Yarkony96,Yarkony01,Yarkony04a,Yarkony04b}).  The excluded
region is therefore a tube whose width is of order $\kappa$
surrounding this curve.  This curve may bifurcate
(\cite{Gordonetal98,Yarkony97b}), and, where it does, the excluded
region surrounds it like the sleeves of a shirt.  The geometry of
shape space and its relation to nuclear configuration space is
explained in detail by \cite{LittlejohnReinsch97}.

The size of the excluded region depends on the nuclear momentum, and
grows in proportion to the square root of the latter.  If we change the
scaling of the nuclear momentum, we find other scalings for the size
of the excluded region.  In the small-amplitude scaling $\Pvec_\alpha$
is of order $\kappa^{-1}$, as explained above.  This means that the
size of the excluded region is of order $\kappa^{3/2}$, or roughly
0.03 atomic units.  Conversely, if we make $\Pvec_\alpha$ larger than
the large-amplitude value of $\kappa^{-2}$, we can make the excluded
region larger.  If we make $\Pvec_\alpha$ large enough, we can make the
size of the excluded region as large as order $\kappa^0=1$, which
means that the excluded region is everywhere and the perturbation
expansion fails everywhere.  This occurs when $\Pvec_\alpha$ is of
order $\kappa^{-4}$, so that the nuclear velocity
$\Pvec_\alpha/M_\alpha$ is of order $\kappa^0=1$.  This means that the
nuclear velocity is of the same order as the electron velocity, at
which point naturally there is no adiabatic separation of time scales.
Such momenta are too large to be of much interest for the usual
chemical processes, but it is nice to see that the theory breaks down
completely where it should.

The generator $G_1$ has the same structure as the one-dimensional,
Landau-Zener transition probability (\cite{Landau32, Zener32,
Stuckelberg32, NikitinUmanskii84}), which can be written as
\begin{equation}
	\exp\left(-\frac{2\pi V_{12}^2}{\hbar v_0 |V'_{11}-V'_{22}|}
	\right).
	\label{LZcoef}
\end{equation}
Apart from the notation and numerical constants, the dimensionless
quantity in the exponential has the same structure as the reciprocal
of $\hbar\,G_{1,nk}$, which is also dimensionless.  To see this we
identify $v_0$ in (\ref{LZcoef}), which refers to the velocity with
which the one-dimensional transition point is passed, with
$\Pvec_\alpha/M_\alpha$; we identify $V_{12}$ in (\ref{LZcoef}), which
is otherwise one half the minimum energy difference between two levels
at the transition point, with $(\epsilon_n-\epsilon_k)/2$; and we
identify $V'_{11}-V'_{22}$ in (\ref{LZcoef}), which is the
$x$-derivative of the difference in the diagonal elements of the
Hamiltonian matrix, with $\matrixelement{x;n} {\nabla_\alpha
H_e}{x;k}$.  Thus, the quantity $\exp[-\pi/(2\hbar|G_{1,nk}|)]$, for
nearest neighbor $k$, behaves something like a local Landau-Zener
transition probability, and the excluded region is where this quantity
is of order unity.

The excluded region defines what we may think of as the ``size of a
conical intersection,'' that is, the distance over which it exerts its
influence in preventing the adiabatic separation of potential energy
surfaces.

\subsection{Diabatic Basis on $\Sspace^\perp(x)$}
\label{diabaticbasis}

When nondegenerate, the adiabatic basis vector $\ket{ax;k}$ is only
determined by (\ref{axkdef}) to within a phase, and when degenerate
the set of degenerate vectors is only determined to within the choice
of an orthonormal frame inside the degenerate eigenspace.  Phase and
frame conventions can be chosen at each and every point $x$ of nuclear
configuration space, and some such choice is implied in the notation
$\ket{ax;k}$, including in the continuum.  But Born-Oppenheimer theory
generates derivatives of the basis vectors, for example, inside the
derivative couplings, so we must ask whether the phase and frame
conventions for the basis $\{\ket{ax;k}\}$ can be made as smooth
functions of $x$ over some region of interest of the nuclear
configuration space.

The answer is no, for topological reasons it is impossible, in
general, to make a smooth assignment of phase and frame conventions
for the vectors of the adiabatic basis, and singularities (places
where the basis vectors are not differentiable with respect to $x$)
are inevitable.  Such singularities occur on surfaces of two different
types.  First, when level $k$ is degenerate with level $k\pm 1$, the
derivatives $\nabla_\alpha\ket{ax;k}$ are not defined.  The manifolds
where this occurs are like the degeneracy manifolds introduced
earlier, except they apply to any $k$, not just $k=n$.  In the
electrostatic model these manifolds have codimension 2.  As we
approach such a manifold along a curve, the frame has a limit, but the
limit depends on the direction of approach so derivatives are not
defined on the manifold itself.  (The singularity is like that of the
orthonormal frame in 3-dimensional space in spherical coordinates,
$(\rvechat,\thetavechat,\phivechat)$, as we approach the $z$-axis.)
The second type of surface upon which the adiabatic basis is singular
consists of codimension-1 manifolds that emanate from the degeneracy
manifolds. These manifolds can be moved about by gauge transformations
like a branch cut in complex variable theory.  The adiabatic basis
vectors suffer a change in sign when crossing such a manifold and so
are not differentiable there.

These manifolds (of both types) proliferate as $k$ increases, and
present serious conceptual complications when derivatives of the
adiabatic basis arise. In addition there is the problem of the
continuum, where the concept of a nearest neighbor breaks down.  For
these reasons we advocate replacing the adiabatic basis on
$\Sspace^\perp(x)$ with a diabatic basis, which is a choice of an
orthonormal frame $\{\ket{dx;k}, k\ne n\}$, with $d$ for ``diabatic,''
that is a smooth function of $x$.  Diabatic bases are usually used
only in the subspace $\Sspace(x)$, but here we are using them in
$\Sspace^\perp(x)$.  The fact that a Hilbert space always possesses a
discrete basis means that the diabatic basis in $\Sspace^\perp(x)$ can
be chosen to be discrete; there is no continuum.  The fact that the
diabatic basis on $\Sspace^\perp(x)$ can be chosen to be a smooth
function of $x$, at least in some region of nuclear configuration
space, is discussed by \cite{LittlejohnRawlinsonSubotnik22}.  (For
example, one can use parallel transport to create such a basis.)  In
addition, one can choose a diabatic basis that is invariant under time
reversal, so that the derivative couplings $\Fvec_{\alpha;kl}(x)$ are
real.

Therefore we now change the definition of the working basis, defining
\begin{equation}
  \ket{x;k}=\begin{cases}
    \ket{ax;n} & k=n,\\
    \ket{dx;k} & \text{otherwise.}
    \end{cases}
    \label{workingbasisdef}
\end{equation}
In the working basis, the electronic Hamiltonian has matrix elements,
\begin{equation}
  \matrixelement{x;k}{H_e(x)}{x;l} =
  \begin{cases}
    \epsilon_n(x) & k=l=n,\\
    W_{kl}(x) & \text{$k\ne n$ and $l \ne n$},\\
    0 & \text{otherwise}.
    \end{cases}
    \label{Wkldef1}
\end{equation}
The derivative couplings are still defined by (\ref{Fkldef}) but are
now taken with respect to the (new) working basis, and the expressions
(\ref{Hmolndefs}) for the Hamiltonian are still valid.

The adiabatic basis is widely used in the literature on
Born-Oppenheimer theory, in which formal sums over an infinite number
of basis states, including the continuum, are of frequent occurrance.
Our scruples over the use of this basis have nothing to do with the
practical impossibility of the numerical evaluation of such sums, but
rather with the singularities of the adiabatic basis as $x$ is varied.
These present such a confusing picture that it is hard to know what
one is really talking about when derivatives with respect to $x$ are
taken.  Our purpose in switching to a discrete, diabatic basis is to
obtain a formalism in which we can be confident that the answers we
obtain are correct.

On the other hand, for a fixed value of $x$ the adiabatic basis states
do form a complete set, and there is no harm in expanding vectors or
operators in terms of them.  Therefore for $k\ne n$ we will consider
it permissible to use the adiabatic basis vectors $\ket{ax;k}$ in
cases where no derivatives of the basis vectors appear and the results
are independent of the phase and frame conventions of those vectors,
for example, in (\ref{R(epsilon,x)def}) below.  In cases where
derivatives of the basis vectors do appear we shall require the
diabatic basis (or the working basis, as we are now defining it).

The fact that the adiabatic basis is complete follows from the
self-adjointness of the electronic Hamiltonian $H_e(x)$.  It has been
claimed that the electronic Hamiltonian is not self-adjoint
(\cite{Sutcliffe12}), but we believe the arguments for this are not
well founded and that the matter is settled by \cite{Jecko14}.

However, since we are using the one adiabatic basis vector
$\ket{ax;n}$ on the privileged subspace, we should address its
singularites. We will assume that we are working in the region where
the Moyal perturbation theory is valid, so that $\Delta x$ is larger
than order $\kappa$ and so that we avoid the degeneracy manifold and
its singularities.  We can also avoid the $-1$ jumps in the phase
of the adiabatic basis if our region is not so large as to encircle a
conical intersection; this can be achieved if our region is simply
connected (\cite{JuanesMarcosAlthorpeWrede05, Althorpe06, Althorpe12,
LittlejohnRawlinsonSubotnik23}).  In the following we will assume that
$\ket{ax;n}$ is free of singularities in the region under
consideration, so that its derivatives are well defined.

\subsection{The Resolvent $R(\epsilon,x)$}
\label{resolvent}

Now that we have changed the working basis, we must return to the
calculation of the generator $G_1$, starting with (\ref{G1criterion}),
and revise it.  The $(nk)$-component of the commutator for $k\ne n$ is
now
\begin{eqnarray}
	[G_1,H_0]_{nk}&=&\sum_p G_{1,np}\,H_{0,pk}
	-\sum_p H_{0,np}\, G_{1,pk}
	=\sum_{p\ne n} G_{1,np}\,W_{pk}(x)
	-\epsilon_n(x) G_{1,nk}
	\nonumber\\
	&=&\sum_{p\ne n}G_{1,np}\bigl[W_{pk}(x)-\epsilon_n(x)\,
	\delta_{pk}\bigr]=-H_{1,nk},
	\label{G1H0commutator}
\end{eqnarray}
where again the kinetic energy part of $H_0$ does not contribute and
where we use (\ref{Wkldef1}) for the potential energy.  In addition,
in the first sum only terms $p\ne n$ contribute since $W_{pk}$ is
block-diagonal, and in the second sum only the term $p=n$ contributes,
for the same reason.  

We now introduce the matrix of the resolvent operator in the working
basis, $R_{kl}(\epsilon,x)$ for $k,l\ne n$, defined by
\begin{equation}
	\sum_{p\ne n} \bigl[\epsilon\,\delta_{kp} -W_{kp}(x)\bigr]
	R_{pl}(\epsilon,x) = \delta_{kl}, \qquad (k,l\ne n),
	\label{Rkldef}
\end{equation}
so that $R(\epsilon,x)$ is basically the inverse of $\epsilon-H_e(x)$
on the space $\Sspace^\perp(x)$.  We also define
$R_{kl}(\epsilon,x)=0$ when $k=n$ or $l=n$, so that $R(\epsilon,x)=0$
on $\Sspace(x)$ and the matrix $R_{kl}(\epsilon,x)$ is
block-diagonal.  The operator $R(\epsilon,x)$ is given explicitly in
the adiabatic basis by
\begin{equation}
	R(\epsilon,x) = \sum_{k\ne n} \frac{
	\ketbra{ax;k}{ax;k}}{\epsilon-\epsilon_k(x)};
	\label{R(epsilon,x)def}
\end{equation}
by $R_{kl}(\epsilon,x)$ we mean the matrix elements of this operator
in the working basis.  This operator is defined as long as $\epsilon$
is not equal to any of the eigenvalues of $H_e(x)$ on
$\Sspace^\perp(x)$.  In particular, $R(\epsilon_n,x)$ is defined over
the region in which we are working, which avoids the degeneracy
manifold.

Now we may solve (\ref{G1H0commutator}) for $G_{1,nk}$,
\begin{eqnarray}
  G_{1,nk} &=& \sum_{l\ne n} H_{1,nl}\,R_{lk}(\epsilon_n,x)
  =-i\sum_{l\ne n} \sum_{\alpha=1}^N \frac{1}{M_\alpha}
  \Pvec_\alpha\cdot \matrixelement{x;n}{\nabla_\alpha}{x;l}
  \matrixelement{x;l}{R(\epsilon_n,x)}{x;k}
  \nonumber\\
  &=& +i\sum_{\alpha=1}^N \frac{1}{M_\alpha}
  \Pvec_\alpha\cdot \bigl(\nabla_\alpha\bra{x;n}\bigr)
  R(\epsilon_n,x)\ket{x;k}, \qquad (k\ne n).
  \label{G1nkdiab}
\end{eqnarray}
In the last step we have used (\ref{Fkldef}) to transfer  the
derivative to $\bra{x;n}$, we have extended the $l$-sum to include the
term $l=n$ which vanishes anyway, and then removed the resolution of the
identity.  

The Feynman-Hellman formula (\ref{FvecFH}) applies only in the
adiabatic basis but we can obtain an equivalent result in the diabatic
basis.  First we note that the resolvent satisfies
\begin{subequations}
  \begin{eqnarray}
    P(x)R(\epsilon,x) &=& R(\epsilon,x)P(x)=0,
    \label{RPeqn}\\
    R(\epsilon,x) &=& Q(x) R(\epsilon,x) = R(\epsilon,x)Q(x) =
    Q(x)R(\epsilon,x)Q(x),
    \label{RQeqn}
    \end{eqnarray}
\end{subequations}
as well as
\begin{equation}
  R(\epsilon,x)\bigl[\epsilon-H_e(x)\bigr] =
  \bigl[\epsilon-H_e(x)\bigr]R(\epsilon,x) = Q(x).
  \label{RHeqn}
\end{equation}
Next we write $[H_e(x)-\epsilon_n(x)]\ket{x;n}=0$ and apply
$\nabla_\alpha$ to obtain
\begin{equation}
  \bigl[\nabla_\alpha H_e(x) - \nabla_\alpha\epsilon_n(x)\bigr]
  \ket{x;n} + \bigl[H_e(x)-\epsilon_n(x)\bigr]\nabla_\alpha
  \ket{x;n}=0.
\end{equation}
We multiply this by $R(\epsilon_n,x)$ to obtain
\begin{equation}
  R(\epsilon_n,x)\,\nabla_\alpha H_e(x)\ket{x;n}
  =Q(x)\nabla_\alpha\ket{x;n},
	\label{gradketeqn}
\end{equation}
where we use (\ref{RQeqn}) and where $R$ moves past the $c$-number
$\nabla_\alpha\epsilon_n(x)$ and annihilates $\ket{x;n}$.  This can be
regarded as a result of first-order perturbation theory, in which
$\delta H_e=\sum_\alpha \xivec_\alpha\cdot\nabla_\alpha H_e(x)$, where
$\xivec_\alpha$ is a small displacement in nuclear configuration
space. It is often needed in Hermitian conjugated form,
\begin{equation}
  \bigl(\nabla_\alpha\bra{x;n}\bigr) Q(x) = \bra{x;n}
  \nabla_\alpha H_e(x)\, R(\epsilon_n,x).
  \label{delalphabraeqn}
\end{equation}
For example, the off-block-diagonal derivative couplings can now be
written
\begin{eqnarray}
  \Fvec_{\alpha;nk}(x) &=& -\bigl(\nabla_\alpha\bra{x;n}\bigr)\ket{x;k}
  =-\bigl(\nabla_\alpha\bra{x;n}\bigr)Q(x)\ket{x;k}
  \nonumber\\
  &=& -\matrixelement{x;n}{\nabla_\alpha H_e(x)\,
    R(\epsilon_n,x)}{x;k}, \qquad (k\ne n).
  \label{FvecFHdiab}
\end{eqnarray}
This is the analog of the Feynman-Hellman formula (\ref{FvecFH}) in the
diabatic basis.    The other off-diagonal block is also useful,
\begin{equation}
	\Fvec_{\alpha;kn}(x) = -\Fvec_{\alpha;nk}(x)^*
	=\matrixelement{x;k}{R(\epsilon_n,x)\,\nabla_\alpha H_e(x)}
	{x;n}, \qquad (k\ne n).
	\label{FvecFHdiabalt}
\end{equation}
  
Similarly, we can replace $R$ by $QR$ in (\ref{G1nkdiab}) and then
substitute (\ref{delalphabraeqn}) to obtain
\begin{equation}
  G_{1,nk} = i\sum_{\alpha=1}^N\frac{1}{M_\alpha} \Pvec_\alpha
  \cdot\matrixelement{x;n}{\nabla_\alpha H_e(x)\,
    R(\epsilon_n,x)^2}{x;k}, \qquad (k\ne n).
  \label{G1nkdiabalt}
\end{equation}
Equations (\ref{G1nkdiab}) and (\ref{G1nkdiabalt}) give us alternative
versions of the $(nk)$-components, for $k\ne n$, of $G_1$ in the
diabatic basis.  The derivation has nowhere relied on dubious
derivatives of the adiabatic basis vectors $\ket{ax;k}$ for $k\ne n$.
The components $G_{1,kn}=-G^*_{1,nk}$ of the other off-diagonal block
are obtained by the anti-Hermiticity of $G_1$.  The block-diagonal
components of $G_1$ vanish, as in the calculation using the adiabatic
basis, so that $G_1$ is purely off-block-diagonal.

The operator $R(\epsilon_n,x)$, when acting on an arbitrary wave
function, has an effect whose order of magnitude is its largest
eigenvalue, which by (\ref{R(epsilon,x)def}) is the nearest neighbor
value of $1/(\epsilon_n-\epsilon_k)$.  Thus (\ref{G1nkdiabalt}) shows
that $G_{1,nk}$ in the diabatic basis behaves like the nearest
neighbor value of $(\epsilon_n-\epsilon_k)^{-2}$, as in the adiabatic
basis.

\subsection{The Second Order Hamiltonian}
\label{secondorderH}

We have found the generator $\Ghat_1$, whose Weyl symbol is given by
(\ref{G1nkdiab}) or (\ref{G1nkdiabalt}), that kills the
off-block-diagonal components of $\Hhat_1$.  To describe this
situation it is convenient to say that an operator or its symbol is
``even'' if it is block-diagonal, and ``odd'' if it is purely
off-block-diagonal.  This terminology is borrowed from the
Foldy-Wouthuysen transformation (\cite{FoldyWouthuysen50,
BjorkenDrell64}), which closely resembles the procedure we are
following here in block-diagonalizing the Born-Oppenheimer
Hamiltonian. An arbitrary symbol or operator matrix is neither even
nor odd but can be uniquely decomposed into its even and odd parts,
which for example in the case of a symbol matrix we write as
$A=A^e+A^o$.  When we multiply matrices of operators or their Weyl
symbols, the product of an even times an even or an odd times an
odd matrix is even, while the product of an odd times an even matrix
or vice versa is odd.  That is, $(AB)^e=A^eB^e+A^oB^o$ and
$(AB)^o=A^eB^o+A^oB^e$.  The same rule applies to commutators of
matrices.

Concerning the operators or symbol matrices discussed so far, $H_0$ is
purely even, $H_0=H_0^e$, and $G_1$ is purely odd, $G_1=G_1^o$.  As for
$H_1$ (see (\ref{Hmol1defWeyl})), it vanishes on the $\Sspace$-block
due to the time-reversal invariance of the basis states, that is,
$H_{1,nn}=0$, but it has nonvanishing elements in the
$\Sspace^\perp$-block.  Altogether, it has both even and odd parts,
$H_1 = H_1^e+H_1^o$.  Likewise we have $H_2=H_2^e+H_2^o$.  The
symbol matrix  $T_1=[G_1,H_0]_*=[G_1^o,H_0^e]_*$ (see (\ref{T1def}))
is purely odd, $T_1=T_1^o$, as are the terms $T_{10}$, $T_{11}$, etc.,
of its $\hbar$-expansion.   The condition that $\Ghat_1$ kill the
off-block-diagonal terms in the Hamiltonian through order $\kappa^2$
can be written as
\begin{equation}
	T_{10}=T_{10}^o =[G_1,H_0] = -H_1^o.
	\label{G1cond}
\end{equation}
This makes $H'_1$ (see (\ref{H'symbolseries2})) purely even,
\begin{equation}
	H'_1 = H_1+T_{10} = H_1 +[G_1,H_0] = H_1^e,
	\label{H'_1iseven}
\end{equation}
which was the point of the first transformation.

As for $H'$ at second order, we can use (\ref{H'symbolseries2}) and
(\ref{G1cond}) to write
\begin{equation}
	H'_2 = H_2 + [G_1,H_1^e] + \frac{1}{2}[G_1,H_1^o] + T_{11},
	\label{H'_2v2}
\end{equation}
which decomposes into even and odd parts,
\begin{subequations}
	\label{H'_2evenodd}
	\begin{eqnarray}
	H^{\prime\,e}_2 &=&
	H_2^e +\frac{1}{2}[G_1,H_1^o],
	\label{H'_2even}\\
	H^{\prime\,o}_2 &=&
	H_2^o + [G_1,H_1^e] +T_{11}.
	\label{H'_2odd}
	\end{eqnarray}
\end{subequations}

The unitary transformation $\Uhat_1=\exp(\hbar \Ghat_1)$ has killed
the off-block-diagonal elements of the Hamiltonian at first order, but
not at second order, as we see from (\ref{H'_2odd}).  To kill these we
apply a second unitary transformation, $\Uhat_2=\exp(\hbar^2\Ghat_2)$,
defining a second transformed Hamiltonian,
\begin{equation}
	\Hhat'' = e^{\hbar^2\Ghat_2} \, \Hhat'\, e^{-\hbar^2\Ghat_2}
	= \Hhat' + \hbar^2 [\Ghat_2,\Hhat'] + \ldots
	\label{U2xfm}
\end{equation}
Converting this operator equation to symbols and expanding in $\hbar$,
we find $H''_0=H'_0$, $H''_1=H'_1$ and
\begin{equation}
	H''_2 =  H'_2 + [G_2,H_0].
	\label{H''expn2}
\end{equation}
Thus, to kill the off-block-diagonal terms in the Hamiltonian to
second order we require
\begin{equation}
	[G_2,H_0] = -H^{\prime\,o}_2,
	\label{G2cond}
\end{equation}
a condition on $G_2$ that may be compared to condition (\ref{G1cond})
on $G_1$.  Combined with (\ref{H''expn2}), (\ref{G2cond}) shows that
$H''_2=H^{\prime\,e}_2$, which is given by (\ref{H'_2even}).

Thus we do not need $G_2$ in order to get $H''_2$, the transformed
Hamiltonian at second order, which is the more interesting object.
Note also that the higher order transformations
($\Uhat_3$, etc.) will not change the Hamiltonian through second
order, so that $\Hbar$ and $H''$ agree through second order (see
(\ref{Hbardef})).  Therefore we will express the answers in terms of
$\Hbar$, not $H''$.  Finally, we are only interested in the
$(nn)$-component of $\Hbar$, which is the effective (scalar)
Hamiltonian driving the wave function on surface $n$.

We will call the Weyl symbol of this scalar Hamiltonian $K$, so that
$K=\Hbar_{nn}$.  (The operator itself is $\Khat$.)  Then we have
$K_0=H_{0,nn}$, which is given by (\ref{H0nndef}), and
$K_1=H_{1,nn}=0$.  As for $K_2=\Hbar_{2,nn}$, it is the
$(nn)$-component of (\ref{H'_2even}), which has two terms.  We write
$K_2=K_{21}+K_{22}$ for these, of which the first term, $K_{21} =
H_{2,nn}$, is the diagonal Born-Oppenheimer correction seen in
(\ref{HBOdef}).  In fact, we have $K=H_{BO}+K_{22}$ to second order,
so that $K_{22}$ is another correction, in addition to the diagonal
Born-Oppenheimer correction.  The new correction $K_{22}$ is
\begin{equation}
	K_{22} = \frac{1}{2}[G_1,H_1^o]_{nn}
	=\frac{1}{2}\sum_{k\ne n} 
	G_{1,nk}\,H_{1,kn} - 
	\frac{1}{2}\sum_{k\ne n} H_{1,nk}\,G_{1,kn}.
	\label{K22def}
\end{equation}
The sums are only over $k\ne n$ because $G_{1,nn}=H_{1,nn}=0$.

We work on the first sum.  The quantity $G_{1,nk}$ is given by
(\ref{G1nkdiabalt}), while $H_{1,kn}$ is given by (\ref{Hmol1defWeyl})
with $l\to n$, into which we substitute (\ref{FvecFHdiabalt}).  This
gives
\begin{equation}
	H_{1,kn}=-i\sum_\beta \frac{1}{M_\beta}\Pvec_\beta\cdot
	\matrixelement{x;k}{R(\epsilon_n,x)\nabla_\beta H_e(x)}
	{x;n}.
	\label{H1knexpr}
\end{equation}
With this expression for $H_{1,kn}$ the first sum in (\ref{K22def})
can be extended to all $k$ since the term $k=n$ vanishes anyway, after
which we have a resolution of the identity.  In this way we evaluate
the first sum in (\ref{K22def}).   When we evaluate the second sum,
including the minus sign, we find that it is equal to the first sum.

Overall, we obtain
\begin{eqnarray}
	K_{22} &=& \sum_{\alpha,\beta}
	\frac{1}{M_\alpha M_\beta}
	\Pvec_\alpha \cdot
	\matrixelement{x;n}{\nabla_\alpha H_e(x)\,
	R(\epsilon_n,x)^3\,\nabla_\beta H_e(x)}{x;n}\cdot \Pvec_\beta
	\nonumber\\
	&=& \sum_{\alpha,\beta}
	\frac{1}{M_\alpha M_\beta}
	\Pvec_\alpha\cdot \bigl(\nabla_\alpha \bra{x;n}\bigr)
	R(\epsilon_n,x) \bigl(\nabla_\beta \ket{x;n}\bigr)
	 \cdot \Pvec_\beta,
	\label{K22def1}
\end{eqnarray}
where in the second form we have used (\ref{RQeqn}) and
(\ref{gradketeqn}).   By inserting resolutions of the identity in the
working basis this can be written
\begin{equation}
	K_{22}=-\sum_{\alpha,\beta} \frac{1}{M_\alpha M_\beta}
	\sum_{kl} [\Pvec_\alpha\cdot\Fvec_{\alpha;nk}(x)]
	R_{kl}(\epsilon_n,x)
	[\Fvec_{\beta;ln}(x)\cdot\Pvec_\beta],
	\label{K22def2}
\end{equation}
where the derivative couplings are evaluated in the working basis.
Finally, transforming  this to the adiabatic basis we obtain
\begin{equation}
	K_{22}=-\sum_{\alpha,\beta}
	\frac{1}{M_\alpha M_\beta}
	\sum_{k\ne n} 
	\frac{[\Pvec_\alpha\cdot\Fvec_{\alpha;nk}(x)]
	[\Fvec_{\beta;kn}(x)\cdot\Pvec_\beta]}
	{\epsilon_n(x)-\epsilon_k(x)},
	\label{K22def3}
\end{equation}
where now the derivative couplings are evaluated in the adiabatic
basis.  This is the result we would have obtained if we had worked
strictly in the adiabatic basis, without worrying about its
singularities.

\subsection{Discussion of $K_{22}$}
\label{K22discussion}

Equations (\ref{K22def1})--(\ref{K22def3}) give the symbol of the
operator $\Khat_{22}$.  The operator itself can be obtained by
replacing $\Xvec_\alpha$ and $\Pvec_\alpha$ by $\Xvechat_\alpha$ and
$\Pvechat_\alpha$, in the order shown, plus commutator terms which are
of higher order in $\hbar$ and can be neglected.  The resulting
operator is Hermitian.  

The term $K_{22}$ should be included in Born-Oppenheimer descriptions
of nuclear motion, but it has usually been neglected, in spite of the
fact that it is of the same order of magnitude as the diagonal
Born-Oppenheimer correction, which has received extensive attention.
(At least, this is true in the large-amplitude ordering.)  Admittedly
both terms are small, and in many applications it may be reasonable to
neglect one or both.  Nevertheless, $K_{22}$ grows with nuclear
momentum while the diagonal Born-Oppenheimer correction does not, and
at sufficiently large nuclear momenta it will become important.

This term has been discussed in depth by \cite{Matyus18}, who calls
it the ``nonadiabatic mass correction.''  The idea is that this
term modifies the expression for the kinetic energy of the nuclei by
adding a contribution due to the electrons that are carried along by
them during nuclear motions.  We refer to \cite{Matyus18} for more
details, mentioning also \cite{Kutzelnigg07}.  Here we add a few
further comments about $K_{22}$.

One is that this term is more difficult to evaluate numerically than
the other terms in the Born-Oppenheimer Hamiltonian.  \cite{Matyus18}
has given ways of evaluating it, basically by inverting the operator
$\epsilon_n-H_e$ numerically.  This requires careful attention to the
choice of the subspace on which the inversion is carried out, and
to numerical issues due to the fact that $\epsilon_n-H_e$ has a
nontrivial kernel.  

Actually the resolvent operator is ubiquitous in higher order
Born-Oppenheimer calculations, appearing not only in $K_{22}$ but also
in matrix elements of the electronic momentum and angular momentum
(\cite{Nafie83,FreedmanNafie86}).  These are needed in studies of
radiative processes such as vibrational circular dichroism
(\cite{StephensLowe85,BuckinghamFowlerGalwas87,Nafie92}), as well as
in expressions for electronic currents
(\cite{Barthetal09,OkuyamaTakatsuka09}).  A method for evaluating such
matrix elements has been implemented by \cite{Stephens85} and similar
ideas have been proposed in a more general context by
\cite{Patchkovskii12}.  Basically, one calls on the fact that matrix
elements involving the resolvent occur in perturbation theory.  By
concocting a perturbation of the right form, one can make the energy
shifts in perturbation theory or their derivatives agree with the
desired matrix elements.

Another remark is that including $K_{22}$, the entire $K=\Hbar_{nn}$, that
is, the Weyl symbol of the operator $\Khat$, is the eigenvalue of the
symbol matrix $H_{kl}$, computed through second order.   Since
$H_{kl}$ is already diagonal to lowest order in $\hbar$ its
eigenvalues to higher order can be computed by means of perturbation
theory, taking into account the off-diagonal contributions from $H_1$
and $H_2$.  At second order this gives both $K_{21}$ and $K_{22}$.
This is interesting in terms of the work of
\cite{LittlejohnFlynn91}, who show that the Weyl symbol of the operator driving
a given mode or polarization of a system of coupled wave equations is
the eigenvalue of the symbol matrix of the wave operator, plus two
corrections.  The first is a correction due to Berry's phase, and the
second, sometimes called the ``no-name'' term
(\cite{EmmrichWeinstein96}), involves the Poisson bracket of the
polarization vectors.  In the Born-Oppenheimer case, the Berry's phase
term vanishes on a single surface, due to time-reversal invariance,
while the Poisson bracket or no-name term is of third order in $\hbar$
and so does not appear.  This gives a complete accounting through
second order of the Hamiltonian $K$ relevant for a single surface in
Born-Oppenheimer theory, including $K_{22}$.

\subsection{Tensor Notation and the Generator $G_2$}
\label{abbrevG2}

We now introduce tensor notation for the kinetic energy metric, as
explained in Appendix~\ref{kemetric}, which is convenient especially
for second-order calculations.  Note that henceforth there will be an
implied summation on repeated indices $\mu$, $\nu$, etc (one lower,
one upper), where $\mu,\nu=1,\ldots,3N$.  Note also that $P_\mu$ (with
a lower or covariant index) is the usual momentum, which is of order
$\kappa^{-2}$, while $P^\mu$ (with an upper or contravariant index) is
the same as the velocity, which is of order $\kappa^2$.  That is,
$P^\mu = {\dot X}^\mu$.  In the new notation
Eqs.~(\ref{HmolndefsWeyl}) become
\begin{subequations}
\label{HmolndefsWeyl1}
\begin{eqnarray}
	H_{0,kl} &=& \left(\frac{1}{2}P^\mu P_\mu\right) \delta_{kl}
	+ W_{kl}(x),
	\label{Hmol0defWeyl1}\\
	H_{1,kl} &=&
	-i\,P^\mu \,F_{\mu;kl}(x) = -i\,P^\mu 
	\matrixelement{x;k}{\partial_\mu}{x;l},
	\label{Hmol1defWeyl1}\\
	H_{2,kl} &=&
	-\frac{1}{2} \sum_p F^\mu{}_{;kp}(x)\, F_{\mu;pl}(x)=
	\frac{1}{2} (\partial^\mu \bra{x;k})(\partial_\mu \ket{x;l}),
	\label{Hmol2defWeyl1}
\end{eqnarray}
\end{subequations}
where we write the derivative couplings as
\begin{equation}
	F_{\mu;kl}(x) = \matrixelement{x;k}{\partial_\mu}{x;l}
	=-(\partial_\mu\bra{x;k})\ket{x;l},
	\label{Fmukldef}
\end{equation}
with
\begin{equation}
	F^\mu{}_{;kl}(x) = G^{\mu\nu}\,F_{\nu;kl}(x).
	\label{F^mudef}
\end{equation}
Notice that $F_{\mu;kl}=O(1)$ while $F^\mu{}_{;kl}=O(\kappa^4)$.
The metric $G^{\mu\nu}$ or $G_{\mu\nu}$ (see (\ref{G_munudef} and
(\ref{Guppermunudef})) is not to be confused with the generators $G_1$,
$G_2$, etc.

In the new notation the Feynman-Hellman formula (\ref{gradketeqn})
becomes
\begin{equation}
	R(\epsilon_n,x)\,[\partial_\mu H_e(x)]\,\ket{x;n}
	=Q(x)\partial_\mu\ket{x;n},
	\label{FHcov}
\end{equation}
and the generator $G_1$ becomes
\begin{equation}
	G_{1,nk} = i\,P^\mu (\partial_\mu\bra{x;n})\,
	R(\epsilon_n,x)\ket{x;k}, \qquad (k\ne n),
	\label{G1nkcov}
\end{equation}
which is equivalent to (\ref{G1nkdiab}), while the diagonal element
$K=\Hbar_{nn}=K_0+\hbar K_1+\hbar^2 K_2+\ldots$ of the diagonalized
Hamiltonian is given by
\begin{subequations}
\label{Kndefs}
\begin{eqnarray}
	K_0 &=& \frac{1}{2} P^\mu P_\mu + \epsilon_n(x),
	\label{K0def}\\
	K_1 &=& 0,
	\label{K1def}\\
	K_2 &=& K_{21} + K_{22}.
	\label{K2def}
	\end{eqnarray}
\end{subequations}
Here $K_{21}=H_{2,nn}$ is the diagonal Born-Oppenheimer correction,
\begin{equation}
	K_{21} = \frac{1}{2}(\partial^\mu \bra{x;n})
	(\partial_\mu \ket{x;n}) 
	=-\frac{1}{2} \matrixelement{x;n}{\partial^\mu\partial_\mu}
	{x;n},
	\label{K21def}
\end{equation}
which is equivalent to the last term in (\ref{HBOdef}), and $K_{22}$
is the extra correction,
\begin{equation}
	K_{22} = P^\mu P^\nu (\partial_\mu \bra{x;n}) R(\epsilon_n,x)
	(\partial_\nu \ket{x;n}),
	\label{K22defcov}
\end{equation}
which is a version of (\ref{K22def1}).  

The second generator $G_2$ is worked out in Appendix~\ref{derivG2}.
It is purely odd (off-block-diagonal) and anti-Hermitian.   Its
$(nk)$-block for $k\ne n$ is given by
\begin{eqnarray}
  G_{2,nk}= &-& P^\mu P^\nu \, \partial_\nu[(\partial_\mu\bra{x;n})
  R(\epsilon_n,x)]R(\epsilon_n,x)\ket{x;k}
  \nonumber\\
  &+& [\partial^\mu\epsilon_n(x)]\,
  (\partial_\mu\bra{x;n})\,R(\epsilon_n,x)^2\,\ket{x;k}
  \nonumber\\
  &+&\frac{1}{2} (\partial_\mu\bra{x;n})
  \partial^\mu[R(\epsilon_n,x)\,\ket{x;k}], \qquad (k\ne n).
  \label{G2soln}
\end{eqnarray}
Notice that the final term contains derivatives of the basis vectors
$\ket{x;k}$ on $\Sspace^\perp$.  The other off-diagonal block is given
by $G_{2,kn}=-G_{2,nk}^*$.  
  
\subsection{Changes of Diabatic Basis in $\Sspace^\perp$}
\label{changebasisSperp}

On $\Sspace^\perp$, the working basis vectors are the same as the
diabatic basis vectors, that is, $\ket{x;k}=\ket{dx;k}$ for $k\ne n$
(see (\ref{workingbasisdef})).  But the diabatic basis on
$\Sspace^\perp$ is highly arbitrary, that is, we can change it
according to
\begin{equation}
  \ket{dx;k} = \sum_{l\ne n} \ket{d'x;l}\,V^\perp_{lk}(x),
	\qquad (k\ne n),
  \label{diabchangebasis}
\end{equation}
where $V^\perp_{lk}$ is a smooth function of $x$.  Here $\ket{dx;k}$
is the old diabatic basis on $\Sspace^\perp$ and $\ket{d'x;k}$ is the
new one.  We restrict $V^\perp_{lk}$ to be unitary so that the new
basis is orthonormal, and restrict it further to be real and
orthogonal so that the new basis is invariant under time reversal (we
are assuming that the old basis is invariant under time reversal).

According to (\ref{psikdef}), under the change of basis
(\ref{diabchangebasis}) the wave function transforms according to 
\begin{equation}
	\psi'_k(x) = \sum_{l\ne n} V^\perp_{kl}(x) \, \psi_l(x),
	\qquad (k\ne n).
	\label{psixfmlaw}
\end{equation}
The wave function $\psi_k$ is not invariant under the change of
diabatic basis on $\Sspace^\perp$, but, as we shall say, it transforms
as a vector.

Physical results cannot depend on the choice of diabatic basis
vectors on $\Sspace^\perp$.  Note that the second order Hamiltonians
$K_{21}$ and $K_{22}$, given by (\ref{K21def}) and (\ref{K22defcov}),
make no reference to the diabatic basis on $\Sspace^\perp$ and thus
are invariant under a change of that basis.  The wave function
$\psi_k$ is not invariant but transforms in a simple manner (as a
vector) under a change of basis on $\Sspace^\perp$.  This makes it
easy to construct invariants, for example, $\sum_{k\ne n}
|\psi_k(x)|^2$.  The components $G_{1,nk}$ of the
generator $G_1$ are not invariant but do transform as a vector in the
index $k$, as seen by (\ref{G1nkcov}).   This means that physical
results will never involve individual components $G_{1,nk}$, but
rather sums over $k\ne n$ of $G_{1,nk}$ times something that
transforms as a vector in $k$, in order to make an invariant.   

On the other hand, it is clear from (\ref{diabchangebasis}) that
derivatives of the diabatic basis vectors in $\Sspace^\perp$, things
that look like $\partial_\mu\ket{x;k}$ for $k\ne n$, have a more
complicated transformation law, one that involves derivatives of the
transformation matrix $V^\perp_{kl}(x)$.  Derivatives of the basis vectors
on $\Sspace^\perp$ do not occur in the components $G_{1,nk}$ of the
generator $G_1$, but they do occur in $G_2$, as shown in
Appendix~\ref{derivG2}.  Such terms complicate the construction of
quantities that are invariant under the change of basis
(\ref{diabchangebasis}).

This complication arises because we are using ordinary derivatives
$\partial_\mu$ instead of a properly defined covariant derivative,
which is the ordinary derivative plus a correction term.   The
correction term is, however, of higher order in $\kappa^2$ than the
ordinary derivative, so the use of covariant derivatives mixes up the
expansion in powers of $\kappa^2$.  We will stick with ordinary
derivatives, to keep the formalism as simple as possible.

\subsection{Transformation of the Wave Function}
\label{transfwavefun}

Following the discussion of Sec.~\ref{dressing}, we write
$\Hbarhat\psibar=E\psibar$ or
$\Hbarhat\psibar=i\hbar\,\partial\psibar/\partial t$ for the
Schr\"odinger equation in the completely dressed representation (with
hats on operators), where $\psibar$ is an infinite-dimensional vector
and $\Hbarhat$ an infinite-dimensional matrix.  But $\Hbarhat$ is
block-diagonal and the wave function $\psibar$ is nonzero only in the
slot $k=n$, so we write $\psibar_k =\delta_{kn}\,f$ where $f$ is the
single wave function on surface $n$, satisfying $\Khat f=Ef$ or $\Khat
f=i\hbar\,\partial f/\partial t$, where $\Khat=\Hbarhat_{nn}$.  The
symbol $K$ of operator $\Khat$ is given by (\ref{Kndefs}).  If we
solve this equation, we obtain $f$ and hence $\psibar$ in the fully
dressed Born-Oppenheimer representation.

The wave function in the original Born-Oppenheimer representation is
given by $\psi=\Uhat^\dagger\psibar$, that is, $\psi_k=\sum_l
(\Uhat^\dagger)_{kl} \psibar_l$.  All the components $\psi_k$ are
nonzero.  The component $k=n$ is the largest and approximately equal
to $f$.  Keeping this one and neglecting all the others is equivalent
to the usual Born-Oppenheimer approximation.  Thus, finding the other
components, $\psi_k$ for $k\ne n$, as well as the corrections to
$\psi_n$, amounts to finding corrections to the Born-Oppenheimer
approximation.  It is feasible to calculate these corrections by hand
to second order, as we shall now do.

Since $\psibar_k=\delta_{nk}\,f$, the transformation equations are
\begin{subequations}
\label{psinkdefs}
\begin{eqnarray}
	\psi_n &=& (\Uhat^\dagger)_{nn}\, f=(\Uhat_{nn})^\dagger f
        =(\Uhat^e_{nn})^\dagger f,
	\label{psinf}\\
	\psi_k &=& (\Uhat^\dagger)_{kn}\, f=(\Uhat_{nk})^\dagger f
        =(\Uhat^o_{nk})^\dagger f,
	\qquad (k\ne n),
	\label{psink}
\end{eqnarray}
\end{subequations}
where, as we see, only the even part of $\Uhat$ contributes to $\psi_n$
while only the odd part contributes to $\psi_k$ for $k\ne n$.  On the
other hand, we have
\begin{equation}
  \Uhat = \ldots \Uhat_2 \Uhat_1 = \ldots \exp(\hbar^2 \Ghat_2)
  \exp(\hbar\Ghat_1) 
  = 1 + \hbar\Ghat_1 +\hbar^2\left(\Ghat_2 + \frac{1}{2}\Ghat_1^2\right) 
  + \ldots,
  \label{Uhatexpn}
\end{equation}
so, taking into account the fact that $\Ghat_1$ and $\Ghat_2$ are
purely odd, we can write (\ref{psinkdefs}) as
\begin{subequations}
\label{psinkdefsv2}
\begin{eqnarray}
  \psi_n &=& f + \frac{\hbar^2}{2} [(\Ghat_1^2)_{nn}]^\dagger f,
  \label{psinfv2}\\
  \psi_k &=& \hbar(\Ghat_{1,nk})^\dagger f
  +\hbar^2 (\Ghat_{2,nk})^\dagger f, \qquad (k\ne n),
  \label{psikfv2}
\end{eqnarray}
\end{subequations}
which are valid through second order.  Note that the first correction
to $\psi_n=f$ occurs at second order. 

We work first on the term involving $\Ghat_1^2$.  Since $\Ghat_1^2$ is
Hermitian (as a matrix of operators) its diagonal elements are
Hermitian (as scalar operators), and the $\dagger$ is unnecessary in
(\ref{psinfv2}).  Converting operators into symbols, we have
\begin{eqnarray}
  (\Ghat_1^2)_{nn} &=& \sum_k \Ghat_{1,nk} \Ghat_{1,kn}
  \longleftrightarrow \sum_k G_{1,nk} * G_{1,kn}
  =-\sum_k |G_{1,nk}|^2 + \ldots
  \nonumber\\
  &=& -P^\mu P^\nu\,
  (\partial_\mu\bra{x;n})\,R(\epsilon_n,x)^2 (\partial_\nu\ket{x;n})
  +\ldots,
  \label{G1squaredeqn}
\end{eqnarray}
where we use $\longleftrightarrow$ as in Appendix~\ref{Moyal}, where
we have used $G_{1,kn}=-G_{1,nk}^*$ and where the ellipsis indicates
higher order terms in the expansion of the $*$-product, which we
neglect since we are working at second order.  In the final step we
have used (\ref{G1nkcov}) and evaluated the sum.  When we convert the
final expression, a Weyl symbol, back into an operator, we are free to
put the factors $\Phat^\mu$, $\Phat^\nu$ in any order, since the
differences among the choices are commutators that are of higher
order in $\hbar$.  With this understanding, we can write
(\ref{psinfv2}) as
\begin{equation}
   \psi_n = f -\frac{\hbar^2}{2}
   (\partial_\mu\bra{x;n})\,R(\epsilon_n,x)^2 (\partial_\nu\ket{x;n})
   \Phat^\mu \Phat^\nu f.
   \label{psinresult}
\end{equation} 

As for (\ref{psikfv2}), we will need
\begin{eqnarray}
   (\Ghat_{1,nk})^\dagger &\longleftrightarrow&
   G_{1,nk}^* = -i\,P^\mu\bra{x;k}\,R(\epsilon_n,x)\,
   (\partial_\mu\ket{x;n})\nonumber\\
   &\longleftrightarrow&
   -i\,\bra{x;k}\,R(\epsilon_n,x)\,
   (\partial_\mu\ket{x;n})\,\Phat^\mu
   -\frac{\hbar}{2}\partial^\mu[
   \bra{x;k}\,R(\epsilon_n,x)\,(\partial_\mu\ket{x;n})],
   \label{G1nkconversion}
\end{eqnarray}
where we use (\ref{G1nkcov}) and a version of
(\ref{f(x)potherorder}).  Since $G_1$ occurs at first order, we must
keep the commutator term (the final term) in (\ref{G1nkconversion}) to
obtain results valid to second order.

A similar treatment of $G_2$ (see (\ref{G2soln})) gives rise to three
terms, one of which is
\begin{equation}
   \frac{1}{2}\{\partial^\mu[\bra{x;k}\,R(\epsilon_n,x)]\}
   (\partial_\mu\ket{x;n}),
\end{equation}
the term that depends on the derivatives of $\bra{x;k}$.  But so does
the commutator term in (\ref{G1nkconversion}), and when the two are
combined the derivatives of $\bra{x;k}$ drop out.   Thus
(\ref{psikfv2}) becomes
\begin{eqnarray}
   \psi_k &=& -i\hbar\,\bra{x;k}\,R(\epsilon_n,x)\,
   (\partial_\mu\ket{x;n})\,\Phat^\mu f 
   + \hbar^2\{-\bra{x;k}\,R(\epsilon_n,x)\,
   \partial_\nu[R(\epsilon_n,x)\,(\partial_\mu\ket{x;n})]\,
   \Phat^\mu \Phat^\nu
   \nonumber\\
   &+& [\partial^\mu\epsilon_n(x)]\, 
   \bra{x;k} \, R(\epsilon_n,x)^2 (\partial_\mu\ket{x;n})
   -\frac{1}{2} \bra{x;k}\,R(\epsilon_n,x)\,
   (\partial^\mu\partial_\mu\ket{x;n})
   \}f, \quad (k\ne n).
   \label{psikresult}
\end{eqnarray}
We see that through second order, $\psi_k$ transforms as a vector, as
it must. 

We can also write out the expansion of the wave function
$\Psi(\Xvec,\rvec)$, the solution of the original molecular
Schr\"odinger equation, (\ref{molScheqnmolrep}) or
(\ref{molScheqnmolreptdep}).   We write
\begin{equation}
   \Psi(\Xvec,\rvec)=\Psi_0(\Xvec,\rvec) + 
   \Psi_1(\Xvec,\rvec) + \ldots,
   \label{Psiexpansion}
\end{equation}
where the subscripts 0, 1, etc., indicate the power of $\kappa^2$.
Then $\Psi_0(\Xvec,\rvec) = \phi_{n}(\Xvec;\rvec)\,f(\Xvec) =
\braket{\rvec}{x;n}f(\Xvec)$, which is the usual Born-Oppenheimer
approximation, and
\begin{eqnarray}
   \Psi_1(\Xvec,\rvec) &=& -i\hbar \sum_{k\ne n} \phi_k(\Xvec;\rvec)\,
   \bra{x;k}\,R(\epsilon_n,x)(\partial_\mu\ket{x;n})\,\Phat^\mu f
   \nonumber\\
   &=&-i\hbar\bra{\rvec}\,R(\epsilon_n,x)(\partial_\mu\ket{x;n})
   \,\Phat^\mu f \nonumber\\
   &=& -i\hbar\sum_{\alpha=1}^N \frac{1}{M_\alpha}
   \bra{\rvec}\,R(\epsilon_n,x)(\nabla_\alpha\ket{x;n})\cdot
   \Pvechat_\alpha f
   \label{Psi1def}
\end{eqnarray}
is the first order correction, written in several different ways.
For brevity we omit the second order correction, but it is easily
written out with the help of (\ref{psinresult}) and
(\ref{psikresult}).

The first order correction to the Born-Oppenheimer wave function was
apparently first derived by \cite{Nafie83}, in the context of small
vibrations, and has been rederived many times since.  Our results
apply in the more general context of large-amplitude motions, and our
second order results (\ref{psinresult}) and (\ref{psikresult}) seem to
be new.

\section{Generalizations and Applications}
\label{generalizationapplication}

In this section we outline the generalization of Moyal perturbation
theory to multisurface problems and we discuss some applications
including the calculation of electronic currents. 

\subsection{Multiple Surfaces}
\label{multiplesurfaces}

For multi-surface problems we define an
index set $I$ of $N_l$ adjacent electronic energy levels,
\begin{equation}
   I=\{k_0,k_0+1,\ldots,k_0+N_l-1\},
   \label{Idef}
\end{equation}
which are regarded as ``strongly coupled,'' at least somewhere in the
region of nuclear configuration space under consideration.  These
levels define a privileged subspace $\Sspace(x)$,
\begin{equation}
   \Sspace(x) = \mathspan \{\ket{ax;k}, k\in I\},
   \label{S(x)defmultisurface}
\end{equation}
which replaces (\ref{S(x)def}). As before, the complementary,
orthogonal subspace is $\Sspace^\perp(x)$.  Most of the formulas
of Sec.~\ref{singlesurface}, which was devoted to single surface
problems in which the privileged subspace was the single level $k=n$,
are generalized to the multi-surface case by replacing $k=n$ and $k\ne
n$ by $k\in I$ and $k\notin I$, respectively.  For example, the
resolvent is still given by (\ref{R(epsilon,x)def}), except now summed
over $k\notin I$, and the projector onto $\Sspace(x)$ becomes
\begin{equation}
   P(x) = \sum_{k\in I} \ketbra{ax;k}{ax;k}.
   \label{P(x)defmultisurface}
\end{equation}
Now we can distinguish between ``internal degeneracies,'' those that
occur among levels $k\in I$, and ``external degeneracies,'' those that
occur between a level $k\in I$ and a level $k\notin I$.  Now the
region of nuclear configuration space must be restricted to exclude
external degeneracies and a margin around them, while internal
degeneracies are allowed. 

Ever since (\ref{workingbasisdef}) we have been using a working basis
which is the diabatic basis on $\Sspace^\perp$.  Now we must
introduce a diabatic basis also on $\Sspace$, that is, a basis
$\ket{dx;k}$ for $k\in I$, so that derivatives of
the basis states are defined.  Now we define the working basis by
\begin{equation}
  \ket{x;k}=\begin{cases}
    \ket{dx;k} & k\in I,\\
    \ket{dx;k} & k\notin I.
    \end{cases}
    \label{workingbasisdefmultisurface}
\end{equation}
To be clear about the notation, there are two diabatic bases, one on
$\Sspace$ and one on $\Sspace^\perp$.   The first is a linear
combination of the adiabatic basis vectors $\ket{ax;k}$ for $k\in I$,
and the second is a linear combination of those for $k\notin I$.  
For the transformation between the diabatic and adiabatic bases on
$\Sspace$ we write
\begin{equation}
    \ket{dx;k} = \sum_{l\in I} \ket{ax;l}\,V_{lk},
    \qquad k\in I,
    \label{diabbasisdef}
\end{equation}
where $V_{lk}$ is an $N_l\times N_l$ orthogonal matrix, 
\begin{equation}
    V_{lk} = \braket{ax;l}{dx;k}.
    \label{Vlkdef}
\end{equation}
Now ``even'' and ``odd'' are defined relative to the sets of indices,
$k\in I$ and $k\notin I$.

The derivation proceeds much as in the case of single surface.  The
generator $G_1$ (that is, its Weyl symbol) is purely odd and satisfies
$[G_1,H_0]=-H_1^o$.  This can be solved for one off-diagonal block,
\begin{equation}
   G_{1,kp}=\sum_{l,m\in I}\sum_{q\notin I}(V^\dagger)_{kl}\,
   V_{lm} \, H_{1,mq}\, R_{qp}(\epsilon_l,x), \qquad 
   \hbox{$k\in I$, $p\notin I$},
   \label{G1defmultisurface1}
\end{equation}
while the other is determined by the anti-Hermiticity of $G_1$.  The
presence of the matrix $V$ in this formula means that effectively we
have carried out the linear algebra of the solution for $G_1$ in the
adiabatic basis, then transformed the result to the working (that is,
diabatic) basis.  The calculation would be formally simpler if we had
defined the working basis to be the adiabatic basis on $\Sspace$, but
we wish to avoid any derivatives of these adiabatic basis vectors since
they are not defined at internal degeneracies (that is, on the seams
of the single surface problem).  Our presentation of the multisurface
problem is designed so that such derivatives never occur (but
derivatives of the diabatic basis states on $\Sspace$ are allowed).

Equation~(\ref{G1defmultisurface1}) raises the question of whether the
solution $G_{1,kp}$ is a smooth function of $x$, since the matrix $V$
is not smooth.  (Matrix $V$ connects the diabatic basis, which is
smooth, with the adiabatic basis, which is not.) To answer this
question we use the logic explained more fully by
\cite{LittlejohnRawlinsonSubotnik22}, that the singularities of the
adiabatic basis are caused by the topological impossibility of making
smooth phase conventions for energy eigenstates in the neighborhood of
a degeneracy, and that correspondingly expressions that are
independent of such phase conventions are smooth.  It is easily seen
that $G_{1,kp}$ given by (\ref{G1defmultisurface1}) is independent of
the phase conventions for the adiabatic basis, and so is a smooth
function of $x$.  A more sophisticated argument can be based on a
contour integral (see Appendix~A of
\cite{LittlejohnRawlinsonSubotnik22}).  Now substituting
(\ref{Hmol1defWeyl1}) into (\ref{G1defmultisurface1}) and carrying out
some manipulations we find
\begin{equation}
   G_{1,kp} = -iP^\mu \sum_{l\in I}
   (V^\dagger)_{kl}\, 
   \matrixelement{ax;l}{[\partial_\mu R(\epsilon_l,x)]}{x;p},
   \qquad \hbox{$k\in I$, $p\notin I$}.
   \label{G1defmultisurface2}
\end{equation}

Results equivalent to (\ref{G1defmultisurface1}) have previously been
reported by \cite{MatyusTeufel19}, who work with an electronic
operator that in our notation could be written as
\begin{equation}
   \sum_{k\in I}\sum_{p\notin I} \ket{x;k}\,G_{1,kp}\,\bra{x;p}
   + \hbox{\rm h.c.}
   \label{MTG1def}
\end{equation}
This is part of their formalism of operator-valued Weyl symbols.  The
use of this formalism does not obviate questions of smoothness, for
example, the projector onto individual energy eigenstates, which
appears in their formulas, is not smooth at internal degeneracies.

The Hamiltonian $\Khat$ on the privileged subspace is an $N_l \times
N_l$ matrix, that is, with indices $k,l\in I$, whose symbol matrix can
be written $K=K_0 + \hbar K_1 + \hbar^2 K_2$, to second order.  As for
$K_0$, it is a version of (\ref{Hmol0defWeyl1}),
\begin{equation}
   K_{0,kl} = \left(\frac{1}{2}P^\mu P_\mu\right)\delta_{kl}
	+ W_{kl},
   \qquad k,l\in I,
   \label{K0kldef}
\end{equation}
where $W_{kl}$ is a full matrix because we are using the diabatic basis
for $k\in I$.  As for $K_1$, it is the same as $H_1$ on the diagonal
block corresponding to $\Sspace$,
\begin{equation}
   K_{1,kl}= -i\,P^\mu 
	\matrixelement{x;k}{\partial_\mu}{x;l},
   \qquad k,l \in I.
   \label{K1kldef}
\end{equation}
Note that in multi-surface problems, $K_1$ has non-vanishing,
off-diagonal components with derivative couplings.  Finally,
$K_2=K_{21}+K_{22}$ consists of two terms, of which $K_{21}$ is a
version of (\ref{Hmol2defWeyl1}),
\begin{equation}
      K_{21,kl} =
	\frac{1}{2} (\partial^\mu \bra{x;k})(\partial_\mu \ket{x;l}),
   \qquad k,l \in I,
   \label{K21kldef}
\end{equation}
which is a multi-surface version of the diagonal Born-Oppenheimer
correction.   The second correction $K_{22}$ can be written 
\begin{equation}
   K_{22,kl} =
     \frac{1}{2}P^\mu P^\nu \sum_{m,n\in I}
     (V^\dagger)_{km}\,
     \matrixelement{ax;m}{[\partial_\mu P(x)]
     [R(\epsilon_m,x)+R(\epsilon_n,x)][\partial_\nu P(x)]}
     {ax;n}\, V_{nl}, \qquad k,l\in I.
     \label{K22multisurface}
\end{equation}
In this expression we have taken derivatives of the projection
operator $P(x)$, which is smooth, but not of the adiabatic basis
states, which are not.  (The projection operator is not to be confused
with the momenta $P^\mu$, $P^\nu$.)  A result equivalent to
(\ref{K22multisurface}) has been previously reported by
\cite{MatyusTeufel19}.

The second generator $G_2$ and the corrections to the wave function
may also be computed.

\subsection{Electronic Momentum}
\label{electronicmomentum}

In single-surface problems the usual (lowest order) Born-Oppenheimer
approximation gives a vanishing result for the matrix elements of
electron momenta (linear and angular) and in the calculation of
electronic currents.  For such problems it is necessary to go to
higher order.  The same applies to the single-surface matrix elements
of any Hermitian electronic operator that is odd under time-reversal.
In this subsection we explain how the issue may be treated within
Moyal perturbation theory.

We begin with the momentum $\pvec_i$ of one of the electrons, which is
odd under time reversal, $T\pvec_i T^\dagger =-\pvec_i$.  This is a
useful practice case, although it has a special feature not shared
with other such operators, namely, the operator $\pvec_i$ can be
represented as an exact time derivative.  That is, let
$\Psi(\Xvec,\rvec,t)$ be the time-dependent wave function of the
molecule.  Then
\begin{equation}
   \matrixelement{\Psi}{\pvec_i}{\Psi}_{Xr} = 
   \frac{m_e}{i\hbar}\matrixelement{\Psi}{[\rvec_i,H]}{\Psi}_{Xr}
   =m_e\frac{d}{dt}\matrixelement{\Psi}{\rvec_i}{\Psi}_{Xr}.
   \label{p=mdr/dt}
\end{equation}
This is exact in the electrostatic model, but if the
Born-Oppenheimer approximation (\ref{BOappx}) is used in it (where
$\Psi$ and $\psi$ are allowed to depend on $t$) then the left-hand
side vanishes while the right-hand side does not.  To reconcile the
two it is necessary to go to higher order in the Born-Oppenheimer
expansion in the evaluation of the left-hand side.  Note that
$\rvec_i$ is even under time reversal, so its matrix element does not
vanish at lowest order.

To evaluate $\matrixelement{\Psi}{\pvec_i}{\Psi}$ we use a version of
(\ref{Psimerule}), 
\begin{equation}
   \matrixelement{\Psi}{A}{\Psi}_{Xr} = 
   \sum_{kl} \matrixelement{\psi_k}{\Ahat_{kl}}{\psi_l}_X=
   \sum_{kl} \matrixelement{\psibar_k}{\Abarhat_{kl}}{\psibar_l}_X,
   \label{Amevariousreps}
\end{equation}
showing how the expectation value of an operator $A$ changes as the
representation moves from the molecular to the original
Born-Oppenheimer to the dressed Born-Oppenheimer (that is, across the
first row of Table~\ref{reptable}).  Here we define $\Abarhat =
\Uhat\Ahat\Uhat^\dagger$, as in (\ref{Hbardef}), using hats to
indicate nuclear operators or matrices thereof.  If the original $A$
is a purely electronic operator then
$\Ahat_{kl}=\matrixelement{x;k}{A}{x;l}$ (see
subsection~\ref{tworepresentations}), which is a purely multiplicative
nuclear operator (that is, a function of $\Xvec$ only).  In this case
there is hardly a difference between the operator $\Ahat_{kl}$ and its
Weyl symbol $A_{kl}$.

We identify the original operator $A$ with $\pvec_i$, and write
$(\pvec_i)$ for the matrix representing it in the original
Born-Oppenheimer representation, so that
\begin{equation}
   (\pvec_i)_{kl} = \matrixelement{x;k}{\pvec_i}{x;l}.
   \label{(p_i)def}
\end{equation}
If we wish to emphasize that this is a nuclear operator, we will write
this matrix as $\widehat{(\pvec_i)}$ or its components as
$\widehat{(\pvec_i)}_{kl}$, but these are purely multiplicative
operators (functions of $\Xvec$ only).   We write
\begin{equation}
    \widehat{(\pvecbar_i)}=\Uhat \widehat{(\pvec_i)} \Uhat^\dagger
    \label{(pbar)def}
\end{equation}
(with an overbar) for the matrix of nuclear operators representing
$\pvec_i$ in the dressed Born-Oppenheimer representation, as in
(\ref{Hbardef}).  Now it is necessary to distinguish the nuclear
operators $\widehat{(\pvecbar_i)}_{kl}$ from their Weyl symbols, which
are $(\pvecbar_i)_{kl}$ (without the hat), because the conjugation by
$\Uhat$ introduces a momentum dependence.

We can write the expectation value of the momentum in three different
representations, moving across the first row of Table~\ref{reptable},
\begin{equation}
   \matrixelement{\Psi}{\pvec_i}{\Psi}_{Xr} =
   \sum_{kl} \matrixelement{\psi_k}{(\pvec_i)_{kl}}{\psi_l}_X =
   \matrixelement{f}{\widehat{(\pvecbar_i)}_{nn}}{f}_X,
   \label{pvec_i3mes}
\end{equation}
where we have used $\psibar_k = \delta_{kn}\,f$ in the dressed
Born-Oppenheimer representation, since we are working on a single
surface.   In this representation the double sum has reduced to a
single term.  

Now there is the question of which representation is best for the
calculation of $\xpecval{\pvec_i}$.  For the molecular representation
we have the wave function (\ref{Psiexpansion}) and (\ref{Psi1def}) and
for the original Born-Oppenheimer representation we have the wave
function (\ref{psinresult}) and (\ref{psikresult}), where $f$ is a
solution of $\Khat f=Ef$ or $\Khat f = i\hbar\partial f/\partial t$.
But we suggest that the fully dressed Born-Oppenheimer representation
is optimal for the calculation, since it does not require us to
transform wave functions.  Instead, it requires us to transform
operators, to obtain the one (scalar) nuclear operator 
\begin{equation}
  \widehat{(\pvecbar_i)}_{nn}  = [\Uhat \widehat{(\pvec_i)}
  \Uhat^\dagger]_{nn}
  = \widehat{(\pvec_i)}_{nn} + \hbar[\Ghat_1,\widehat{(\pvec_i)}]_{nn}
  +\ldots,
  \label{pvecbariexpansion}
\end{equation}
or, in terms of Weyl symbols,
\begin{equation}
  (\pvecbar_i)_{nn} = (\pvec_i)_{nn} + \hbar[G_1,(\pvec_i)]_{nn} 
  +\ldots
  \label{pvecbariexpansion1}
\end{equation}
The first term of this expansion vanishes,
\begin{equation}
  (\pvec_i)_{nn} = \matrixelement{x;n}{\pvec_i}{x;n} = 0,
  \label{pvecinn=0}
\end{equation}
because of time-reversal (this is the vanishing of the expectation
value to lowest order).  With the help of (\ref{G1nkcov}) the
commutator in the correction term is easily worked out,  giving
altogether 
\begin{equation}
   (\pvecbar_i)_{nn} = i\hbar \,P^\mu[
   (\partial_\mu\bra{x;n})R(\epsilon_n,x)\pvec_i\ket{x;n}-
   \bra{x;n}\pvec_i R(\epsilon_n,x)(\partial_\mu\ket{x;n})],
   \label{pvecinncorrec}
\end{equation}
valid to first order.  Actually, due to time reversal, the two terms
shown (including the minus sign in the second term) are equal to one
another, but it is convenient to leave the answer as shown.

Now we may use the fact that
$\pvec_i=m_e(d/dt)\rvec_i=(m_e/i\hbar)[\rvec_i,H_e]$ is an exact time
derivative (this is like (\ref{p=mdr/dt}) but using $H_e$ instead of
the full molecular Hamiltonian $H$).  With this substitution the first
matrix element in (\ref{pvecinncorrec}) becomes
\begin{equation}
  \frac{m_e}{i\hbar} (\partial_\mu\bra{x;n})R(\epsilon_n,x)
  (\epsilon_n-H_e)\rvec_i\ket{x;n}.   
  \label{pvecicommutatorsubstn}
\end{equation}
Treating the second matrix element similarly and using (\ref{RHeqn})
and $(\partial_\mu\bra{x;n})Q(x) =
\partial_\mu\bra{x;n}$ we can rewrite (\ref{pvecinncorrec}) as
\begin{equation}
  (\pvecbar_i)_{nn} = m_eP^\mu \partial_\mu(
  \matrixelement{x;n}{\rvec_i}{x;n}).
  \label{pvecinnrvecform}
\end{equation}
Now using (\ref{symmf(x)psymbol}) to convert the Weyl symbol into an
operator, we can use (\ref{pvec_i3mes}) to write
\begin{eqnarray}
  \xpecval{\pvec_i} &=& \frac{m_e}{2} \int dX\, f(\Xvec,t)^*\,
  [\Phat^\mu (\partial_\mu\matrixelement{x;n}{\rvec_i}{x;n})+
  (\partial_\mu\matrixelement{x;n}{\rvec_i}{x;n})\Phat^\mu]\,f(\Xvec,t)
  \nonumber\\
  &=& m_e\int dX\, J_n^\mu(\Xvec,t)\,
  \partial_\mu\matrixelement{x;n}{\rvec_i}{x;n},
  \label{<p_i>J_n}
\end{eqnarray}
where $J_n^\mu(\Xvec,t)$ is the nuclear current,
\begin{equation}
  J_n^\mu(\Xvec,t) = f(\Xvec,t)^* \left(\frac{-i\hbar\partial^\mu}{2}\right)
            f(\Xvec,t) + \hbox{\rm c.c.}
  \label{J_ndef}
\end{equation}
The current satisfies the continuity equation,
\begin{equation}
  \partial_\mu J_n^\mu(\Xvec,t) + \frac{\partial\rho_n(\Xvec,t)}
  {\partial t}=0,
  \label{Jncontinuity}
\end{equation}
where the nuclear probability density is
\begin{equation}
   \rho_n(\Xvec,t) = |f(\Xvec,t)|^2.
   \label{rhondef}
\end{equation}
Thus we can integrate (\ref{<p_i>J_n}) by parts and discard boundary
terms (assuming $f(\Xvec,t)$ is normalizable) to obtain
\begin{equation}
   \xpecval{\pvec_i}= -m_e\int dX \, [\partial_\mu J_n^\mu(\Xvec,t)]
   \matrixelement{x;n}{\rvec_i}{x;n}
   =m_e\frac{d}{dt}\int dX \,
   \rho_n(\Xvec,t)\,\matrixelement{x;n}{\rvec_i}{x;n}
   =m_e\frac{d}{dt}\xpecval{\rvec_i},
   \label{<p_i><r_i>}
\end{equation}
thereby recovering (\ref{p=mdr/dt}).

We have introduced the usual definitions of the nuclear probability
density and current and invoked the continuity equation that they
satisfy, but we are working in the dressed Born-Oppenheimer
representation and these operations should be justified.  That is,
one should take the usual definitions and properties in the molecular
representation and track them through the changes of representation
across the first row of Table~\ref{reptable}.  It turns out that what
we have done is correct through the order to which we are working but
for brevity we omit the details.

The integral over $X$ on the right-hand side of (\ref{pvec_i3mes}) is
nominally taken over all of configuration space, but the
transformation to the dressed representation on a single surface is
only valid outside the excluded region, that is, away from
degeneracies.  Thus, the manipulations presented are valid if $f$
effectively has support in the allowed region, but if the wave
function extends to the neighborhood of a degeneracy then we must use
a multi-surface approach.  In multisurface problems the expectation
value of the electronic momentum does not vanish at lowest order and
the answers cannot be expressed so simply in terms of nuclear
currents.  Time reversal forces the diagonal matrix elements,
$\matrixelement{x;k}{\pvec_i}{x;l}$ for $k=l$, to vanish, but not the
off-diagonal elements ($k\ne l$).  If both $f_k$ and $f_l$ are nearly
zero in the excluded region then the $X$ integral in
$\matrixelement{f_k}{(\pvec_i)_{kl}}{f_l}_X$ for $k\ne l$ contains an
integrand that is a rapidly oscillating function times a slowly
varying one.  In this case the integral is nearly zero, and the
off-diagonal elements do not contribute.  But if both $f_k$ and $f_l$
extend to the neighborhood of the degeneracy, then the off-diagonal
elements will make a contribution, nominally at lowest order.  On the
other hand, speaking of a two-surface problem, the off-diagonal matrix
elements of the momentum vanish at internal degeneracies and are
proportional to the energy difference as we move away from those
degeneracies.  This should be true over the extent of the excluded
region, which, as explained, is of order $\kappa$.  Overall, the
$\kappa$-ordering of the expectation value of the electronic momentum
displays a somewhat complicated picture in multi-surface problems.
Such problems are currently of considerable interest, and this is a
topic that deserves further development.

\subsection{Electronic Current}
\label{electroniccurrent}
  
We turn now to the electronic probability current, which also vanishes
at lowest order in the Born-Oppenheimer expansion.  We show this by
expressing it as the expectation value of an Hermitian operator that
is odd under time reversal.   We continue with the convention of using
hats on nuclear operators, and omitting them on the corresponding Weyl
symbols.   We do not use hats on electronic operators, but in this
subsection it is important to distinguish between electronic operators
and corresponding quantities that are $c$-numbers.  

In this subsection we use index $\sigma=1,\ldots,3N_e$ to label
the coordinates in the $3N_e$-dimensional electronic configuration
space, writing $r^\sigma$ for the components of the vectors $\rvec_i$,
$i=1,\ldots,N_e$ and $p_\sigma = -i\hbar\partial/\partial r^\sigma$
for the corresponding momenta.   We employ an implied summation over
repeated indices $\sigma$.

We define a purely electronic operator $\rho_{r_0}$, parameterized
by a $c$-number $r_0$ representing a point of the electronic
configuration space,
\begin{equation}
   \rho_{r_0} = \delta(r-r_0).
   \label{rhosubr0def}
\end{equation}
Really $r_0$ is a $3N_e$-vector of $c$-numbers, the components
$r_0^\sigma$ of $r_0$, while $r$ represents the $3N_e$-dimensional
vector of electronic operators whose components are multiplication by
$r^\sigma$.   Then we define the electronic probability density
$\rho(r_0,t)$ by 
\begin{equation}
  \rho(r_0,t)=\matrixelement{\Psi}{\rho_{r_0}}{\Psi}_{Xr}
  = \int dX\, |\Psi(\Xvec,\rvec_0,t)|^2,
  \label{rhoofr0def}
\end{equation}
which is the usual definition.  Note that $\rho_{r_0}$ is an operator
while $\rho(r_0,t)$ is a $c$-number.

Now we compute the time derivative,
\begin{eqnarray}
  \frac{\partial\rho(r_0,t)}{\partial t} &=&
  \frac{d}{dt}\matrixelement{\Psi}{\rho_{r_0}}{\Psi}_{Xr}=
  \frac{1}{i\hbar}\matrixelement{\Psi}{[\rho_{r_0},H]}{\Psi}_{Xr}
  \nonumber\\
  &=& \frac{1}{2m_e}\matrixelement{\Psi}{[p_\sigma \delta_\sigma(r-r_0)
  +\delta_\sigma(r-r_0)p_\sigma]}{\Psi}_{Xr},
  \label{drhodteqn}
\end{eqnarray}
where $H$ is the molecular Hamiltonian (\ref{Hmoldef}), where we have
worked out the commutator, and where $\delta_\sigma(r-r_0)$ means
$(\partial/\partial r^\sigma)\delta(r-r_0)$.  The commutator comes
entirely from the electronic kinetic energy in the Hamiltonian, which
can be written as $p_\sigma p_\sigma/2m_e$.  But since 
\begin{equation}
  \frac{\partial}{\partial r^\sigma}\delta(r-r_0) =
  -\frac{\partial}{\partial r^\sigma_0}\delta(r-r_0),
  \label{deltarr0ident}
\end{equation}
we can write
\begin{equation}
   \frac{\partial\rho(r_0,t)}{\partial t} = 
   -\frac{\partial J^\sigma(r_0,t)}{\partial r_0^\sigma}
   \label{electroncontinuity}
\end{equation}
where
\begin{equation}
  J^\sigma(r_0,t) = \matrixelement{\Psi}{J^\sigma_{r_0}}{\Psi}_{Xr}
  \label{Jsigma(r_0)def}
\end{equation}
and
\begin{equation}
  J^\sigma_{r_0} = \frac{1}{2m_e}[p_\sigma \delta(r-r_0) 
  +\delta(r-r_0)p_\sigma].
  \label{Jsigmasubr0def}
\end{equation}
Note that $J^\sigma_{r_0}$ is an operator parameterized by $r_0$ while
$J^\sigma(r_0,t)$ is the usual electronic current, which is a
$c$-number.  \cite{Diestler12} has obtained results equivalent to
(\ref{Jsigma(r_0)def}) and (\ref{Jsigmasubr0def}) by performing an {\em
ad hoc} quantization of the classical expression for the current.

Equation~(\ref{Jsigma(r_0)def}) expresses the electronic current as
the expectation value of a purely electronic operator
(\ref{Jsigmasubr0def}) that is odd under time reversal, so the result
vanishes to lowest order of the Born-Oppenheimer expansion.  We can
treat this operator exactly as we did $\pvec_i$ in
subsection~\ref{electronicmomentum}, except that $J^\sigma_{r_0}$,
unlike $\pvec_i$, cannot be expressed as an exact time derivative.
That is, we write $(J^\sigma_{r_0})$ for the matrix of operators
representing the electronic current in the Born-Oppenheimer
representation, with components $(J^\sigma_{r_0})_{kl}$.  These are
functions of $\Xvec$ and if we wish to emphasize that they are nuclear
operators we can place a hat
over them, writing $\widehat{(J^\sigma_{r_0})}$ for the matrix or
$\widehat{(J^\sigma_{r_0})}_{kl}$ for its components.   These are then
transformed  to the dressed Born-Oppenheimer representation according to
$\widehat{(\Jbar^\sigma_{r_0})}=\Uhat \widehat{(J^\sigma_{r_0})}
\Uhat^\dagger$, and we write $(\Jbar^\sigma_{r_0})$ for the
corresponding matrix of Weyl symbols (without the hat).  Finally, the
$nn$-component of this is the effective (nuclear) operator
representing the electronic current on the given surface $n$; we give
this a special name, writing
\begin{equation}
   \Ihat^\sigma_{r_0} = \widehat{(\Jbar^\sigma_{r_0})}_{nn},
   \label{Ihatcurrentdef}
\end{equation}
or $I^\sigma_{r_0}$ for the corresponding Weyl symbol.   Then we have
\begin{equation}
   I^\sigma_{r_0} = -2i\hbar \,P^\mu
   \bra{x;n}J^\sigma_{r_0}R(\epsilon_n,x)(\partial_\mu\ket{x;n}),
   \label{I^sigma_{r_0}eqn}
\end{equation}
which is derived just like (\ref{pvecinncorrec}) except that the two
equal terms in the result have been combined.  Then we have
\begin{equation}
   J^\sigma(r_0,t) = \matrixelement{f}{\Ihat^\sigma_{r_0}}{f}_X
   =-2i\hbar\int dX\, J^\mu_n(\Xvec,t)\,
   \bra{x;n}J^\sigma_{r_0}R(\epsilon_n,x)(\partial_\mu\ket{x;n}),
   \label{ecurrentintegral}
\end{equation}
which is derived just like (\ref{<p_i>J_n}).  

Results equivalent to (\ref{ecurrentintegral}) have been derived
previously by \cite{Nafie83} and \cite{Patchkovskii12}, although it
takes some extrapolation to compare our results with Nafie's because
he worked in the small-amplitude regime and expanded his results about
an equilibrium.  What Nafie calls ``complete adiabatic'' versions of
operators such as the dipole and current are the same as our dressed
operators.  He emphasizes that these have a dependence on nuclear
momentum.  Nafie employs an {\em ad hoc} quantization and
dequantization of nuclear operators, roughly equivalent to our use of
Weyl symbols, and he has presented some differential equations
satisfied by the current operators in their dequantized versions
(\cite{Nafie20}).  \cite{Patchkovskii12} allows large-amplitude
motions, as we do.  He evaluates the matrix element in
(\ref{ecurrentintegral}) by inserting a resolution of the identity in
the adiabatic basis just before the resolvent operator.  This gives
\begin{eqnarray}
   J^\sigma(r_0,t) &=& \frac{\hbar^2}{m_e} \int dX\,
   J^\mu_n(\Xvec,t)\, \sum_{k\ne n} \frac{\matrixelement{ax;k}
   {\partial_\mu}{ax;n}}
   {\epsilon_n(x)-\epsilon_k(x)}
   \nonumber\\
   &&\quad \times \left[
   \frac{\partial\phi_{an}^*(x;r_0)}{\partial r_{0\sigma}}
   \phi_{ak}(x;r_0)-
   \phi_{an}^*(x;r_0)\frac{\partial\phi_{ak}(x;r_0)}{\partial
     r_{0\sigma}}\right],
   \label{Patchkovskiicurrent}
\end{eqnarray}
which is essentially the expression given by him.

The sum over all adiabatic states engages the continuum and is not
useful for direct evaluation, but in any case it is unlikely that one
will want to compute the $3N_e$-dimensional electronic current.
Instead, reduced versions obtained by integrating over all electron
coordinates but one are more useful.  See the remarks in
subsection~\ref{K22discussion} on the evaluation of matrix elements
involving the resolvent.  \cite{Diestler13} evaluates the sum in
(\ref{Patchkovskiicurrent}) by replacing the energy denominator
$\epsilon_n-\epsilon_k$ by an appropriately chosen average energy
$\Delta E$.  

If the current operator $J^\sigma_{r_0}$ were an exact time derivative
we could use the trick employed with the electronic momentum $\pvec_i$
to get rid of the resolvent; but it is not.  On the other hand, the
divergence $\partial_\mu J^\mu_{r_0}$ of the current operator is an
exact time derivative.  If we use this fact to eliminate the resolvent
operator, we just recover the continuity equation.  This is a trivial
observation but it emphasizes a point made by \cite{Barthetal09}, that
normally one is interested in the integral of the current over
surfaces that extend to infinity.  The evaluation of these does not
require the use of the resolvent operator.

In the special case that all the nuclei are displaced by the same
displacement $\xivec$ (a 3-vector), it is clear that the electrons are
just carried along by the nuclei and that the expression for the
electronic current should simplify.   This is another case in which we
can get rid of the resolvent operator.  To treat this problem we must
make a digression into the behavior of the basis states $\ket{x;k}$
under translations, including their phase conventions.

\section{Translational Degrees of Freedom, in Greater Depth}
\label{translationaldofsec}

We return to the subject of the translational degrees of freedom,
showing that their elimination can be achieved by means of a unitary
transformation and showing how the rectilinear motion of the center of
mass is manifested in the Born-Oppenheimer and dressed
Born-Oppenheimer representations.  See also our treatment of rotations
in \cite{LittlejohnRawlinsonSubotnik23}.

\subsection{Transformation of $H_e(x)$ Under Translations}
\label{He(x)translations}

 Let $\xivec$ be a displacement vector in 3-dimensional
space, and let $\xi=(\xivec,\xivec,\ldots,\xivec)$ be the corresponding
displacement vector in the $3N$-dimensional nuclear configuration
space.   Also, let $\rvec+\xivec$ stand for the collection
$(\rvec_1+\xivec,\ldots,\rvec_{N_e}+\xivec)$.  Then the translational
invariance of the electronic Hamiltonian is expressed by
\begin{equation}
   H_e(x+\xi;\rvec+\xivec,\pvec)=H_e(x;\rvec,\pvec).
   \label{Hetranslationallyinvariant}
\end{equation}
This is a statement about the functional form of $H_e(x;\rvec,\pvec)$,
which depends only on vector differences among $\Xvec_\alpha$ and
$\rvec_i$.  

We define electronic, nuclear, and total translation operators by
\begin{equation}
   T_e(\xivec)=e^{-i\xivec\cdot\pvec_t/\hbar},
   \qquad
   T_n(\xivec)=e^{-i\xivec\cdot\Pvec_t/\hbar},
   \qquad
   T(\xivec)=T_e(\xivec)T_n(\xivec)=
   e^{-i\xivec\cdot(\Pvec_t+\pvec_t)/\hbar}.
   \label{TeTnTdefs}
\end{equation}
See (\ref{ptdef}) and (\ref{Ptdef}) for definitions of $\pvec_t$ and
$\Pvec_t$.  These are in the molecular representation.  The electronic
translation operators satisfy
\begin{equation}
   T_e(\xivec)\,\rvec_i\,T_e(\xivec)^\dagger = \rvec_i-\xivec,
   \qquad
   T_e(\xivec)\,\pvec_i \,T_e(\xivec)^\dagger = \pvec_i.
   \label{ripitranslationconjugation}
\end{equation}
Thus we have
\begin{equation}
    T_e(\xivec) \,H_e(x;\rvec,\pvec) \,T_e(\xivec)^\dagger =
    H_e(x;\rvec-\xivec,\pvec) = H_e(x+\xi;\rvec,\pvec),
    \label{Heconjugationtranslations1}
\end{equation}
where in the first equality the conjugation by $T_e(\xivec)$ does
nothing to $x$, which is just the set of parameters of the electronic
Hamiltonian, and where the second equality follows from
(\ref{Hetranslationallyinvariant}).  We abbreviate this by writing
\begin{equation}
    T_e(\xivec)\,H_e(x) \,T_e(\xivec)^\dagger = H_e(x+\xi),
    \label{Heconjugationtranslations}
\end{equation}
which is the transformation law of the electronic Hamiltonian under
electronic translations.

\subsection{The Translational Fiber Bundle}
\label{translationalfiberbundle}

\begin{figure}
\includegraphics[scale=0.5]{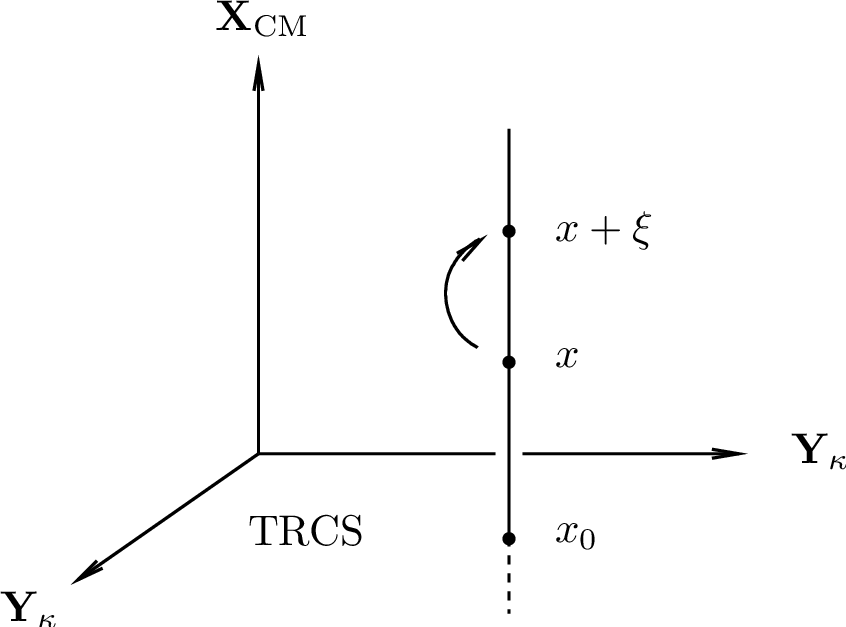}%
\caption{\label{bundle}The translational fiber bundle.}%
\end{figure}

The transformation law (\ref{Heconjugationtranslations}) has a
geometrical interpretation.  A translation acts on the nuclear
configuration space according to $x\mapsto x+\xi$, or, in coordinates,
$(\Xvec_1,\ldots,\Xvec_N) \mapsto
(\Xvec_1+\xivec,\ldots,\Xvec_N+\xivec)$.  In the alternative
coordinate system on the nuclear configuration space (see
subsection~\ref{translationaldof}) this becomes $(\Xvec_{\rm
CM},\Yvec_\kappa) \mapsto (\Xvec_{\rm CM}+\xivec,\Yvec_\kappa)$, that
is, the center of mass is translated and the translationally invariant
Jacobi vectors do not change.  Nuclear configuration space in this
coordinate system is illustrated in Fig.~\ref{bundle}, in which the
vertical $\Xvec_{\rm CM}$-axis is really a 3-dimensional subspace,
while the complementary plane, upon which the Jacobi vectors
$\Yvec_\kappa$ are coordinates, has dimension $3N-3$.  We will call
this plane the ``translation-reduced configuration space'' (TRCS in
the figure).  These two subspaces are orthogonal to one another in the
kinetic energy metric.  

According to (\ref{Heconjugationtranslations}), conjugation by
$T_e(\xivec)$ maps the electronic Hamiltonian at nuclear configuration
$x$ into that at configuration $x+\xi$.  This calls attention to the
surface swept out when the entire translation group is allowed to act
on $x$, that is, the set of all points $x+\xi$ for fixed $x$ when
$\xivec$ runs over $\Reals^3$.  It is the set of points that can be
reached from $x$ by applying translation operators.  It is otherwise
the {\it orbit} of the point $x$ under the action of the translation
group; it is a 3-dimensional subspace parallel to the $\Xvec_{\rm
CM}$-axis.  As drawn it is not a vector subspace because it does not
pass through the origin, but it does pass through a unique point of
the translation-reduced configuration space, labeled $x_0$ in the
figure, whose $\Xvec_{\rm CM}$ coordinate is 0.  The action of the
translation group foliates nuclear configuration space into a
$(3N-3)$-parameter family of 3-dimensional orbits, which are the
fibers of the translational fiber bundle.  In this context, ``fibers''
and ``orbits'' are the same thing; we will call them the
``translational fibers.''  Any point in a given fiber can be reached
from any other point of the same fiber by means of a translation
operator.  The $3N-3$ parameters labeling the translational fibers are
the components of the Jacobi vectors $\Yvec_\kappa$.  The
translation-reduced configuration space is an example of a {\it
section} of the fiber bundle, that is, it is a surface of
complementary dimensionality that cuts through the fibers,
intersecting each at a unique point.  The section plays the role of an
initial value surface.

\subsection{Transformation of Basis Vectors}
\label{transformationbasisFkl}

The phase and frame conventions for the adiabatic basis vectors
$\ket{ax;k}$, assumed to be invariant under time reversal, are
discussed in subsection~\ref{diabaticbasis}.  Thus there is the
question of the phase and frame conventions for these basis vectors as
we vary $x$.  We would like these vectors to be smooth functions of
$x$.

Suppose $\ket{ax_0;k}$ is an eigenvector of $H_e(x_0)$ at point $x_0$
on the translation-reduced configuration space, so that $H_e(x_0)
\ket{ax_0;k} = \epsilon_k(x_0)\ket{ax_0;k}$. Let us define
$\ket{ax;k}$ at other points of the same translational fiber by
\begin{equation}
   \ket{ax;k} = T_e(\xivec)\ket{ax_0;k},
   \label{initvalbasis}
\end{equation}
where $x=x_0+\xi$.  Then $\ket{ax;k}$ is automatically an eigenstate
of $H_e(x)$, 
\begin{equation}
   H_e(x)\ket{ax;k} = \epsilon_k(x_0)\ket{ax;k},
   \label{translatedeigenbasis}
\end{equation}
with an eigenvalue that has not changed.  That is, the eigenvalue is
invariant under translations,
\begin{equation}
   \epsilon_k(x)=\epsilon_k(x+\xi).
   \label{evaluetransformation}
\end{equation}
The use of translation operators to define the phase of the electronic
eigenstates, as in (\ref{initvalbasis}), makes those eigenstates
smooth functions of $x$ along the translational fibers.  If we make a
smooth choice of phase conventions for $\ket{x_0;k}$ for $x_0$ in some
region of the translation-reduced configuration space of
dimensionality $3N-3$, and then extend those definitions along
translational fibers according to (\ref{initvalbasis}), then we will
have smooth conventions over a region of dimensionality $3N$ of the
full nuclear configuration space.

Now let $x_1=x_0+\xi_0$, let $\xi_1$ be another displacement,  and consider
\begin{eqnarray}
   T(\xivec_1)\ket{ax_1;k} &=& T(\xivec_1)T(\xivec_0)\ket{ax_0;k}
   = T(\xivec_1+\xivec_0)\ket{ax_0;k}
   \nonumber\\
   &=& \ket{a,x_0+\xi_0+\xi_1;k} 
   =\ket{a,x_1+\xi_1;k}.
   \label{compositionoftranslations}
\end{eqnarray}
With a change of notation this is
\begin{equation}
   T(\xivec)\ket{ax;k} = \ket{a,x+\xi;k},
   \label{abasistranslationlaw}
\end{equation}
which is like (\ref{initvalbasis}) except that it applies to any point
$x$ on a translational fiber, not just the initial point $x_0$. 

If we assume that the matrix that transforms us from the adiabatic
to the diabatic basis is independent of $\Xvec_{\rm CM}$, then the
transformation law (\ref{abasistranslationlaw}) also applies to the
diabatic basis or any working basis we construct.   This means that
the transformation matrix is constant along translational fibers and
that it depends only on the Jacobi vectors $\Yvec_\kappa$.   Then we
have
\begin{equation}
   T(\xivec)\ket{x;k} = \ket{x+\xi;k},
   \label{basistranslationlaw}
\end{equation}
for all electronic bases that we consider.   This is the
transformation law for electronic basis states under translations.

\subsection{Molecular Momentum in Various Representations}
\label{molecularmomentumvariousreps}

Equation~(\ref{basistranslationlaw}) leads to an important result when
we let the displacement vector $\xivec$ be infinitesimal.  In that
case we can expand the translation operator $T_e(\xivec)$ (see
(\ref{TeTnTdefs})) and the state $\ket{x+\xi;k}$, obtaining
\begin{equation}
   \left[1-\frac{i}{\hbar}\xivec\cdot\pvec_t\right]\ket{x;k} =
   \ket{x;k}+ \sum_{\alpha=1}^N \xivec\cdot\nabla_\alpha\ket{x;k},
\end{equation}
or, since $\xivec$ is arbitrary,
\begin{equation}
   \pvec_t\ket{x;k} = i\hbar\sum_{\alpha=1}^N
   \nabla_\alpha\ket{x;k}.
\end{equation}
Now writing $\Pvec_\alpha=-i\hbar\nabla_\alpha$ for the operator that
acts on the parametric $x$-dependence of $\ket{x;k}$ and using
(\ref{Ptdef}), we can express the result as
\begin{equation}
   (\Pvec_t+\pvec_t)\ket{x;k}=0.
   \label{Pt+ptbasis=0}
\end{equation}

Equation~(\ref{Pt+ptbasis=0}) leads to an important conclusion.  In
the molecular representation the operator $\Pvec_{\rm
CM}=\Pvec_t+\pvec_t$ is the total momentum of the system, which is
conserved.  Let us allow this operator to act on a molecular wave
function $\Psi(\Xvec,\rvec)$, which we expand according to
(\ref{psikdef}).  This gives
\begin{eqnarray}
   \Pvec_{\rm CM} \Psi(\Xvec,\rvec) &=& \sum_k
   [(\Pvec_t+\pvec_t)\psi_k(\Xvec)]\phi_k(x;\rvec)+
   \psi_k(\Xvec)[(\Pvec_t+\pvec_t)\phi_k(x;\rvec)]
   \nonumber\\
   &=& \sum_k[\Pvec_t \psi_k(\Xvec)]\phi_k(x;\rvec),
\end{eqnarray}
where we distribute $\Pvec_t+\pvec_t$, a first-order, linear
differential operator, over the two factors according to the product
rule, where the second major term vanishes according to
(\ref{Pt+ptbasis=0}), and where the operator $\pvec_t$ in the first
major term does not contribute since $\psi_k(\Xvec)$ is independent of
$\rvec$.   

We summarize this by writing out the transformation of $\Pvec_{\rm
CM}$ from the molecular to the Born-Oppenheimer representation,
\begin{equation}
   \Pvec_{\rm CM} = \Pvec_t+\pvec_t \longleftrightarrow
   \Pvechat_t\,\delta_{kl},
   \label{PCMinBOrep}
\end{equation}
where we put a hat on the nuclear operator that appears in the
Born-Oppenheimer representation.  In other words, in the
Born-Oppenheimer representation, the operator that looks like the
total nuclear momentum actually represents the total momentum of the
molecule, including that of the electrons.  Moreover, this is exact.
This in turn implies that
\begin{equation}
   T(\xivec)=e^{-i\xivec\cdot(\Pvec_t+\pvec_t)/\hbar}
   \longleftrightarrow \widehat{T_n(\xivec)}\,\delta_{kl}=
   e^{-i\xivec\cdot\Pvechat_t/\hbar}\,\delta_{kl}
   \label{translation2reps}
\end{equation}
(see (\ref{TeTnTdefs})).  That is, in the Born-Oppenheimer
representation what looks like a purely nuclear translation is
actually a translation of the whole molecule, including the electrons.

\subsection{Transformation of Derivative Couplings Under Translations}
\label{TransformationFkl}

Finally, we consider how the derivative couplings transform under
translations, that is, we ask how $\Fvec_{\alpha;kl}(x+\xi)$ is related to
$\Fvec_{\alpha;kl}(x)$.   As above we let
$\xi=(\xivec,\xivec,\ldots,\xivec)$.   In addition we let
$\eta=(\etavec_1,\etavec_2,\ldots,\etavec_N)$ be an infinitesimal
displacement in which the component vectors $\etavec_\alpha$ are
allowed to be different.   Then
\begin{equation}
   \ket{x+\eta;l}=\ket{x;l}+\sum_{\alpha=1}^N \etavec_\alpha\cdot
   \nabla_\alpha\ket{x;l},
   \label{ketetaexpn}
\end{equation}
so that
\begin{equation}
   \braket{x;k}{x+\eta;l} = \delta_{kl} +
   \sum_{\alpha=1}^N \etavec_\alpha\cdot\Fvec_{\alpha;kl}(x).
   \label{Fkletaexpn}
\end{equation}
But $x$ is a dummy variable in this equation, so we can replace it by
$x+\xi$.   If we do,  the left-hand side does not change, since by
(\ref{basistranslationlaw}) we have
\begin{equation}
   \braket{x+\xi;k}{x+\eta+\xi;l}=
   \matrixelement{x;k}{T_e(\xivec)^\dagger T_e(\xivec)}{x+\eta;l}=
   \braket{x;k}{x+\eta;l}.
\end{equation}
Therefore the right-hand side of (\ref{Fkletaexpn}) does not change,
either, and since the $\etavec_\alpha$ are arbitrary we have
\begin{equation}
   \Fvec_{\alpha;kl}(x) = \Fvec_{\alpha;kl}(x+\xi).
   \label{Fklinvarianttranslations}
\end{equation}
That is, the derivative couplings are invariant under nuclear
translations, they are constant along translational fibers and they
are independent of $\Xvec_{\rm CM}$, depending only on the Jacobi
vectors $\Yvec_\kappa$.  This means that, regarded as nuclear
operators, that is, functions of $\Xvec$ or $\Xvechat$, they commute
with both $\Pvechat_t$ and nuclear translations
$\widehat{T_n(\xivec)}$, where we are careful to put hats on nuclear
operators that appear in the Born-Oppenheimer representation. 

Nuclear translation operators satisfy the commutation relations,
\begin{equation}
   \widehat{T_n(\xivec)}\,\Xvechat_\alpha\,
   \widehat{T_n(\xivec)}^\dagger
   = \Xvechat_\alpha-\xivec, \qquad
   \widehat{T_n(\xivec)}\,\Pvechat_\alpha\,\widehat{T_n(\xivec)}^\dagger
   = \Pvechat_\alpha,
   \label{Tncommrels}
\end{equation}
so from (\ref{Ghat1soln}) we see that the generator $\Ghat_1$ commutes
with nuclear translations and hence with $\Pvechat_t$.  More
precisely, all operators are represented by matrices, but the matrices
for $\Pvechat_t$ and $\widehat{T_n(\xivec)}$ are multiples of the
identity, so each component $\Ghat_{1,kl}$ commutes with both
$\Pvechat_t$ and $\widehat{T_n(\xivec)}$.   The same, it can be shown,
applies to all the higher order generators $\Ghat_n$ for any $n>1$.

This in turn means that all the commutators and iterated commutators
in the dressing of $\Pvechat_t$ vanish (see (\ref{HtoH'}) for the
series of commutators), so that the dressing transformation does
nothing to $\Pvechat_t$ or to the nuclear translation operators
$\widehat{T_n(\xivec})$.  Thus, the total linear momentum of the
molecule, represented by $\Pvec_{\rm CM}=\Pvec_t+\pvec_t$ in the
molecular representation, is represented by $\Pvechat_t\,\delta_{kl}$
in the Born-Oppenheimer and in all dressed Born-Oppenheimer
representations, to all orders.  Moreover, this is exact.

We can summarize this situation in a notation like that used above for
the electronic momentum and current.  We transform the molecular
operator $\Pvec_{\rm CM}=\Pvec_t+\pvec_t$ to the Born-Oppenheimer
representation, writing $\widehat{(\Pvec_{\rm CM})}$ for the resulting
matrix of nuclear operators with matrix elements $\widehat{(\Pvec_{\rm
CM})}_{kl}=\Pvechat_t\,\delta_{kl}$.  The corresponding matrix of Weyl
symbols is $(\Pvec_{\rm CM})$ (without the hat) with components
$(\Pvec_{\rm CM})_{kl}=\Pvec_t\,\delta_{kl}$.  We define the dressed
versions of these by the matrix $\widehat{(\Pvecbar_{\rm CM})}=\Uhat
\widehat{(\Pvec_{\rm CM})}\Uhat^\dagger$ with components
$\widehat{(\Pvecbar_{\rm CM})}_{kl}=\Pvechat_t\,\delta_{kl}$,
or as Weyl symbols, $(\Pvecbar_{\rm CM})_{kl} = \Pvec_t\,\delta_{kl}$.

\subsection{The Molecular Center of Mass}
\label{molecCOM}

We can treat the molecular center of mass $\Rvec_{\rm CM}$  similarly.
We write $\Rvec_{\rm CM} = \Rvec_{1,{\rm CM}} + \Rvec_{2,{\rm CM}}$
for the two terms seen in (\ref{RCMdef}).   When we transform these to
the Born-Oppenheimer representation, we obtain
\begin{subequations}
\label{RCM12BOdefs}
\begin{eqnarray}
   (\Rvec_{1,{\rm CM}})_{kl} &=& \left(\frac{1}{M_{\rm mol}}
   \sum_{\alpha=1}^N M_\alpha \Xvec_\alpha\right)\delta_{kl}, 
   \label{RCM1BOdef}\\
   (\Rvec_{2,{\rm CM}})_{kl} &=& \frac{m_e}{M_{\rm mol}}
   \sum_{i=1}^{N_e} \matrixelement{x;k}{\rvec_i}{x;l}.
   \label{RCM2BOdef}
\end{eqnarray}
\end{subequations}
These are purely multiplicative nuclear operators, that is, functions
of $\Xvec$ only, and if we wish to emphasize their operator aspect we
can put hats over them.  Note that $\Rvec_{2,{\rm CM}}$ is of order
$\kappa^4$ in comparison to $\Rvec_{1,{\rm CM}}$.   

We transform the center of mass to the dressed Born-Oppenheimer
representation according to $\widehat{(\Rvecbar_{\rm CM})} = \Uhat
\widehat{(\Rvec_{\rm CM})}\Uhat^\dagger$, working with Weyl symbols
and carrying out the expansion through order $\hbar^2$.  We start with
$(\Rvec_{1,{\rm CM}})$.  Since this is a multiple of the identity, the
first correction term is
\begin{equation}
  \hbar[G_1,(\Rvec_{1,{\rm CM}})]_{*,kl} = i\hbar^2
  \{G_{1,kl},\Rvec_{1,{\rm CM}}\}+O(\hbar^4),
  \label{RCM1correc}
\end{equation}
which is of order $\hbar^2$ as shown, and all other corrections are of
order $\hbar^3$ or higher.  The bracket shown is the Poisson bracket
and the series is a version of (\ref{starbracketdef}).  But $G_1$ is
purely off-block-diagonal, so through order $\hbar^2$, the corrections
are, too.   The $(nk)$-components of these corrections, for $k\ne n$,
are, according to (\ref{G1nkdiab}) and (\ref{RCM1BOdef}),
\begin{equation}
   \frac{\hbar^2}{M_{\rm mol}} \sum_{\alpha=1}^N (\nabla_\alpha
   \bra{x;n})R(\epsilon_n,x)\ket{x;k}.
   \label{R1CMcorrec}
\end{equation}
But
\begin{equation}
   \sum_{\alpha=1}^N \nabla_\alpha\ket{x;n} = 
   \frac{i}{\hbar}\Pvec_t\ket{x;n}=
   -\frac{i}{\hbar}\pvec_t\ket{x;n}=
   -\frac{m_e}{\hbar^2}\sum_{i=1}^{N_e}(\epsilon_n-H_e)
   \rvec_i\ket{x;n},
   \label{EOMusefulsubstitution}
\end{equation}
so with this substitution the resolvent in (\ref{R1CMcorrec}) is
canceled and the correction becomes
\begin{equation}
  -\frac{m_e}{M_{\rm mol}}\sum_{i=1}^{N_e}
  \matrixelement{x;n}{\rvec_i}{x;k}.
\end{equation}
This exactly cancels the off-diagonal contribution of
(\ref{RCM2BOdef}), and we see that $(\Rvecbar_{\rm CM})$ is purely
even through order $\hbar^2$.  In other words, the dressing
transformation that block-diagonalizes the Hamiltonian also
block-diagonalizes the molecular center-of-mass position, at least
through order $\hbar^2$.

We believe this must be true to all orders.  When the equation of
motion $(d/dt)\Rvec_{\rm CM}=(1/i\hbar)[\Rvec_{\rm CM},H] =\Pvec_{\rm
CM}/M_{\rm mol}$, which is valid in the molecular representation, is
transformed to the dressed Born-Oppenheimer representation it becomes
\begin{equation}
   \frac{1}{i\hbar}[\widehat{(\Rvecbar_{\rm CM})},\Hbarhat]=
   \frac{\Pvechat_t}{M_{\rm mol}} \, I,
\end{equation}
where $I$ is the identity matrix.  But since $\Hbarhat$ and the
identity are purely even, the odd part of $\widehat{(\Rvecbar_{\rm
CM})}$ must commute with $\Hbarhat$ and therefore be a vector constant
of motion.  The only vector constants of motion are the linear and
angular momentum, both of which are purely even in the dressed
Born-Oppenheimer representation.  Therefore the odd part of
$\widehat{(\Rvecbar_{\rm CM})}$ cannot be proportional to these, and
the only choice left is zero.

Assuming this is true, the center of mass of the molecule $\Rvec_{\rm
CM}$ is represented on surface $n$ by a scalar operator (that is, not
a matrix), which is the one component $\widehat{(\Rvecbar_{\rm
CM})}_{nn}$.  This is just as the Hamiltonian is represented by the
operator $\Khat$ and the total momentum of the molecule $\Pvec_{\rm
CM}$ is represented by $\Pvechat_t$.   Therefore we must have an exact
equation of motion for the center-of-mass position in the dressed
Born-Oppenheimer representation,
\begin{equation}
   \frac{d (\Rvecbar_{\rm CM})_{nn}}{dt}=
   \frac{1}{i\hbar}
   [(\Rvecbar_{\rm CM})_{nn},K]_* = 
   \frac{\Pvec_t}{M_{\rm mol}},
   \label{dRCMnn/dt}
\end{equation}
here expressed in terms of Weyl symbols.  But, as we have shown, the
diagonal elements $(\Rvecbar_{\rm CM})_{nn}$ and $(\Rvec_{\rm
CM})_{nn}$ are the same through order $\hbar^2$, so we can write
\begin{equation}
   (\Rvecbar_{\rm CM})_{nn} = (\Rvecbar_{1,{\rm CM}})_{nn} +
   (\Rvecbar_{2,{\rm CM}})_{nn} =
   \frac{1}{M_{\rm mol}} \sum_{\alpha=1}^N
   M_\alpha \Xvec_\alpha + 
   \frac{m_e}{M_{\rm mol}} \sum_{i=1}^{N_e}
   \matrixelement{x;n}{\rvec_i}{x;n}.
   \label{RCMonsurfacen}
\end{equation}
This is the explicit form for the molecular center of mass on surface
$n$ in the dressed Born-Oppenheimer representation, valid through
order $\kappa^4$.

The molecular center of mass must exhibit uniform, rectilinear motion,
which means that it must satisfy the equation of motion
(\ref{dRCMnn/dt}).  We now check that it does.  We write the symbol of
the Hamiltonian as $K=K_0+\hbar K_1 +\hbar^2(K_{21}+K_{22})$, where
$K_0=H_{0,nn}$ is given by (\ref{H0nndef}), where $K_1=0$, where
$K_{21}=H_{2,nn}$ is the diagonal Born-Oppenheimer correction seen in
(\ref{HBOdef}), and where $K_{22}$ is given by (\ref{K22def1}).   The
leading term is 
\begin{equation}
   \frac{1}{i\hbar}[(\Rvecbar_{1,{\rm CM}})_{nn},K_0]_* =
   \{ (\Rvecbar_{1,{\rm CM}})_{nn},K_0 \} =
   \frac{\Pvec_t}{M_{\rm mol}},
\end{equation}
which is the right answer, so the correction terms must cancel.  In
all these calculations the Moyal $*$-bracket reduces to a Poisson
bracket.  One correction term is
\begin{equation}
   \frac{1}{i\hbar}[(\Rvecbar_{2,{\rm CM}})_{nn},K_0]_* =
   \{ (\Rvecbar_{2,{\rm CM}})_{nn},K_0 \} =
   \frac{m_e}{M_{\rm mol}} \sum_{i=1}^{N_e}
   \sum_{\alpha=1}^N \frac{1}{M_\alpha}
   \Pvec_\alpha\cdot\nabla_\alpha
   (\matrixelement{x;n}{\rvec_i}{x;n}).
   \label{(d/dt)RCM1stcorrec}
\end{equation}
The contribution from $K_{21}$ vanishes, and that from $K_{22}$ is
\begin{equation}
   \frac{i}{\hbar}[(\Rvecbar_{1,{\rm CM}})_{nn},\hbar^2 K_{22}]_*
   =\{ (\Rvecbar_{1,{\rm CM}})_{nn},\hbar^2 K_{22} \}
   =\frac{2}{M_{\rm mol}} \sum_{\alpha,\beta}
   \frac{1}{M_\beta}(\nabla_\alpha\bra{x;n})R(\epsilon_n,x)
   (\nabla_\beta\ket{x;n})\cdot\Pvec_\beta.
\end{equation}
But with the aid of (\ref{EOMusefulsubstitution}) this is shown to be
equal to the first correction (\ref{(d/dt)RCM1stcorrec}), with the
opposite sign.   

Thus the corrections do cancel, and the equation (\ref{dRCMnn/dt}) of
uniform, rectilinear motion is shown to be valid through order
$\kappa^4$.   Notice that the correction $K_{22}$ to the Hamiltonian
is required for this result.

\subsection{Translational Component of the Current}
\label{translationalcompscurrent}

We return to the point raised at the end of
subsection~\ref{electroniccurrent}, that it should be possible to
simplify the translational component of the current.   We return to
the expression (\ref{I^sigma_{r_0}eqn}) for the Weyl symbol of the
operator that represents the electronic current on a single surface $n$
in the dressed Born-Oppenheimer representation.  We note that
\begin{equation}
   P^\mu \partial_\mu = \sum_{\alpha=1}^N \frac{1}{M_\alpha}
   \Pvec_\alpha\cdot\nabla_\alpha = 
   \frac{1}{M_t} \Pvec_t \cdot \frac{\partial}{\partial\Xvec_{\rm CM}}
   +\sum_{\kappa=1}^{N-1} \frac{1}{\mu_\alpha}
   \Qvec_\kappa\cdot\frac{\partial}{\partial\Yvec_\kappa},
   \label{Pdotdeltransformation}
\end{equation}
where we use Appendix~\ref{kemetric} and where in the second step we
carry out the coordinate transformation $\Xvec_\alpha \to (\Xvec_{\rm
CM},\Yvec_\kappa)$ on the nuclear configuration space (see
subsection~\ref{translationaldof}).   The first term on the right
gives  the translational component of the current; we note that
\begin{equation}
   \frac{\partial}{\partial\Xvec_{\rm CM}} = 
   \sum_{\alpha=1}^N \nabla_\alpha,
   \label{d/dX_CMexpansion}
\end{equation}
and we define the translational component of the current operator
$I^\sigma_{r_0}$ by
\begin{equation}
   I^\sigma_{{\rm CM},r_0} =
   -\frac{2i\hbar}{M_t} \Pvec_t\cdot
   \matrixelement{x;n}{J^\sigma_{r_0}R(\epsilon_n,x)\sum_{\alpha=1}^N
   \nabla_\alpha}{x;n}.
   \label{IsigmaCMr0def}
\end{equation}
We substitute (\ref{EOMusefulsubstitution}) into this, thereby
eliminating the resolvent and obtaining
\begin{equation}
   I^\sigma_{{\rm CM},r_0} =
   \frac{2im_e}{\hbar M_t} \Pvec_t\cdot\matrixelement{x;n}
   {J^\sigma_{r_0}Q(x)\sum_{i=1}^{N_e}\rvec_i}{x;n}.
   \label{ICMintermed}
\end{equation}
Since $\matrixelement{x;n}{J^\sigma_{r_0}}{x;n}=0$ the projector
$Q(x)$ can be dropped.  The current is a vector in the electronic
configuration space with $3N_e$ components that we can organize as a
collection of $N_e$ 3-vectors, one associated with each electron.  For
that purpose let us replace $\sigma$ by the double index $(jJ)$, as in
Appendix~\ref{kemetric}.  We also substitute the definition
(\ref{Jsigmasubr0def}) of $J^\sigma_{r_0}$.  Then we have
\begin{equation}
   I^{jJ}_{{\rm CM},r_0}=\frac{i}{\hbar
   M_t}\sum_{i=1}^{N_e}\sum_{I=1}^3
   P_t^I \matrixelement{x;n}{[p_{jJ}\delta(r-r_0)+\delta(r-r_0)p_{jJ}]
   r_{iI}}{x;n}.
   \label{IjJCMintermed}
\end{equation}
The operator in the matrix element can be written
\begin{equation}
   p_{jJ}\delta(r-r_0)r_{iI} + \delta(r-r_0)r_{iI}p_{jJ}
   -i\hbar \delta(r-r_0)\delta_{ij}\,\delta_{IJ},
   \label{current3terms}
\end{equation}
whereupon the first two terms constitute an Hermitian operator that is
odd under time-reversal, whose $nn$-matrix element vanishes.  Only the
last term in (\ref{current3terms}) contributes, and we obtain
\begin{equation}
   I^{jJ}_{{\rm CM},r_0} = \frac{P^J_t}{M_t} |\phi_n(x;r_0)|^2.
   \label{IjJCMr0result}
\end{equation}
The current vector is independent of $j$, that is, it is the same for
all the electrons, and it is given by the velocity of the nuclear
center of mass $\Pvec_t/M_t$ times the electronic probability density
at $r_0$.   This is just what we expect for the translational
component of the current.

\subsection{Translational Components of the Derivative Couplings}
\label{translcompsderivcouplings}

Let us return to the original Born-Oppenheimer representation (the
middle entry of the first row of Table~\ref{reptable}), for which the
Hamiltonian is given by (\ref{HmolndefsWeyl}) in Weyl symbol form.
The first order Hamiltonian (\ref{Hmol1defWeyl}) is essentially the
nuclear velocity contracted against the derivative couplings, which
are the components of a differential form on nuclear configuration
space.  The linear differential operator that appears in $H_1$ is the
same as in (\ref{Pdotdeltransformation}), which allows us to pick out
the translational component of the derivative couplings.   We define
\begin{equation}
   \Fvec_{{\rm CM},kl}(x) = \matrixelement{x;k}
   {\frac{\partial}{\partial\Xvec_{\rm CM}}}{x;l},
   \qquad
   \Fvec_{\kappa;kl}(x) = \matrixelement{x;k}
   {\frac{\partial}{\partial\Yvec_\kappa}}{x;l},
   \label{FCM,Fkappadefs}
\end{equation}
and we note that the center-of-mass component can be written in
several different ways,
\begin{equation}
   \Fvec_{{\rm CM},kl}(x) = \matrixelement{x;k}
   {\sum_{\alpha=1}^N \nabla_\alpha}{x;l} =
   \sum_{\alpha=1}^N \Fvec_{\alpha,kl}(x)=
   \frac{i}{\hbar}\matrixelement{x;k}{\Pvechat_t}{x;l} =
   -\frac{i}{\hbar}\matrixelement{x;k}{\pvec_t}{x;l},
   \label{FCMforms}
\end{equation}
where we use (\ref{Pt+ptbasis=0}).  

Thus the translational component of $H_{1,kl}$ is 
\begin{equation}
   -i\frac{\Pvec_t}{M_t}\cdot\Fvec_{{\rm CM},kl}(x).
   \label{H1kltranslational}
\end{equation}
This component sometimes causes confusion, especially when the basis
vectors $\ket{x;k}$ (some of them, at least) are the adiabatic basis,
since it implies transitions between potential energy surfaces when
the nuclear motion is purely translational.  This is most clear in
models in which the nuclear coordinates are treated as classical
variables, since in that case $\Pvec_t/M_t$ is the nuclear velocity
and the term (\ref{H1kltranslational}) becomes simply
$-i\matrixelement{x;k} {(d/dt)} {x;l}$.  For example, it would appear
that in the uniform, rectilinear motion of a hydrogen atom we have
transitions between the $1s$ and $2p$ levels, which does not make
sense physically.  This leads to the suggestion, which we now pursue,
of eliminating the translational component (\ref{H1kltranslational})
of $H_1$ by means of a unitary transformation.  We explain the
hydrogen-atom paradox in subsection~\ref{unitarycoordinate}.

To this end, let the working basis $\ket{x;k}$ be an arbitrary
discrete basis.  We do not assume that it diagonalizes or
block-diagonalizes the electronic Hamiltonian $H_e(x)$, and there is
no privileged subspace or concept of even or odd operators.  As above,
we let $\Ghat_1$ be a generator and we transform the Hamiltonian
according to $\Hhat'=\Uhat_1 \Hhat
\Uhat_1^\dagger$, where $\Uhat_1=\exp(\hbar\Ghat_1)$.   Upon 
translating to Weyl symbols and expanding to first order, this gives
\begin{equation}
   H'_1=H_1+[G_1,H_0],
   \label{H'_1def}
\end{equation}
which is expressed in terms of matrices of Weyl symbols.   We wish to
choose $G_1$ so that
\begin{equation}
   [G_1,H_0]_{kl} = H'_{1,kl}-H_{1,kl} = i\frac{\Pvec_t}{M_t}\cdot
   \Fvec_{{\rm CM},kl}(x)=-\frac{iN_em_e}{\hbar^2 M_t}\Pvec_t\cdot
   \matrixelement{x;k}{[\rvec_{\rm CM},H_e(x)]}{x;l},
   \label{G1condtransldof}
\end{equation}
for all $k,l$, to remove the translational component of $H_1$.  In the
last step we have used (\ref{FCMforms}) and $\pvec_t= N_e m_e
(d/dt)\rvec_{\rm CM}$.

The left-hand side of (\ref{G1condtransldof}) can be written
$[G_1,W]_{kl}$ where $W_{kl}=\matrixelement{x;k}{H_e(x)}{x;l}$, since
the kinetic energy part of $H_0$, a multiple of the identity, does not
contribute, while the right-hand side is proportional to $[(\rvec_{\rm
CM}),W]_{kl}$, where $(\rvec_{\rm CM})_{kl} =
\matrixelement{x;k}{\rvec_{\rm CM}}{x;l}$ is the matrix representing
$\rvec_{\rm CM}$ in the Born-Oppenheimer representation.  Therefore a
solution of (\ref{G1condtransldof}) is given by
\begin{equation}
   G_{1,kl} = -\frac{iN_e m_e}{\hbar^2 M_t}\Pvec_t \cdot
   \matrixelement{x;k}{\rvec_{\rm CM}}{x;l}.
   \label{solnG1bad}
\end{equation}
We can add to this any matrix that commutes with $W$, of which a
multiple of the identity is an obvious choice.

The procedure being carried out here, the elimination of the
translational components of the derivative couplings, bears some
relation to the introduction of electron translation factors in
electronic structure calculations (\cite{Delos81,
FatehiAlguireShaoSubotnik11}), which we will explore in future
publications. In that context, the electron translation factors are
also not unique, although a meaningful choice can be made in an atomic
orbital basis.

One obvious shortcoming of (\ref{solnG1bad}) is that the right-hand
side does not commute with translations.  We wish $G_1$ to be
translationally invariant, so that $U_1$ will map the old Hamiltonian
$H$, which is translationally invariant, into a new one $H'$ with the
same property.  Also, making $G_1$ translationally invariant means
that $\Pvec_t \delta_{kl}$, the matrix representing the total linear
momentum of the molecule, is not changed by the transformation $U_1$.
To show that the right-hand side of (\ref{solnG1bad}) is not
translationally invariant we conjugate it with a nuclear translation
operator, whereupon $\Pvec_t$ goes into itself and the matrix element
transforms according to
\begin{eqnarray}
   T_n(\xivec) \matrixelement{x;k}{\rvec_{\rm CM}}{x;l} 
   T_n(\xivec)^\dagger &=& \matrixelement{x-\xi;k}{\rvec_{\rm CM}}
   {x-\xi;l} = \matrixelement{x;k}{T_e(\xivec) 
   \rvec_{\rm CM}T_e(\xivec)^\dagger}{x;l}\nonumber\\
   &=&
   \matrixelement{x;k}{\rvec_{\rm CM}}{x;l} - \xivec\,\delta_{kl},
   \label{rCMmetranslation}
\end{eqnarray}
where we use (\ref{Tncommrels}), (\ref{basistranslationlaw}) and
(\ref{ripitranslationconjugation}).   

Now for a given $x$ let $x_0$ be the point on the translation-reduced
configuration space on the same translational fiber as $x$, as
illustrated in Fig.~\ref{bundle}.  This means that $x$ and $x_0$ are
connected by a translation; let $\xivec$ be the vector such that
$T(\xivec)x_0=x$, that is, $x=x_0+\xi$.  Then $\xivec$ is the same as
the $\Xvec_{\rm CM}$ coordinate of $x$, and we have
\begin{equation}
   \matrixelement{x_0;k}{\rvec_{\rm CM}}{x_0;l} =
   \matrixelement{x;k}{\rvec_{\rm CM}}{x;l} - \Xvec_{\rm CM}\,
   \delta_{kl}.
   \label{rCMtranslinv}
\end{equation}
The final term is a multiple of the identity, so we obtain a solution
of (\ref{G1condtransldof}) that is also translationally invariant if
we replace (\ref{solnG1bad}) by
\begin{equation}
   G_{1,kl} = -\frac{iN_e m_e}{\hbar^2 M_t}\Pvec_t \cdot
   (\matrixelement{x;k}{\rvec_{\rm CM}}{x;l}-
   \Xvec_{\rm CM}\,\delta_{kl}).
   \label{solnG1good}
\end{equation}

\subsection{An Exact Unitary Operator for Translations}
\label{exactunitary}

Although in the practice of perturbation theory we expand the unitary
operator $\Uhat_1=\exp(\hbar\Ghat_1)$ in a power series, in fact for
the generator shown in (\ref{solnG1good}) it is possible to carry out
the unitary transformation $\Uhat_1$ exactly.   That is, we can give
exact formulas for the action of $\Uhat_1$ on wave functions and for
its action by conjugation on operators.   This is easiest to see if we
transform (\ref{solnG1good}) to the molecular representation,
whereupon we find
\begin{equation}
   G_1=-\frac{iN_e m_e}{\hbar^2 M_t}\,(\Pvec_t+\pvec_t)
   \cdot(\rvec_{\rm CM}-\Xvec_{\rm CM}),
   \label{G1molecrepn}
\end{equation}
where we use (\ref{PCMinBOrep}).  To be clear about the notation,
(\ref{solnG1good}) gives the Weyl symbols of the components of the
matrix that represents the operator in the Born-Oppenheimer
representation.  When these are converted into operators, the operator
$\Pvechat_t$ can be placed before or after the factor $\rvec_{\rm
CM}-\Xvec_{\rm CM}$, which is translationally invariant.  When these
operators are converted to the molecular representation, using the
rules explained in subsection~\ref{tworepresentations}, we obtain
(\ref{G1molecrepn}).  We omit the hats on operators when working in
the molecular representation, but $G_1$ in (\ref{G1molecrepn}) is an
operator.  Recall that $\Pvec_{\rm CM} = \Pvec_t+\pvec_t$ in this
representation (see (\ref{PCMdef})).  For the rest of this subsection
we work in the molecular representation.

Let us write the unitary operator as $U_1(\lambda) = \exp(\hbar G_1) =
\exp(-i\lambda g_1/\hbar)$, where
\begin{equation}
  g_1=\Pvec_{\rm CM}\cdot(\rvec_{\rm CM} -\Xvec_{\rm CM})
  \label{g1def}
\end{equation}
and where 
\begin{equation}
   \lambda=\lambda_0=\frac{N_em_e}{M_t}.
   \label{lambdadeforig}
\end{equation}
Let $A=A(0)$ be an arbitrary operator and let 
\begin{equation}
    A(\lambda) = U_1(\lambda) A(0) U_1(\lambda)^\dagger.
    \label{AconjbyU1}
\end{equation}
We can solve for $A(\lambda)$ by promoting $\lambda$ into a variable
and solving the differential equation,
\begin{equation}
   \frac{dA(\lambda)}{d\lambda} = -\frac{1}{i\hbar}[A,g_1],
   \label{dA/dlambda}
\end{equation}
a version of the Heisenberg equations of motion.   

For example, evaluating the commutators $[\rvec_i,g_1]$ and
$[\Xvec_\alpha,g_1]$ we find
\begin{equation}
   \frac{d\rvec_i(\lambda)}{d\lambda}=
   \frac{d\Xvec_\alpha(\lambda)}{d\lambda} = 
   -(\rvec_{\rm CM}-\Xvec_{\rm CM}).
   \label{riXalphaeoms}
\end{equation}
As derived, the vector $\rvec_{\rm CM}-\Xvec_{\rm CM}$ on the
right-hand side should be evaluated at $\lambda$, but, in fact, it is
easy to show that this vector is independent of $\lambda$ so it can be
evaluated anywhere, such as $\lambda=0$.   Thus we have the solutions,
\begin{subequations}
\label{riXalphasolns}
\begin{eqnarray}
   \rvec_i(\lambda) &=& \rvec_i -\lambda(\rvec_{\rm CM}
    -\Xvec_{\rm CM}),
   \label{rilambdasoln}
   \\
   \Xvec_\alpha(\lambda) &=& \Xvec_\alpha -\lambda(\rvec_{\rm CM}
    -\Xvec_{\rm CM}),
   \label{Xalphalambdasoln}
\end{eqnarray}
\end{subequations}
where on the right-hand side all vectors are understood to be
evaluated at $\lambda=0$.  

Similarly, we find
\begin{equation}
   \frac{d\pvec_i(\lambda)}{d\lambda} = 
   \frac{1}{N_e}\Pvec_{\rm CM}, \qquad
   \frac{d\Pvec_\alpha(\lambda)}{d\lambda} =
   -\frac{M_\alpha}{M_t}\Pvec_{\rm CM}.
   \label{piPalphaeoms}
\end{equation}
As before, we find that $\Pvec_{\rm CM}$ is independent of $\lambda$.
The solutions are
\begin{subequations}
\label{piPalphasolns}
\begin{eqnarray}
   \pvec_i(\lambda) &=& \pvec_i + \frac{\lambda}{N_e}\Pvec_{\rm CM},
   \label{pilambdasoln}\\
   \Pvec_\alpha(\lambda) &=& \Pvec_\alpha -
   \frac{\lambda M_\alpha}{M_t}\Pvec_{\rm CM},
   \label{Palphalambdasoln}
\end{eqnarray}
\end{subequations}
where all vectors on the right-hand side are evaluated at $\lambda=0$.

\subsection{Eliminating $\Fvec_{{\rm CM},kl}$ to All Orders}
\label{eliminatingFCM}

In this subsection we continue working in the molecular
representation, omitting hats on nuclear operators.

The transformation laws (\ref{riXalphasolns}) and
(\ref{piPalphasolns}) allow us to evaluate $U_1(\lambda)\,H\,
U_1(\lambda)^\dagger$, where $H$ is the molecular Hamiltonian
(\ref{Hmoldef}).  For $\lambda=\lambda_0$ this transformation should
eliminate the translational components of the derivative couplings to
order $\kappa^2$.  We find that the potential energy does not change,
\begin{equation}
    U_1(\lambda)\,V_{\rm Coul}(\Xvec,\rvec)\,U_1(\lambda)^\dagger
    =V_{\rm Coul}(\Xvec,\rvec),
    \label{UlambdaVCoul}
\end{equation}
since it depends only on vector differences among $\Xvec_\alpha$ and
$\rvec_i$.  

As for the kinetic energy, we find
\begin{eqnarray}
   &&U_1(\lambda)\left(\sum_{\alpha=}^N \frac{\Pvec_\alpha^2}{2M_\alpha}
   +\sum_{i=1}^{N_e}\frac{\pvec_i^2}{2m_e}\right)U_1(\lambda)^\dagger
   =\frac{1}{2M_t}(\Pvec_{\rm CM}-\pvec_t)^2 
   + \sum_{\kappa=1}^{N-1}
   \frac{\Qvec_\kappa^2}{2\mu_\kappa}+\sum_{i=1}^{N_e}
   \frac{\pvec_i^2}{2m_e}\nonumber\\
   &&\qquad +\lambda\left[\left(\frac{1}{N_em_e}
   +\frac{1}{M_t}\right)\pvec_t\cdot\Pvec_{\rm CM}
   -\frac{\Pvec_{\rm CM}^2}{M_t}\right]
   +\lambda^2\left(\frac{1}{N_em_e}+\frac{1}{M_t}\right)
   \frac{\Pvec_{\rm CM}^2}{2},
   \label{UlambdaKE}
\end{eqnarray}
where we have transformed the nuclear kinetic energy according to
(\ref{nuclearKExfm}).  In this result we have used $\Pvec_t=\Pvec_{\rm
CM}-\pvec_t$ to eliminate $\Pvec_t$, and we note that $\Pvec_{\rm CM}
\longleftrightarrow \Pvec_t\,\delta_{kl}$ and $\pvec_t
\longleftrightarrow i\hbar \Fvec_{{\rm CM},kl}$, the latter of which
follows from (\ref{FCMforms}).   

Thus in the molecular representation the terms that will produce the
translational component (\ref{H1kltranslational}) of $H_{1,kl}$ in the
Born-Oppenheimer representation are those proportional to
$\pvec_t\cdot\Pvec_{\rm CM}$.  The coefficient of these terms in
(\ref{UlambdaKE}) is
\begin{equation}
   -\frac{1}{M_t} + \lambda\left(\frac{1}{N_em_e}+\frac{1}{M_t}
    \right).
    \label{PCMdotptcoef}
\end{equation}
If we set $\lambda=\lambda_0=N_em_e/M_t$, the coefficient of this term
is reduced by a factor of order $\kappa^4$, which is better than
expected, since $U_1$ was only supposed to eliminate this term through
order $\kappa^2$.  But we see that if we choose the slightly different
value,
\begin{equation}
    \lambda=\lambda_1=\frac{N_em_e}{M_{\rm mol}}
    =\frac{N_em_e}{M_t + N_em_e}
    \label{lambda1def}
\end{equation}
then the translational component of the derivative couplings is
eliminated exactly (to all orders in $\kappa$). 

With this choice (\ref{lambda1def}) of $\lambda$ (and henceforth
dropping the 1-subscript on $\lambda$ and $U$), we find for the
transformed molecular Hamiltonian,
\begin{equation}
   U(\lambda)\,H \,U(\lambda)^\dagger =
   \frac{\Pvec_{\rm CM}^2}{2M_{\rm mol}} +
   \sum_{\kappa=1}^{N-1} \frac{\Qvec_\kappa^2}{2\mu_\kappa}
   +\sum_{i=1}^{N_e} \frac{\pvec_i^2}{2m_e}
   +\frac{\pvec_t^2}{2M_t} + V_{\rm Coul}(\Xvec,\rvec).
   \label{UlambdaHmol}
\end{equation}
Apart from the meanings of the symbols, the result is the same as the
molecular Hamiltonian seen in (\ref{HmolTRdef}), (\ref{HTRdef}) and
(\ref{HTRedef}).   In particular, the fourth term of
(\ref{UlambdaHmol}) is the mass polarization term.   Our
unitary transformation $U(\lambda)$ reproduces the effects of the
coordinate transformation discussed in
subsection~\ref{translationaldof}, the purpose of which was to
separate the translational degrees of freedom.   

With the value (\ref{lambda1def}) of $\lambda$ we note that
(\ref{RCMdef}) can be written $\Rvec_{\rm CM} = (1-\lambda)\Xvec_{\rm
CM} + \lambda \rvec_{\rm CM}$.  Thus the transformation equations
(\ref{riXalphasolns}) and (\ref{piPalphasolns}) become
\begin{eqnarray}
  U(\lambda)\,\rvec_i\,U(\lambda)^\dagger &=&
  \rvec_i-(\Rvec_{\rm CM}-\Xvec_{\rm CM}),
  \qquad U(\lambda)\,\pvec_i\,U(\lambda)^\dagger =
  \pvec_i+\frac{m_e}{M_{\rm mol}}\Pvec_{\rm CM},
  \nonumber\\
  U(\lambda)\,\Xvec_\alpha\,U(\lambda)^\dagger &=&
  \Xvec_\alpha-(\Rvec_{\rm CM}-\Xvec_{\rm CM}),
  \qquad U(\lambda)\,\Pvec_\alpha\,U(\lambda)^\dagger =
  \Pvec_\alpha-\frac{N_em_e\,M_\alpha}{M_t\,M_{\rm mol}}\Pvec_{\rm CM}.
  \label{Ulambdatransformations}
\end{eqnarray}
The unitary transformation itself becomes
\begin{equation}
    U(\lambda)=\exp\left[-\frac{i}{\hbar}
    \Pvec_{\rm CM}\cdot(\Rvec_{\rm CM}-\Xvec_{\rm CM})\right].
    \label{Ulambdaexact}
\end{equation}
An example of this transformation appeared in \cite{Cederbaum13}, who
used it in his study of translational invariance in the method of
exact factorization.

\subsection{Unitary Transformation as a Coordinate Transformation}
\label{unitarycoordinate}

Equations~(\ref{Ulambdatransformations}) look like a coordinate
transformation taking us from variables $\rvec_i$, $\Xvec_\alpha$,
etc., on the right to variables $U(\lambda)\rvec_i
U(\lambda)^\dagger$, etc., on the left.  It is convenient to call the
variables on the right the ``new'' variables and those on the left the
``old'' ones.  This seems counter-intuitive but it is the
interpretation needed to obtain the transformation law
(\ref{HHprimexfm}) given below.  We note that the transformation of
the momenta follows from that of the coordinates, via the chain rule
for the operators $\pvec_i = -i\hbar\,\partial/\partial\rvec_i$, etc.

We will denote the new coordinates with a prime and the old without a
prime.  We obtain the coordinate transformation in this language if we
replace $\rvec_i$, etc., on the right side of
(\ref{Ulambdatransformations}) with $\rvec'_i$, etc., and then replace
$U(\lambda)\rvec_i U(\lambda)^\dagger$, etc., on the left with simply
$\rvec_i$, etc.  This gives us the old variables as functions of the
new.  Inverting these equations to solve for the new variables as
functions of the old, we obtain
\begin{eqnarray}
  \rvec'_i &=&
  \rvec_i+(\Rvec_{\rm CM}-\Xvec_{\rm CM}),
  \qquad \pvec'_i =
  \pvec_i-\frac{m_e}{M_{\rm mol}}\Pvec_{\rm CM},
  \nonumber\\
  \Xvec'_\alpha &=&
  \Xvec_\alpha+(\Rvec_{\rm CM}-\Xvec_{\rm CM}),
  \qquad \Pvec'_\alpha=
  \Pvec_\alpha+\frac{N_em_e\,M_\alpha}{M_t\,M_{\rm mol}}\Pvec_{\rm CM},
  \label{inversetransformation}
\end{eqnarray}
where we note that $\Rvec'_{\rm CM} - \Xvec'_{\rm CM} = \Rvec_{\rm CM}
-\Xvec_{\rm CM}$ and $\Pvec'_{\rm CM} = \Pvec_{\rm CM}$.

Now if we call (\ref{Hmoldef}) the ``old'' Hamiltonian $H$, a function
of the old variables $\rvec_i$, $\Xvec_\alpha$, etc., and
(\ref{UlambdaHmol}) the ``new'' Hamiltonian $H'$, regarded as a
function of the new variables $\rvec'_i$, $\Xvec'_\alpha$, etc., then
we have
\begin{equation}
   H(\Xvec,\Pvec,\rvec,\pvec) = H'(\rvec',\pvec',\Xvec',\Pvec'),
   \label{HHprimexfm}
\end{equation}
in which the primed variables are functions of the unprimed ones as
shown in (\ref{inversetransformation}) and vice versa.

Alternatively, we may regard (\ref{Ulambdatransformations}) as a
dressing transformation, taking us from the lab molecular
representation to the lab, translation-reduced representation (down the
first column of Table~\ref{reptable}).  The two points of view are
complementary; in the dressing point of view the operators do not
change but the wave functions do, while in the
coordinate-transformation point of view it is the opposite.  The
dressing point of view has some interesting features, for example, in
the translation-reduced, molecular representation the operator that
looks like the center of mass of the nuclei is physically the center
of mass of the molecule, since
\begin{equation}
    U(\lambda)\,\Rvec_{\rm CM}\,U(\lambda)^\dagger = \Xvec_{\rm CM}.
    \label{RCMXCMxfm}
\end{equation}
This reminds us of the fact that in the Born-Oppenheimer
representation, the operator that looks like the nuclear angular
momentum actually includes the electronic angular momentum, a main
point of \cite{LittlejohnRawlinsonSubotnik23}.   See also the end of
subsection~\ref{molecularmomentumvariousreps} for the situation
regarding the total momentum of the molecule.

Having achieved the translation-reduced molecular representation (the
lower left corner of Table~\ref{reptable}) we can move to the right
(across the lower row) in much the same manner as we did previously
along the upper row, except that the the electronic Hamiltonian is now
the translation-reduced version (\ref{HTRedef}) instead of the
original version (\ref{Hedef}).  In particular, the potential energy
surfaces and adiabatic basis states are those of $H_{{\rm TR}e}$ and
not $H_e$, the former of which contains the mass-polarization terms.
In the case of hydrogen, mentioned in
subsection~\ref{translcompsderivcouplings}, the mass-polarization
term provides the correction between the true electron mass and the
reduced mass; the latter is needed for the correct energy
eigenvalues.  Naturally, if we use the wrong electronic Hamiltonian
then there will be transitions among the surfaces, even for uniform
translational motion of the atom.

Once we have reached the dressed, translation-reduced representation
(the lower right-hand corner of Table~\ref{reptable}) we will be in a
position to separate the rotational degrees of freedom.  This involves
a coordinate transformation on the translation-reduced configuration
space, taking us from the $\Yvec_\kappa$ coordinates to orientational
(Euler angle) and shape coordinates.  This is often described as the
construction of the kinetic energy operator on the internal space.
There is an extensive literature on this procedure for single-surface
problems in the electrostatic model, for example,
\cite{WangCarrington00}.   \cite{Kendrick18} has given a partial 
analysis of the case of two coupled surfaces in the electrostatic
model in his study of low-temperature scattering.  The separation of
rotational degrees of freedom in the case of fine structure effects
has apparently not been worked out, but, as noted by
\cite{LittlejohnRawlinsonSubotnik23}, in the odd-electron case the
nuclear Born-Oppenheimer wave function is a 2-component
pseudo-spinor.  We will report on this problem in more detail in the
future.

\section{Conclusions}
\label{conclusions}

This article is a part of a series in which we hope to introduce a
geometrical flavor to Born-Oppenheimer theory and to reveal some new
results, building on older work.  The latter includes
\cite{LittlejohnReinsch97}, which dealt with rotations in the $n$-body
problem but not specifically with Born-Oppenheimer theory.  That
article presented a rather complete explanation of the rotational
fiber bundle structure of the nuclear configuration space.  The first
article of the new series was \cite{LittlejohnRawlinsonSubotnik22},
which was devoted to diabatic bases.  This was necessary preparation
since it is impossible to talk about the theory without diabatic
bases, and because the existing literature on the subject has several
shortcomings.  The second, \cite{LittlejohnRawlinsonSubotnik23}, dealt
with angular momentum and its conservation within Born-Oppenheimer
theory.  This article, the third, has explained Moyal theory for the
separation of nuclear and electronic degrees of freedom, and
translational invariance.  Future articles will treat the separation
of rotational degrees of freedom, which practically speaking means the
construction of kinetic energy operators on the internal space.  We
also plan an article on the interaction between molecular point groups
and the geometry of the nuclear configuration space, especially the
rotational fiber bundle.  We feel that this is an important step in
the development of a proper geometrical understanding of aspects of
the Jahn-Teller effect and the bifurcation of degeneracy manifolds.
Finally, we plan a parallel treatment of some of the same issues in
the context of hybrid classical-quantum models, which are currently
popular in surface hopping and related algorithms.  For such models
the unitary transformations implemented by the Moyal star product are
replaced by canonical transformations on a phase space that
incorporates both classical and quantum variables.

\begin{acknowledgments}
JES was supported by U.S. Department of Energy (DoE), Office of
Science, Office of Basic Energy Sciences, under Award
No. DE-SC0019397.
\end{acknowledgments}

\appendix

\section{Weyl-Moyal Formalism}
\label{Moyal}

Original articles and those that review the Wigner-Weyl-Moyal
formalism include \cite{Weyl27, Wigner32, Groenewold46, Moyal49,
BalazsJennings84, Littlejohn86, McDonald88, OsbornMolzahn95}.  The
notation in this appendix is independent of that of the rest of the
paper. 

\subsection{The Weyl Symbol Correspondence}
\label{weylsymbolcorresp} 

We work with wave functions $\psi(x)$, where $x=(x_1,\ldots,x_n)\in
\Reals^n$.   The conjugate momentum is $p=(p_1,\ldots,p_n)$.
Operators are indicated by hats, for example, $\xhat_i=$
multiplication by $x_i$ and $\phat_i=-i\hbar\,\partial/\partial x_i$,
when acting on wave functions $\psi(x)$.  Classical quantities or
$c$-numbers are indicated without the hat.

If $\Ahat$ is an operator then we define the {\it Weyl symbol} of
$\Ahat$ as the function $A(x,p)$ on the classical phase space given by
\begin{equation}
	A(x,p)=\int ds\, e^{-ip\cdot s/\hbar}\,
	\matrixelement{x+s/2}{\Ahat}{x-s/2},
	\label{Weylsymboldef}
\end{equation}
where $s\in\Reals^n$ and where the matrix element shown is the kernel
of the integral operator $\Ahat$ in the $x$-representation, that is,
\begin{equation}
	(\Ahat \psi)(x) = \int dx' \,
	\matrixelement{x}{\Ahat}{x'}\,\psi(x').
	\label{Akerdef}
\end{equation}
Equation~(\ref{Weylsymboldef}) specifies the {\it Weyl transform}
$A(x,p)$ of the operator $\Ahat$.  The inverse Weyl transform is
given by
\begin{equation}
	\matrixelement{x}{\Ahat}{x'} =
	\int \frac{dp}{(2\pi\hbar)^n} \,
	e^{ip\cdot(x-x')/\hbar} \,
	A\Bigl(\frac{x+x'}{2},p\Bigr).
	\label{inverseWeyl}
\end{equation}
We denote the correspondence between operators and their Weyl symbols
by $\Ahat\longleftrightarrow A(x,p)$.   This is not the same usage of
the symbol $\longleftrightarrow$ as in other parts of this paper, but
what the notation has in common is that it represents a one-to-one
correspondence between two sets of objects, in this case, operators
and functions on phase space. 

As special cases, we note that the operators ${\hat 1}$ (the identity
operator), $\xhat_i$ and $\phat_i$ have Weyl symbols $1$, $x_i$ and
$p_i$, respectively. 

No information is lost by mapping operators to their Weyl symbols or
vice versa.   In particular, there is no ``ordering ambiguity'' on
passing from symbols (objects that are nominally classical) to
operators.   For example, we have
\begin{subequations}
\label{xipjsymbols}
\begin{eqnarray}
	\xhat_i\phat_j &\longleftrightarrow& x_ip_j 
	+\frac{i\hbar}{2}\,\delta_{ij},\\
	\phat_j\xhat_i &\longleftrightarrow& x_ip_j
	-\frac{i\hbar}{2}\,\delta_{ij},
\end{eqnarray} 
\end{subequations}
which shows that the ordering of operators is indicated by terms in
the symbol that are of higher order in $\hbar$.  A useful
generalization of (\ref{xipjsymbols}) is
\begin{subequations}
\label{f(x)psymbols}
\begin{eqnarray}
	f(\xhat)\phat_i &\longleftrightarrow&
	f(x)p_i + \frac{i\hbar}{2}
	\frac{\partial f(x)}{\partial x_i},\\
	\phat_i f(\xhat) &\longleftrightarrow&
	f(x)p_i -\frac{i\hbar}{2}
	\frac{\partial f(x)}{\partial x_i}.
        \label{f(x)potherorder}
\end{eqnarray}
\end{subequations}

If $\Ahat$ and $\Bhat$ are operators such that $\Ahat=\Bhat^\dagger$,
then the corresponding symbols satisfy
\begin{equation}
	A(x,p) = B(x,p)^*.
	\label{WeylHermconj}
\end{equation}
In particular, the Weyl symbol of a Hermitian operator is real.  A
special case of this follows from (\ref{f(x)psymbols}),
\begin{equation}
	\frac{1}{2}[\phat_i f(\xhat) + f(\xhat) \phat_i]
	\longleftrightarrow f(x)p_i.
	\label{symmf(x)psymbol}
\end{equation}

\subsection{The $*$-Product}
\label{starproduct}

If $\Ahat$, $\Bhat$ and $\Chat$ are operators such that
$\Chat=\Ahat\Bhat$, then we write the corresponding relation between
the symbols as
\begin{equation}
	C(x,p)=A(x,p) * B(x,p),
	\label{stardef}
\end{equation}
where $*$ is the ``Moyal star product.''   It has an expansion in
powers of $\hbar$,
\begin{equation}
	C(x,p) = \sum_{n=0}^\infty \frac{1}{n!}
	\Bigl(\frac{i\hbar}{2}\Bigr)^n 
	\{A,B\}_n,
	\label{Moyalexpn}
\end{equation}
where
\begin{equation}
	\{A,B\}_n = A(x,p)\left(
	\frac{
	\begin{array}{c}
	\leftarrow\\
	\noalign{\vskip-9pt}
	\partial
	\end{array}
	}{\partial x}
	\cdot
	\frac{
	\begin{array}{c}
	\rightarrow\\
	\noalign{\vskip-9pt}
	\partial
	\end{array}
	}{\partial p}-
	\frac{
	\begin{array}{c}
	\leftarrow\\
	\noalign{\vskip-9pt}
	\partial
	\end{array}
	}{\partial p}
	\cdot
	\frac{
	\begin{array}{c}
	\rightarrow\\
	\noalign{\vskip-9pt}
	\partial
	\end{array}
	}{\partial x}
	\right)^n B(x,p).
	\label{Moyalstarexpansion}
\end{equation} 
The arrows indicate the direction in which the partial derivatives
act. The first few of these brackets are
\begin{subequations}
\label{fewMoyalbrackets}
	\begin{eqnarray}
	\{A,B\}_0 &=& AB,
	\nonumber\\
	\{A,B\}_1 &=& \{A,B\} = 
	\frac{\partial A}{\partial x}\cdot
	\frac{\partial B}{\partial p} -
	\frac{\partial A}{\partial p}\cdot
	\frac{\partial B}{\partial x},
	\nonumber\\
	\{A,B\}_2 &=& \sum_{ij}\left(
	\frac{\partial^2 A}{\partial x_i \partial x_j}
	\frac{\partial^2 B}{\partial p_i \partial p_j}
	- 2\frac{\partial^2 A}{\partial x_i \partial p_j}
	\frac{\partial^2 B}{\partial p_i \partial x_j}
	+\frac{\partial^2 A}{\partial p_i \partial p_j}
	\frac{\partial^2 B}{\partial x_i \partial x_j}\right),
	\nonumber
	\end{eqnarray}
\end{subequations}
etc.  The zeroth order bracket is the ordinary product and the first
order bracket is the ordinary Poisson bracket.   If no subscript is
given, the Poisson bracket is implied.   The $n$-th order bracket is
symmetric or antisymmetric in the exchange of the operands as $n$ is
even or odd, respectively.  Writing out the first few terms of
(\ref{Moyalstarexpansion}), we have
\begin{equation}
	A*B = AB + \frac{i\hbar}{2} \{A,B\} -
	\frac{\hbar^2}{8}
	\{A,B\}_2 +\ldots. 
	\label{A*Bfewterms}
\end{equation}
We also define the ``Moyal bracket,''
\begin{equation}
	[A,B]_* = A*B - B*A = i\hbar \{A,B\} 
	-\frac{i\hbar^3}{24}\{A,B\}_3 + \ldots,
	\label{starbracketdef}
\end{equation}
which is the Weyl symbol of the commutator $[\Ahat,\Bhat]$. 

\subsection{Matrices of Operators and Weyl Symbols}
\label{matricesWeyl}

In the main body of the paper we make frequent use of commutators of
matrices of operators.  We now make some comments about what happens
to such commutators when translated into symbols.   Let $\Ahat_{kl}$
and $\Bhat_{kl}$ be two matrices of operators.  By the commutator
$\Chat=[\Ahat,\Bhat]$, we mean  the matrix of operators,
\begin{equation}
  \Chat_{kl}=[\Ahat,\Bhat]_{kl} = \sum_p \bigl(\Ahat_{kp}\,\Bhat_{pl} -
  \Bhat_{kp}\,\Ahat_{pl}\bigr).
  \label{matrixcommutdef}
\end{equation}
When this is translated into symbols, we obtain
\begin{eqnarray}
  C_{kl} &=& [A,B]_{*kl}=\sum_p\bigl(A_{kp}*B_{pl} -B_{kp}*A_{pl}\bigr)
  =\sum_p\bigl(A_{kp}\,B_{pl}-B_{kp}\,A_{pl}\bigr)
  \nonumber\\
  &+& \frac{i\hbar}{2}
  \sum_p\bigl(\{A_{kp},B_{pl}\}-\{B_{kp},A_{pl}\}\bigr)
  -\frac{\hbar^2}{8}
  \sum_p\bigl(\{A_{kp},B_{pl}\}_2-\{B_{kp},A_{pl}\}_2\bigr)
  +\ldots
  \label{matrixcommutexpand}
\end{eqnarray}
This equation defines the Moyal bracket $C=[A,B]_*$ of matrices of
symbols.  Notice that the lowest, order $\hbar^0$ term in the
expansion is otherwise just the commutator $[A,B]$ of the symbol
matrices, in which symbols are multiplied as $c$-numbers.  That is, the
first term of $[A,B]_*$ is $[A,B]$ (without the star).  Notice also
that in the case of scalar operators, the expansion of the symbol of
$[\Ahat,\Bhat]$ involves only odd powers of $\hbar$, as shown by
(\ref{starbracketdef}), whereas in the case of matrices of operators,
all powers occur, including the power $\hbar^0$.

\section{Coordinates on Nuclear Configuration Space}
\label{coordinatesnuclearCS}

The coordinate transformation $\Xvec_\alpha \to (\Xvec_{\rm
  CM},\Yvec_\kappa)$ was introduced in
subsection~\ref{translationaldof}, where the definitions of
$\Xvec_{\rm CM}$ and $\Pvec_t$ were given (in (\ref{XCMdef}) and
(\ref{Ptdef})).  

The new coordinates $\Yvec_\kappa$, $\kappa=1,\ldots,N-1$, are a set
of $N-1$ translationally invariant vectors defined by an $(N-1)\times
N$ matrix $B_{\kappa\alpha}$,
\begin{equation}
   \Yvec_\kappa = \sum_{\alpha=1}^N B_{\kappa\alpha}\,\Xvec_\alpha,
   \qquad \kappa=1,\ldots,N-1.
   \label{Yvecdef}
\end{equation}
The matrix $B_{\kappa\alpha}$ must satisfy
\begin{equation}
    \sum_{\alpha=1}^N B_{\kappa\alpha}=0,
    \label{Bistranslatinv}
\end{equation}
in order that the vectors $\Yvec_\kappa$ be translationally
invariant. 

We require that the $\Yvec_\kappa$ be Jacobi vectors (\cite{Delves60,
Hirschfelder69, AquilantiCavalli86, Gattietal98}).  This means that
the matrix $B_{\kappa\alpha}$ satisfies
\begin{equation}
   \sum_{\alpha=1}^N
   \frac{B_{\kappa\alpha}\,B_{\lambda\alpha}}{M_\alpha}=
   \frac{\delta_{\kappa\lambda}}{\mu_\kappa},
   \label{Bkappaalphaisorthog}
\end{equation}
which is a kind of an orthogonality condition on $B_{\kappa\alpha}$
with respect to the kinetic energy metric.  Here the $\mu_\kappa$,
$\kappa=1,\ldots,N-1$, are reduced masses associated with the Jacobi
vectors $\Yvec_\kappa$.  The inverse coordinate transformation
$(\Xvec_{\rm CM},\Yvec_\kappa) \to \Xvec_\alpha$ is given by
\begin{equation}
   \Xvec_{\alpha}=\Xvec_{\rm CM} + \frac{1}{M_\alpha}
   \sum_{\kappa=1}^{N-1} \mu_\kappa \, B_{\kappa\alpha}\, 
   \Yvec_\kappa.
   \label{X=X(XCM,Y)}
\end{equation}

The matrix $B_{\kappa\alpha}$ satisfies a kind of inverse
orthogonality relation,
\begin{equation}
    \frac{M_\alpha\,M_\beta}{M_t} + \sum_{\kappa=1}^{N-1}
    \mu_\kappa\,B_{\kappa\alpha}\,B_{\kappa\beta}=
    M_\alpha\,\delta_{\alpha\beta}.
    \label{Bkappaalphainverseorthog}
\end{equation}
This may be proven by writing out the $N\times N$ matrix of the linear
transformation $\Xvec_\alpha \to (\Xvec_{\rm CM},\Yvec_\kappa)$ and
the one for its inverse, $(\Xvec_{\rm CM},\Yvec_\kappa) \to
\Xvec_\alpha$, and multiplying the two matrices in both orders.  

For example, in the case of the water molecule, we let atoms 1 and 2
be hydrogens and atom 3 be the oxygen.  Then we define Jacobi vectors
$\Yvec_1 = \Xvec_2-\Xvec_1$ and
$\Yvec_2=\Xvec_3-(\Xvec_1+\Xvec_2)/2$, which implies
\begin{equation}
    B_{\kappa\alpha}=\left(
	\begin{array}{ccc}
	-1 & 1 & 0 \cr
         -1/2 & -1/2 & 1
	\end{array}\right).
    \label{Bkappaalphaexample}
\end{equation}
These imply reduced masses $\mu_1 = M_{\rm H}/2$, and 
\begin{equation}
    \frac{1}{\mu_2} = \frac{1}{2M_{\rm H}} + \frac{1}{M_{\rm O}}.
    \label{mu2example}
\end{equation}

The momenta conjugate to $\Xvec_\alpha$ are $\Pvec_\alpha$, and those
conjugate to $(\Xvec_{\rm CM},\Yvec_\kappa)$ are
$(\Pvec_t,\Qvec_\kappa)$.   The momenta acting on wave functions are
defined by $\Pvec_\alpha=-i\hbar\partial/\partial\Xvec_\alpha$,
$\Pvec_t = -i\hbar\partial/\partial\Xvec_{\rm CM}$, and
$\Qvec_\alpha=-i\hbar\partial/\partial\Yvec_\kappa$.  One way to
transform the momenta, which is consistent with the chain rule, is to
demand the equality of the differential forms,
\begin{equation}
  \sum_{\alpha=1}^N \Pvec_\alpha\cdot d\Xvec_\alpha =
  \Pvec_t\cdot d\Xvec_{\rm CM} + \sum_{\kappa=1}^{N-1}
  \Qvec_\kappa\cdot d\Yvec_\kappa.
  \label{PdQ=pdq}
\end{equation}
Substituting the coordinate transformation (\ref{X=X(XCM,Y)}) into this
gives (\ref{Ptdef}), showing that $\Pvec_t$ is the total nuclear
momentum, and it gives
\begin{equation}
   \Qvec_\kappa = \sum_{\alpha=1}^N \frac{\mu_\kappa}{M_\alpha}
   B_{\kappa\alpha}\,\Pvec_\alpha.
   \label{Qkappadef}
\end{equation}
Substituting the the inverse transformation (\ref{XCMdef}) and
(\ref{Yvecdef}) gives
\begin{equation}
   \Pvec_\alpha = \frac{M_\alpha}{M_t}\Pvec_t +
   \sum_{\kappa=1}^{N-1}B_{\kappa\alpha}\,\Qvec_\kappa.
   \label{Palphaxfm}
\end{equation}

\section{Coordinates on the Molecular Configuration Space}
\label{coordinatesmolecularcs}

The transformation on the molecular configuration space proceeds in
two steps, $(\Xvec_\alpha,\rvec_i) \to (\Xvec_{\rm
CM},\Yvec_\kappa,\rvec_i) \to (\Rvec_{\rm
CM},\Yvec_\kappa,\svec_i)$. The first step is explained in
subsection~\ref{translationaldof} and
Appendix~\ref{coordinatesnuclearCS}.   In the second step $\Rvec_{\rm
CM}$, the molecular center of mass, is given by (\ref{RCMdef}) and 
$\svec_i$, the electron position relative to the nuclear center of
mass, is given by (\ref{sidef}), while $\Yvec_\kappa$, the nuclear
Jacobi coordinates, do not change.   The inverse transformation is
given by
\begin{equation}
   \Xvec_{\rm CM} = \Rvec_{\rm CM}-\frac{m_e}{M_{\rm mol}}
   \sum_{i=1}^{N_e}\svec_i,
   \label{XCMinverse}
\end{equation}
and $\rvec_i=\svec_i+\Xvec_{\rm CM}$.

The momenta conjugate to $(\Rvec_{\rm CM},\Yvec_\kappa,\svec_i)$ are
$(\Pvec_{\rm CM},\Qvec_\kappa,\qvec_i)$, which are functions of the
momenta $(\Pvec_t,\Qvec_\kappa,\pvec_i)$ after the first coordinate
transformation.  Of these $\Pvec_{\rm CM}$,
the total molecular momentum, is given by (\ref{PCMdef}) and
$\Qvec_\kappa$ does not change from the first transformation (it is
given by (\ref{Qkappadef})).  As for $\qvec_i$, it is given by
\begin{equation}
   \qvec_i=\pvec_i-\frac{m_e}{M_{\rm mol}}(\Pvec_t+\pvec_t).
   \label{qidef}
\end{equation}
The inverse transformations are
\begin{equation}
   \Pvec_t = \frac{M_t}{M_{\rm mol}}\Pvec_{\rm CM} -
   \sum_{i=1}^{N_e}\qvec_i,
   \label{Ptinverse}
\end{equation}
and
\begin{equation}
   \pvec_i = \frac{m_e}{M_{\rm mol}}\Pvec_{\rm CM}+\qvec_i.
   \label{piinverse}
\end{equation}  

\section{Covariant Notation for  the Kinetic Energy Metric}
\label{kemetric}

We write $X_{\alpha I}$ for the $I$-th component of Jacobi vector
$\Xvec_\alpha$, for $I=1,2,3$, and we write $x^\mu = X^\mu = X_{\alpha
I}$, where $\mu=(\alpha I)$ and $\mu=1,\ldots,3N-3$.  The kinetic
energy can be expressed in terms of the metric,
\begin{equation}
	G_{\mu\nu} = G_{(\alpha I)(\beta J)} = M_\alpha \,
	\delta_{\alpha\beta}\, \delta_{IJ},
	\label{G_munudef}
\end{equation}
that is, the kinetic energy itself is $(1/2)G_{\mu\nu}\,\xdot^\mu
\xdot^\nu$, where here and throughout this article we use the
summation convention on indices $\mu$, $\nu$, etc.  The tensor
$G_{\mu\nu}$ (with lower indices) is the covariant metric tensor.  The
contravariant metric  tensor (with upper indices) is the inverse of
(\ref{G_munudef}),
\begin{equation}
	G^{\mu\nu} = G^{(\alpha I)(\beta J)} =
	\frac{1}{M_\alpha}\,\delta_{\alpha\beta} \,
	\delta_{IJ},
   \label{Guppermunudef}
\end{equation}
so that $G^{\mu\sigma}\,G_{\sigma\nu} = \delta^\mu_\nu$.  

As for the momenta, we define $P_{\mu}=P_{(\alpha I)}$ as the $I$-th
component of the momentum $\Pvec_\alpha$, and we define $P^\mu$ (with
an upper index) as
\begin{equation}
	P^\mu = G^{\mu\nu}\,P_\nu = \frac{P_{\alpha I}}{M_\alpha},
	\label{P^mudef}
\end{equation}
where $\mu=(\alpha I)$, so that $P^\mu$, with an upper or
contravariant index, is the velocity $V^\mu=\xdot^\mu$.   Similarly we
define  the derivative operators,
\begin{equation}
	\partial_\mu = \frac{\partial}{\partial x^\mu},
	\qquad
	\partial^\mu = G^{\mu\nu}\,\partial_\nu,
	\label{partialmudef}
\end{equation}
so that as an operator, the momentum is $\Phat_\mu = -i\hbar
\,\partial_\mu$.  Equations (\ref{P^mudef}) and (\ref{partialmudef})
are examples of raising an index with the metric tensor.  

The kinetic energy can be expressed in terms of the momenta by
\begin{equation}
	\frac{1}{2}\sum_{\alpha=1}^N
	\frac{\Pvec_\alpha^2}{M_\alpha}
	= \frac{1}{2} G^{\mu\nu}\, P_\mu P_\nu
	= \frac{1}{2} P^\mu P_\mu.
	\label{keintermsofP}
\end{equation}

\section{Derivation of $G_2$}
\label{derivG2}

In this Appendix we use the working basis (\ref{workingbasisdef}) and
for brevity we write the resolvent $R\bigl(\epsilon_n(x),x\bigr)$ as
$R(x)$.  

The condition on $G_2$, that it kill the second-order, off-block
diagonal terms in $H'$, is expressed by (\ref{G2cond}), where
$H^{\prime\,o}_2$ is given by (\ref{H'_2odd}).  As in the solution for
$G_1$, we find that (\ref{G2cond}) is consistent with the
anti-Hermiticity of $G_2$ and that it determines the odd
(off-block-diagonal) part of $G_2$, We set the even part to zero so
that $G_2$ is purely odd.  The solution for the $(nk)$-block, $k\ne
n$, is
\begin{equation}
  G_{2,nk}= \sum_{l\ne n}
  (H^{\prime\,o}_2)_{nl}\,R_{lk}(\epsilon_n,x), \qquad (k\ne n),
  \label{G2nksoln}
\end{equation}
which may be compared to (\ref{G1nkdiab}).  The other off-diagonal
block is given by anti-Hermiticity, $G_{2,kn}=-G_{2,nk}^*$, $k\ne n$.
To find $G_2$, we will first find $(H^{\prime\,o}_2)_{nk}$ for $k\ne
n$ and then apply (\ref{G2nksoln}).

According to (\ref{H'_2odd}) there are three terms in
$(H^{\prime\,o}_2)_{nk}$ for $k\ne n$.  The first is
\begin{equation}
  H_{2,nk} =
  \frac{1}{2}(\partial^\mu\bra{x;n})(\partial_\mu\ket{x;k}),
	\qquad (k\ne n),
  \label{H2nk}
\end{equation}
a special case of (\ref{Hmol2defWeyl1}).  This term represents a
purely multiplicative operator, that is, it is independent of the
nuclear momenta $P_\mu$.  Notice that it involves derivatives
$\partial_\mu\ket{x;k}$ of the diabatic basis on $\Sspace^\perp$,
whose problematical aspects are discussed in
Sec.~\ref{changebasisSperp}.  The derivatives $\partial^\mu\ket{x;n}$
of the basis vector on $\Sspace$ present no such difficulties.  The latter
can be expressed in terms of the basis vector $\ket{x;n}$ itself, via
the Feynman-Hellman formula (\ref{FHcov}).  Similar formulas can be
worked out for higher derivatives of the basis vector $\ket{x;n}$ on
$\Sspace$.

For the second term of (\ref{H'_2odd}) we require
\begin{equation}
  [G_1,H^e_1]_{nk} = \sum_l G_{1,nl}\,H^e_{1,lk}
  -\sum_l H^e_{1,nl}\, G_{1,lk}, \qquad (k\ne n),
  \label{H'_2odd2ndterm}
\end{equation}
where the first sum can be restricted to $l\ne n$ since $G_{1,nn}=0$
and where the second sum can be restricted to $l=n$ since $H^e_1$ is
even.  In fact, the second sum vanishes, since $H_{1,nn}=0$.   As for
the first sum, we use (\ref{G1nkcov}) and (\ref{Hmol1defWeyl1}),
whereupon the sum can be extended to $l=n$ and a resolution of the
identity removed.   The result is
\begin{equation}
  [G_1,H^e_1]_{nk} = P^\mu P^\nu (\partial_\mu\bra{x;n})
  R(x) (\partial_\nu \ket{x;k}),
  \qquad (k\ne n),
  \label{H'2odd2ndtermv2}
\end{equation}
another expression  that involves derivatives of $\ket{x;k}$ on
$\Sspace^\perp$ but now the symbol of a second-order differential
operator.   

We see that $(H^{\prime\,o}_2)_{nk}$ contains terms that are independent
of the momentum $P^\mu$, that is, constant in the momentum, and terms
that are quadratic in the momentum.  In fact, it turns out that this
is all it contains, so we will write $(H^{\prime\,o}_2)_{nk} =
(H^{\prime\,o}_2)^C_{nk} + (H^{\prime\,o}_2)^Q_{nk}$ for the two
parts.  Then (\ref{H2nk}) is a contribution to
$(H^{\prime\,o}_2)^C_{nk}$ and (\ref{H'2odd2ndtermv2}) is a
contribution to $(H^{\prime\,o}_2)^Q_{nk}$.
  
For the third term of (\ref{H'_2odd}) we require $T_{11,nk}$ for $k\ne
n$, which by (\ref{T11def}) can be written
\begin{equation}
  T_{11,nk} = \frac{i}{2}\sum_l \{G_{1,nl},H_{0,lk}\}
  +\frac{i}{2}\sum_l \{G_{1,lk},H_{0,nl}\},
  \label{H'_2odd3rdtermv2}
\end{equation}
where we have used the antisymmetry of the Poisson bracket in the
second sum.  It is not hard to see that the kinetic energy part of
$H_0$ (see (\ref{Hmol0defWeyl1})) produces a contribution to
$T_{11,nk}$ that is quadratic in $P^\mu$ while the potential energy
produces one that is independent of $P^\mu$, so we shall write
$T_{11,nk}=T^C_{11,nk} + T^Q_{11,nk}$ for the two contributions.

The kinetic energy part of $H_0$ is a multiple of the identity which
allows the sums in (\ref{H'_2odd3rdtermv2}) to be done, whereupon they
turn out to be equal.  Thus we find
\begin{equation}
  T^Q_{11,nk} = i\{G_{1,nk}, (1/2)P^\mu P_\mu\}=
  -P^\mu P^\nu \partial_\nu [(\partial_\mu \bra{x;n})
  R(x)\ket{x;k}], \qquad (k\ne n),
  \label{H'_2odd3rdtermQ}
\end{equation}
where we have used (\ref{G1nkcov}) and evaluated the Poisson bracket.
When this is added to (\ref{H'2odd2ndtermv2}) we obtain the
quadratic part of $(H^{\prime\,o}_2)_{nk}$,
\begin{equation}
	(H^{\prime\,o}_2)^Q_{nk} =
	-P^\mu P^\nu\, \partial_\nu [(\partial_\mu \bra{x;n})
        R(x)]\ket{x;k}. \qquad (k\ne n),
  \label{H'_2oddQ}
\end{equation}
in which all derivatives of the basis vectors $\ket{x;k}$ on
$\Sspace^\perp$ have disappeared.  The tensor contracting against
$P^\mu P^\nu$ is not symmetric in $(\mu\nu)$ but it may be
replaced by its symmetric part.
	
As for the potential energy contribution to (\ref{H'_2odd3rdtermv2}),
we find that the first sum in (\ref{H'_2odd3rdtermv2}) can be
restricted to $l\ne n$ and the second to $l=n$, since $H_0$ is purely
even.  This gives
\begin{eqnarray}
	T^C_{11,nk}&=& -\frac{1}{2}\sum_{l\ne n} \{P^\mu
	(\partial_\mu\bra{x;n})R(x)\ket{x;l},
	W_{lk}(x)\}
	-\frac{1}{2}\{P^\mu(\partial_\mu\bra{x;n})R(x)
	\ket{x;k},\epsilon_n(x)\}
	\nonumber\\
	&=& \frac{1}{2}\sum_{l\ne n} (\partial_\mu\bra{x;n})
	R(x)\ket{x;l}[\partial^\mu W_{lk}(x)]
	+\frac{1}{2}(\partial_\mu\bra{x;n})R(x)\ket{x;k}
	[\partial^\mu\epsilon_n(x)]
	\nonumber\\
	&=& -\frac{1}{2}\sum_{l\ne n} (\partial_\mu\bra{x;n})
	R(x)\ket{x;l} \partial^\mu[\epsilon_n(x)
	\,\delta_{lk}-W_{lk}(x)]
	\nonumber\\
	&&\qquad +(\partial_\mu\bra{x;n})R(x)
	\ket{x;k}[\partial^\mu \epsilon_n(x)],
	\qquad (k\ne n),
	\label{TC11nk}
\end{eqnarray}
where in the first step we have evaluated the Poisson brackets and in
the second we have added and subtracted a term to make the derivative
of the inverse of the resolvent appear.	 

Now $(H^{\prime\,o}_2)^C_{nk}$ is the sum of (\ref{H2nk}) and the two
terms on the right of (\ref{TC11nk}).  We expand the derivative
appearing there,
\begin{eqnarray}
  && \partial^\mu[\epsilon_n(x)\,\delta_{lk}-W_{lk}(x)]=
  \partial^\mu\left\{\matrixelement{x;l}{[\epsilon_n(x)-H_e(x)]}{x;k}
  \right\}\nonumber\\
  &&\qquad =(\partial^\mu\bra{x;l})[\epsilon_n(x)-H_e(x)]\ket{x;k}
  +\matrixelement{x;l}{\left\{\partial^\mu[\epsilon_n(x)-H_e(x)]\right\}}
  {x;k}\nonumber\\
  &&\qquad +\bra{x;l}[\epsilon_n(x)-H_e(x)](\partial^\mu\ket{x;k}),
  \label{leftmiddleright}
\end{eqnarray}
giving what we will call the ``left,'' ``middle'' and ``right''
contributions to the first term on the right of (\ref{TC11nk}).  The
right contribution to that term is
\begin{equation}
  -\frac{1}{2}\sum_{l\ne n}
  (\partial_\mu\bra{x;n})R(x)\ket{x;l}
  \bra{x;l}[\epsilon_n(x)-H_e(x)](\partial^\mu\ket{x;k})=
  -\frac{1}{2}(\partial_\mu\bra{x;n})(\partial^\mu\ket{x;k},
  \label{rightreduction}
\end{equation}
where we have extended the $l$ sum to include $l=n$ (since
$R(x)\ket{x;n}=0$), removed the resolution of the identity, applied
(\ref{RHeqn}) and used
$(\partial_\mu\bra{x;n})Q(x)=\partial_\mu\bra{x;n}$.  We see that the
right contribution cancels (\ref{H2nk}).  

We simplify the middle contribution to the first term on the right in
(\ref{TC11nk}) by first removing the same resolution of the identity
and then applying $\partial^\mu$ to the identity (\ref{RHeqn}).  Thus
the middle contribution becomes 
\begin{eqnarray}
  && -\frac{1}{2}(\partial_\mu\ket{x;n})\,R(x)\,
  \{\partial^\mu[\epsilon_n(x)-H_e(x)]\}\ket{x;k}
  \nonumber\\
  &&\qquad =-\frac{1}{2}(\partial_\mu\bra{x;n})[\partial^\mu Q(x)]\ket{x;k}
  +\frac{1}{2} (\partial_\mu\bra{x;n})[\partial^\mu R(x)]
  [\epsilon_n(x)-H_e(x)]\ket{x;k}.
  \label{middlereduction}
\end{eqnarray}
The term involving $\partial^\mu Q(x) = -\partial^\mu P(x) =
-(\partial^\mu\ket{x;n})\bra{x;n} - \ket{x;n}(\partial^\mu\bra{x;n})$
vanishes since $\braket{x;n}{x;k}=0$ and
$(\partial_\mu\bra{x;n})\ket{x;n} = -F_{\mu;nn}=0$. Altogether, we now
have
\begin{eqnarray}
  (H^{\prime\,o}_2)^C_{nk} &=&
  [\partial^\mu\epsilon_n(x)]\,(\partial_\mu\bra{x;n})
  \,R(x)\ket{x;k}\nonumber\\
  &&\qquad -\frac{1}{2}\sum_{l\ne n}(\partial_\mu\bra{x;n})R(x)
  \ket{x;l}\,(\partial^\mu\bra{x;l})[\epsilon_n(x)-H_e(x)]
  \ket{x;k}\nonumber\\
  &&\qquad +\frac{1}{2}(\partial_\mu\bra{x;n})\,[\partial^\mu R(x)]\,
  [\epsilon_n(x)-H_e(x)]\ket{x;k}, \qquad (k\ne n).
  \label{H'odd2nkv3}
\end{eqnarray}

Now we may return to (\ref{G2nksoln}) and compute $G_2$.   It also has
constant and quadratic contributions, referring to the dependence on
$P^\mu$.  The execution of the sum in (\ref{G2nksoln}) amounts to
multiplying the operators seen in the matrix elements in $(H^{\prime\,
o}_2)_{nk}$ on the right by $R(x)$, for example, (\ref{H'_2oddQ})
gives the quadratic contribution to $G_2$ as
\begin{equation}
  (G^Q_2)_{nk} = -P^\mu P^\nu\,
    \partial_\nu[(\partial_\mu\bra{x;n})\, R(x)]\,R(x)\ket{x;k},
    \qquad (k\ne n),
  \label{G2Q}
\end{equation}
while the first term on the right in (\ref{H'odd2nkv3}) gives the
contribution $[\partial^\mu\epsilon_n(x)]\,(\partial_\mu\bra{x;n})
\,R(x)^2\ket{x;k}$ to $(G^C_2)_{nk}$.   As for the second and third
terms on the right in (\ref{H'odd2nkv3}), applying the sum in
(\ref{G2nksoln}) effectively cancels the factor of
$[\epsilon_n(x)-H_e(x)]$, first replacing it by $Q(x)$ which then
disappears since $Q(x)\ket{x;k}=\ket{x;k}$.   Thus the third term on
the right in (\ref{H'odd2nkv3}) gives the contribution to
$(G^C_2)_{nk}$,
\begin{equation}
	\frac{1}{2}(\partial_\mu\bra{x;n})\,[\partial^\mu R(x)]
	\ket{x;k}.
	\label{GC2thirdcontrib}
\end{equation}
As for the second term on the right in (\ref{H'odd2nkv3}), it gives
\begin{equation}
	-\frac{1}{2}\sum_{l\ne n}(\partial_\mu\bra{x;n})R(x)
  \ket{x;l}\,(\partial^\mu\bra{x;l})\ket{x;k}
	=\frac{1}{2}(\partial_\mu\bra{x;n})R(x)\,
	(\partial^\mu\ket{x;k}).
	\label{GC2secondcontrib}	
\end{equation}
This contribution to $G_2$ contains derivatives of the basis vectors
$\ket{x;k}$ on $\Sspace^\perp$ that do not cancel.   These pieces of
$G_2$ are assembled and the final solution is displayed in (\ref{G2soln}).

\bibliography{../../rl.bib}

\end{document}